\documentclass[fleqn,usenatbib]{mnras}

\usepackage{newtxtext,newtxmath}

\usepackage[T1]{fontenc}

\DeclareRobustCommand{\VAN}[3]{#2}
\let\VANthebibliography\thebibliography
\def\thebibliography{\DeclareRobustCommand{\VAN}[3]{##3}\VANthebibliography}

\usepackage{graphicx}	
\usepackage{amsmath}	
\usepackage{physics}
\usepackage{bm}
\usepackage{comment}
\usepackage{pdflscape}
\usepackage[dvipsnames]{xcolor}
\usepackage{xspace}

\defcitealias{Osato2023}{Paper~I}

\newcommand{\Msun}{\mathrm{M}_{\sun}}

\newcommand{\hMpc}{h^{-1} \, \mathrm{Mpc}}

\newcommand{\HA}{H$\alpha$\xspace}
\newcommand{\OII}{[\ion{O}{ii}]\xspace}

\title[Clustering of ELGs with IllustrisTNG -- II]
{Clustering of emission line galaxies with IllustrisTNG -- II.
cosmology challenge with anisotropic correlation functions and ELG-halo connections}

\author[K. Osato and T. Okumura]{
Ken Osato$^{1,2,3,4}$\thanks{E-mail: ken.osato@chiba-u.jp}
and Teppei Okumura$^{5,4}$
\\
$^{1}$Center for Frontier Science, Chiba University,
1-33 Yayoi-cho, Inage-ku, Chiba 263-8522, Japan\\
$^{2}$Department of Physics, Graduate School of Science, Chiba University, 1-33 Yayoi-cho, Inage-ku, Chiba 263-8522, Japan\\
$^{3}$RIKEN Center for Advanced Intelligence Project, 1-4-1 Nihonbashi, Chuo-ku, Tokyo 103-0027, Japan\\
$^{4}$Kavli Institute for the Physics and Mathematics of the Universe, The University of Tokyo Institutes for Advanced Study,\\
5-1-5 Kashiwanoha, Kashiwa-shi, Chiba, 277-8583, Japan\\
$^{5}$Academia Sinica Institute of Astronomy and Astrophysics,
No. 1, Section 4, Roosevelt Road, Taipei 10617, Taiwan
}

\date{Accepted XXX. Received YYY; in original form ZZZ}

\pubyear{2025}

\begin{document}
\label{firstpage}
\pagerange{\pageref{firstpage}--\pageref{lastpage}}
\maketitle

\begin{abstract}
Emission line galaxies (ELGs) are the primary tracers of the large-scale structures of the Universe
in ongoing Stage-IV cosmological spectroscopic surveys,
which aim to measure the clustering statistics at higher redshifts $z \simeq 1.5 \text{--} 2$
with unprecedented precision.
In this study, we construct realistic mock ELG samples with IllustrisTNG hydrodynamical simulations
and stellar population synthesis framework.
In order to validate the modelling of clustering, we measure the anisotropic correlation functions of mock ELGs
and infer the linear growth rate, which is one of key cosmological parameters in galaxy clustering.
As a control sample, we construct the mass-limited subhalo samples with the same number density as ELGs.
The isotropic correlation functions in real space for both samples do not differ significantly.
However, the quadrupole moment of the anisotropic correlation function,
which is sensitive to the velocity of galaxies, is suppressed for ELGs, potentially
due to the infalling motion of ELGs towards the centre of the hosting halos.
The smaller amplitude leads to the underestimation of the linear growth rate
and implies the velocity bias between ELGs and dark matter.
When the analysis is limited to large scales $(\gtrsim 15 \, h^{-1} \, \mathrm{Mpc})$,
the parameter bias vanishes.
Next, we investigate the ELG-halo connection through the phase-space distribution of satellite ELGs
within hosting halos and galactic conformity of star formation activity.
The infalling motion is further confirmed by the phase-space distribution relative to the host halo,
and this dynamics of ELGs challenges the assumption that the radial distribution of satellites
follows that of dark matter.
\end{abstract}




\section{Introduction}
\label{sec:introduction}
In the standard paradigm of cosmology, the density fluctuations are seeded in the early Universe
as quantum fluctuations during the inflationary era,
and a myriad of structures are hierarchically formed through gravitational instability
\citep{White1978,White1991,Cole2000}.
The gravitational evolution is driven primarily by dark matter,
and thus, the large-scale structure (LSS) of the Universe is one of the key probes into
the nature of dark matter and gravity theory \citep[for a review, see][]{Weinberg2013}.
Furthermore, in the late-time Universe, the expansion of the space-time
accelerates due to dark energy.
To investigate the geometry of the Universe, the sound horizon of cosmic plasma,
which is referred to as baryon acoustic oscillation (BAO) scale, has played an important role.
BAO is the oscillatory feature imprinted in the matter distribution and
serves as a cosmic standard ruler.
Through the statistical measurement of the BAO scale,
the physical nature of dark energy can be robustly addressed \citep{Eisenstein2005,Aubourg2015}.

The foremost challenge in observations of LSS is that dark matter is not visible from telescopes.
Hence, in the real measurements of the LSS,
galaxies are widely used as tracers of the underlying matter distribution,
which is the direct consequence of gravitational evolution from density perturbations.
On large scales, the approximation that
the galaxy number density contrast is linearly proportional to the matter density contrast holds well.
On the other hand, the relationship between matter and galaxy, i.e., \textit{galaxy-halo connection}
\citep[for a review, see][]{Wechsler2018}, becomes complex on small scales.
In practice, halo occupation distribution \citep[HOD;][]{Berlind2002,Kravtsov2004}
is the most popular model to connect galaxies from host dark matter halos.
HOD describes the probability distribution of the number of galaxies
as a function of the mass of the halo.
This approach is successful in reproducing the two-point correlation functions of galaxies
down to non-linear scales \citep{Zehavi2004}.
For example, \citet{Zheng2005} proposed a five-parameter HOD, and in \citet{Zheng2009}, the HOD is applied
to model the correlation function of luminous red galaxies \citep[LRGs;][]{Eisenstein2001}
observed by Sloan Digital Sky Survey.
LRGs are selected based on colours and magnitudes so that the selected galaxies are
a homogenous population with similar stellar mass and
their photometric redshift is accurately estimated with $4000 \, \text{\AA}$ break
\citep{Padmanabhan2007}.
A similar idea is brought to the sample selection of CMASS and LOWZ samples
\citep{Anderson2012, Reid2016} in Baryon Oscillation Spectroscopic Survey
\citep[BOSS;][]{Dawson2013}.
Since LRGs are massive and old systems, the number density field of LRGs
is highly biased against the matter density field, and the internal motion is well virialised.
Therefore, the HOD and resultant clustering of LRGs are less sensitive to galaxy formation physics.

On the other hand, the next generation of spectroscopic surveys,
which are referred to as Stage-IV spectroscopic surveys, including
Dark Energy Spectroscopic Instrument \citep[DESI;][]{DESI2022}, \textit{Euclid} \citep{EuclidI},
Subaru Prime Focus Spectrograph \citep[PFS;][]{Takada2014}, and
Roman Space Telescope \citep{Spergel2015,Akeson2019},
will target a different galaxy population: emission line galaxies (ELGs).
ELGs are blue and star-forming galaxies characterised by the bright line emission, such as
H$\alpha$ ($\lambda \, 6564.62 \, \text{\AA}$),
[\ion{O}{ii}] ($\lambda \lambda \, 3727.09 \, \text{\AA}, 3729.88 \, \text{\AA}$),
and [\ion{O}{iii}] ($\lambda \, 5008.24 \, \text{\AA}$).
These line emissions are produced from nebular gas irradiated
by massive OB-type stars \citep{Charlot2001,Kewley2019}.
ELGs have been utilised to diagnoise the cosmic star formation history
since the line emission reflects the star formation activity
on a short time scale up to $\sim 10 \, \mathrm{Myr}$ \citep{Byler2018}
corresponding to the lifetime of OB-type stars.
The reason why ELGs are the main target in the future surveys is that
these surveys will measure galaxy clustering at higher redshifts $(z \simeq 1.5 \text{--} 2)$,
where the star formation activity is at its peak
\citep{Madau2014,Forster2020}, i.e., \textit{cosmic noon}.
At this epoch, the expected number density of ELGs is high, and thus,
an accurate measurement of clustering statistics is possible.
In addition, bright line emission helps to determine the redshift accurately.
However, clustering of ELGs is not well studied compared with LRGs and
the nature of ELGs is expected to be quite different from that of LRGs.
In contrast to LRGs, ELGs are likely to be hosted by less massive and younger halos
and undergo infall towards the halo.
Hence, clustering of ELGs is subject to velocity bias, environment effects, and assembly bias.

Recently, several pilot observations targeting ELGs at $z \simeq 1 \text{--} 2$
have been conducted: HiZELS \citep{Geach2008,Sobral2012},
FastSound \citep{Tonegawa2015,Okada2016,Okumura2016}, and
narrow band observations with Subaru Hyper Suprime-Cam \citep{Hayashi2018,Hayashi2020,Okumura2021,Ishikawa2025}.
Subsequently, large spectroscopic survey programmes covering wider areas, e.g.,
eBOSS \citep{Tamone2020,Raichoor2021,deMattia2021}
and DESI \citep{DESI2024III,DESI2024V,DESI2024VI,DESI2024VII},
have measured clustering statistics of ELGs and constrained cosmological parameters
with two-point correlations.

To elucidate ELG clustering and ELG-halo connection from these observations,
empirical approaches such as HOD and subhalo abundance matching (SHAM)
have been widely utilised.
The observations of ELGs so far indicate that extensions beyond HOD are demanded
to improve the accuracy of modelling ELG clustering,
e.g., different functional forms to incorporate the infalling component
\citep{Geach2012,Gao2022},
adding physical parameters beyond halo mass
\citep{Hearin2016,Hadzhiyska2023a,Hadzhiyska2023b},
modifications to velocity distribution of satellite ELGs \citep{Avila2020,Rocher2023},
and constraining HOD parameters jointly with luminosity functions \citep{Ishikawa2025}.
Similarly, SHAM needs to be modified so that the star-formation rate (SFR) is properly modelled
\citep{Favole2022,Ortega-Martinez2024,Ortega-Martinez2025},
because the ELG sample is close to an SFR-limited sample
rather than a stellar mass-limited sample, which standard SHAM deals with.

As observations of ELGs advance, there is also substantial progress
in numerical and theoretical studies of ELG clustering with
hydrodynamical simulations \citep{Hadzhiyska2021,Yuan2022,Yuan2025} and
semi-analytic simulations \citep{GonzalezPerez2018,Orsi2018,Favole2020,GonzalezPerez2020}.
These simulations enable in-depth investigation of ELG-halo connection
since various halo properties are directly accessible.
Based on these simulations, it is found that
a large fraction of ELGs lie in filamentary structures and undergo infall toward
more massive halos,
and ELGs are hosted by younger and less concentrated halos \citep{Yuan2022},
which is the signature of galaxy assembly bias.

To scrutinise cosmology with ELGs and the ELG-halo connection,
we have constructed a high-fidelity ELG mock catalogue
based on IllustrisTNG simulations \citep{Nelson2019}
coupled with stellar population synthesis (SPS) code \texttt{P\'{E}GASE3} \citep{Fioc2019}
in our previous work \citep[][hereafter, \citetalias{Osato2023}]{Osato2023}.
In \texttt{P\'{E}GASE3}, nebular line emissions are derived from the precomputed table
by photo-ionization code \texttt{CLOUDY} \citep{Ferland2017},
which simulates synthesised spectra in the star-forming clouds.
The scheme is useful to interpret various line emissions of star-forming galaxies \citep{Gutkin2016}.
The nebular emissions are subject to dust extinction due to dust in birth clouds \citep{Charlot2000}
in addition to inter-stellar medium \citep[for a review, see][]{Salim2020}.
Since the dust formation and evolution is analytically modelled in \texttt{P\'{E}GASE3},
the extinction effect is properly taken into account.
We first construct a precomputed table of $\mathrm{H}\alpha$
and [\ion{O}{ii}] line emissions in the grids of metallicity and stellar age.
Then, we assign emissions from each stellar particle by looking up the table,
and sum all the contributions from member stellar particles in each simulated galaxy.
The mock ELG samples show agreement with observed line luminosity functions.
We have investigated HOD and projected correlation functions of mock ELGs in \citetalias{Osato2023}.
Recently, \citet{Rapoport2025} presented an approach to simulate $\mathrm{H}\alpha$ emission
with hydrodynamical simulations based on a simple radiative transfer model,
which is complementary to our approach utilising the SPS model.

In this work, we extend the analysis presented in \citetalias{Osato2023}
to address cosmology with ELGs and the ELG-halo connection.
First, we measure anisotropic correlation functions of ELGs in redshift space,
which are primary statistics to constrain cosmological parameters and gravity models.
Since ELGs entail dynamics within halos different from LRGs, i.e.,
infalling along filaments, that feature propagates to cosmological statistics.
Hence, it is not clear whether the current modelling of cosmological statistics
can infer unbiased cosmological parameters.
Thus, since the cosmological parameters are given in simulations,
we can conduct \textit{cosmology challenge} analysis to confirm
that the inferred cosmological parameters can be reproduced without bias.
This analysis is critical to validate the theoretical template of ELG clustering.
Next, we address ELG-halo connection with the mock ELG catalogues.
Since properties of halos hosting ELGs are readily available in simulations,
we closely investigate ELG-halo connection and identify the elements
which incur the difference in cosmological statistics.

This paper is organised as follows.
In Section~\ref{sec:simulations}, we outline how mock ELG catalogues are
built from IllustrisTNG hydrodynamical simulations.
In Section~\ref{sec:theory}, theoretical modelling
of anisotropic correlation functions is explained.
We present the results of the cosmology challenge in Section~\ref{sec:cosmo_challenge}
and ELG-halo connection in Section~\ref{sec:connection}
We conclude in Section~\ref{sec:conclusions}.
In Appendix~\ref{sec:galaxy_bias}, the theoretical modelling
of galaxy power spectra in real space is described.
In Appendix~\ref{sec:conversion}, auxiliary analytic expressions for multipole expansion
are presented. For cosmology challenge, the results with \OII ELG sample are discussed
in the main text and the results with \HA ELG sample are shown in Appendix~\ref{sec:ha_results},
Throughout this paper, we assume the flat $\Lambda$ cold dark matter Universe
and following cosmological parameters adopted in IllustrisTNG simulations:
the baryon density $\Omega_\mathrm{b} = 0.0486$,
the matter density $\Omega_\mathrm{m} = 0.3089$,
the cosmological constant density $\Omega_\Lambda = 1-\Omega_\mathrm{m} = 0.6911$,
the scaled Hubble parameter $h = 0.6774$,
the tilt of the primordial perturbations $n_\mathrm{s} = 0.9667$, and
the amplitude of matter fluctuation at $8 \, h^{-1} \, \mathrm{Mpc}$ $\sigma_8 = 0.8159$.

\section{Construction of mock ELG samples}
\label{sec:simulations}

\subsection{Modelling emission lines in hydrodynamical simulations}
First, we describe the scheme to compute emission line luminosities
for each simulated galaxy in hydrodynamical simulations.
For more details, refer to Section~2 of \citetalias{Osato2023}.

To construct mock ELG catalogues, we employ IllustrisTNG hydrodynamical simulations
\citep{Nelson2018,Pillepich2018b,Springel2018,Naiman2018,Marinacci2018,Nelson2019}.
The simulations are run with the moving mesh code \texttt{AREPO} \citep{Springel2010}
and various galaxy formation processes, such as star formation and feedback effects
due to supernovae and active galactic nuclei are implemented as a subgrid model
\citep{Weinberger2017,Pillepich2018a}.
The subgrid model parameters are calibrated so that the observations of galactic properties,
e.g., stellar mass functions, can be reproduced.
Among IllustrisTNG simulations with different box sizes and resolutions,
TNG300-1 simulation is utilised as the fiducial one.
The simulation volume is $(205 \, h^{-1} \, \mathrm{Mpc})^3$
and the number of dark matter particles and gas cells is $2500^3$,
which corresponds to the mass resolution of
$m_\mathrm{DM} = 5.9 \times 10^7 \, h^{-1} \, \Msun$
and $m_\mathrm{gas} = 1.4 \times 10^6 \, h^{-1} \, \Msun$
for dark matter and gas, respectively.
Halos and their substructures, i.e., subhalos, are identified with \texttt{SubFind}
algorithm \citep{Springel2001}.
As terminology, a simulated \textit{galaxy} corresponds to a \textit{subhalo} in simulations.

Next, we compute emission line luminosities by post-processing simulation outputs.
The stellar component of a simulated galaxy is expressed as an assembly of
stellar particles, each of which records the formation time, metallicity,
and the mass at the formation time.
For each stellar particle, line luminosity is calculated with the SPS code \texttt{P\'{E}GASE3}
\citep{Fioc2019}, where nebular emission is derived from the precomputed results of \texttt{CLOUDY}
\citep{Ferland2017}.
Thus, we construct a table of line emissions ($\mathrm{H}\alpha$ and [\ion{O}{ii}])
with respect to the stellar age and metallicity.
For the total line luminosity, we sum the contribution from each stellar particle,
which is bilinearly interpolated for the stellar age and metallicity from the table,
weighted with the mass.
Note that nebular line emission is subject to dust extinction
and the extinction must be incorporated.
Otherwise, the number density of ELGs is highly overestimated.
\texttt{P\'{E}GASE3} adopts a phenomenological dust evolution
model \citep[see Section~3.2 of][]{Fioc2019},
where the dust mass is proportional to the mass of inter-stellar medium.
Though the model is calibrated with Milky Way observations,
the resultant line luminosity functions match well with observations
at high redshifts $z = 1 \text{--} 2$ \citepalias[see Figure~2 of][]{Osato2023}.

\subsection{Emission line galaxy and mass-limited subhalo control samples}
\begin{figure}
  \includegraphics[width=\columnwidth]{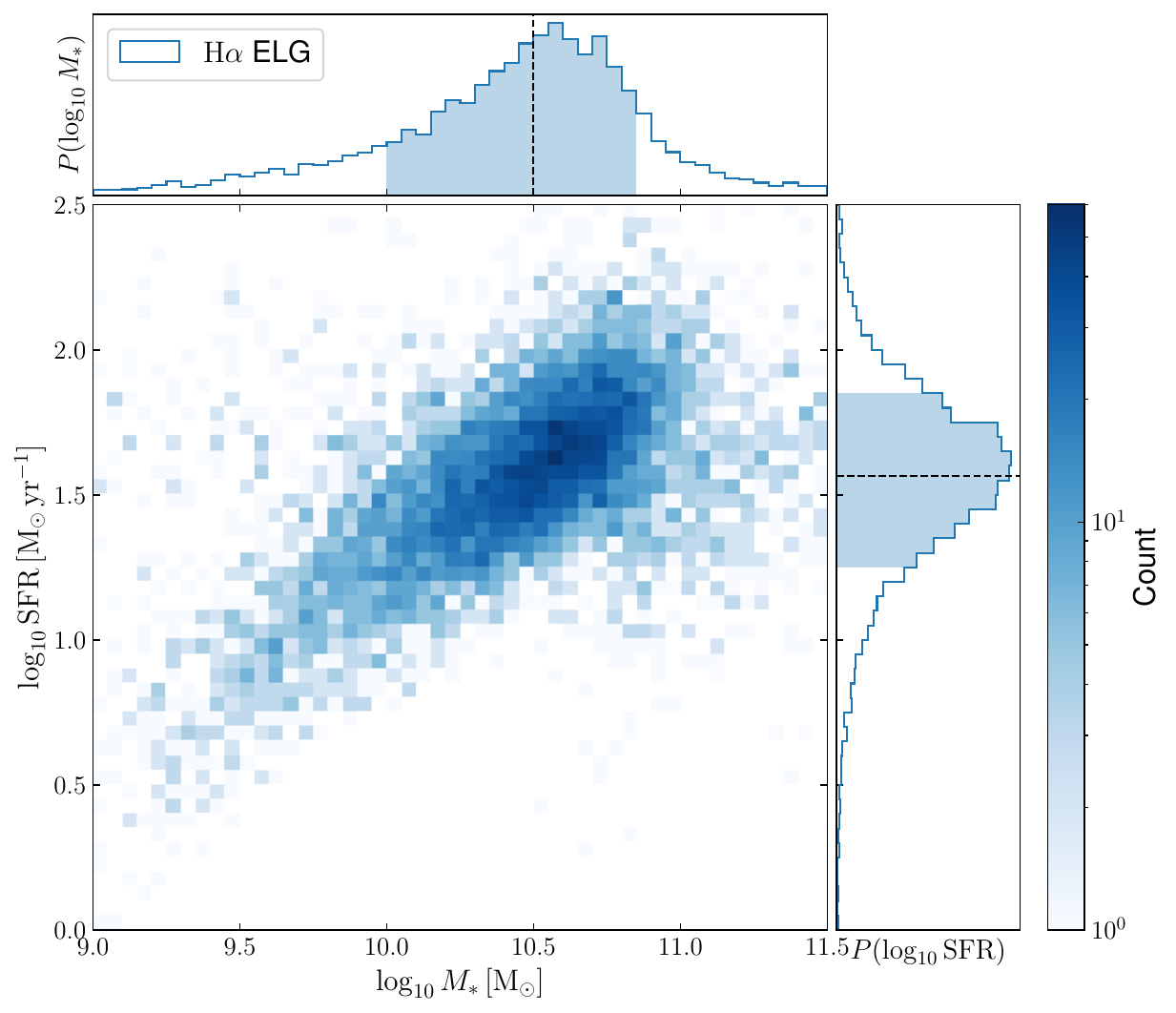}
  \includegraphics[width=\columnwidth]{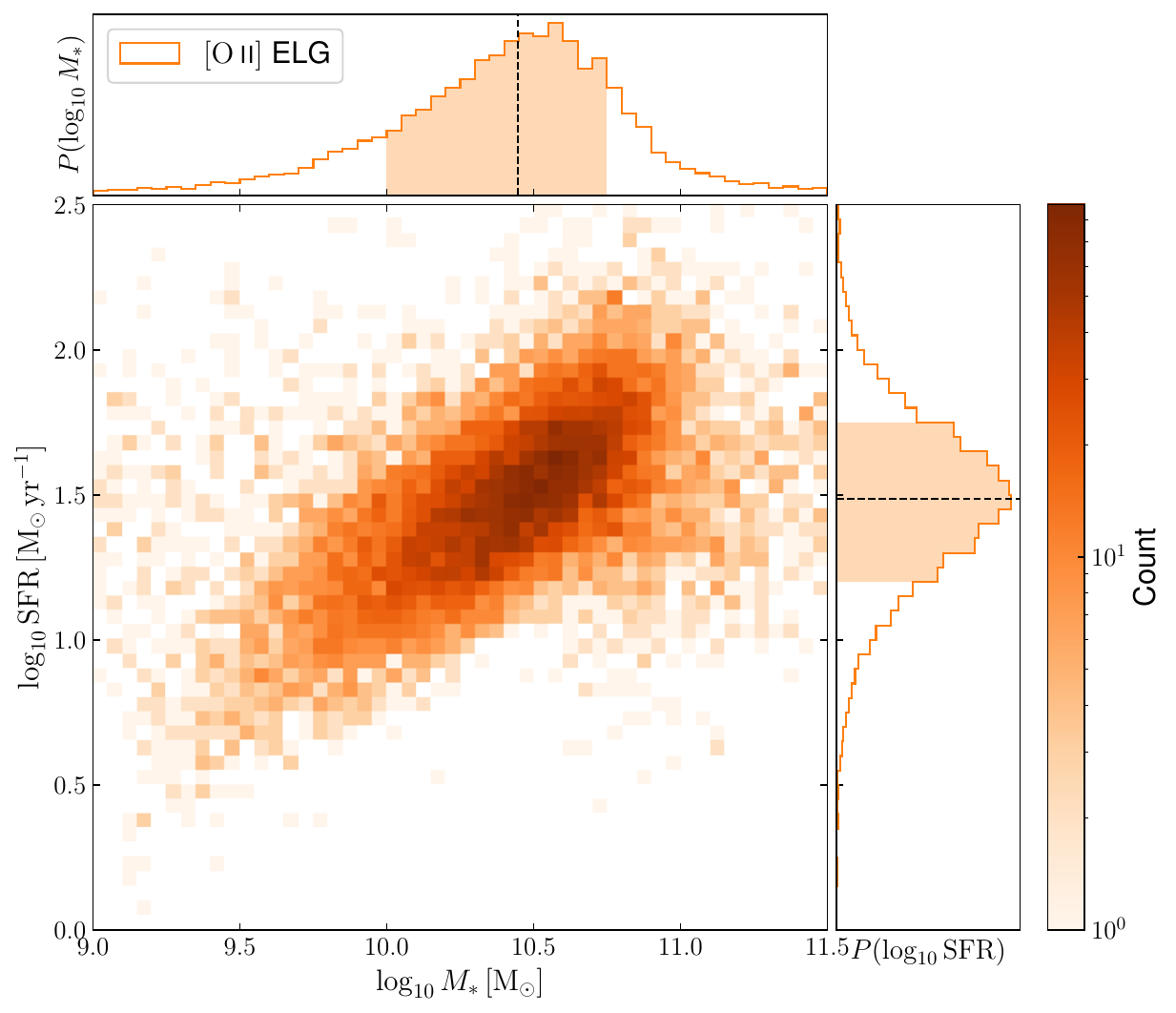}
  \caption{The distribution of stellar mass and SFR for \HA ELG (upper panel) and \OII ELG (lower panel) samples selected based on line luminosities.
  The marginalised distributions of stellar mass and SFR are shown
  in the top and right panels, respectively.
  The black dahsed lines correspond to the median,
  and the shaded regions indicate the 16th and 84th percentiles of the distributions.}
  \label{fig:Mstar_vs_SFR}
\end{figure}

First, we construct two target ELG samples selected with
\HA and \OII line luminosities
from the parent catalogue described in the previous section.
These are strong line emissions from star-forming nebulae and
are used to construct galaxy catalogues in the Stage-IV survey:
\HA for \textit{Euclid}, and \OII for DESI and PFS.
Hereafter, only the catalogue at the redshift $z = 1.5$ is considered,
which is a typical redshift for Stage-IV spectroscopic surveys.
In this work, galaxies with the line luminosity higher than
$2 \times 10^{42} \, \mathrm{erg} \, \mathrm{s}^{-1}$ or
$1 \times 10^{42} \, \mathrm{erg} \, \mathrm{s}^{-1}$
are defined as \HA or \OII ELGs, respectively.
The corresponding flux limit is
$6.71 \times 10^{-17} \, \mathrm{erg} \, \mathrm{s}^{-1} \, \mathrm{cm}^{-2}$
for \HA ELGs and
$1.34 \times 10^{-16} \, \mathrm{erg} \, \mathrm{s}^{-1} \, \mathrm{cm}^{-2}$
for \OII ELGs.
These flux limits are chosen to mimic the expected flux limit of \textit{Euclid} \HA
and PFS \OII ELGs.
The resultant \HA and \OII ELG samples
consist of 6,864 and 11,001 galaxies,
and the comoving number density is $7.97 \times 10^{-4} \, (h^{-1} \, \mathrm{Mpc})^{-3}$
and $1.28 \times 10^{-3} \, (h^{-1} \, \mathrm{Mpc})^{-3}$, respectively.
In the spectroscopic surveys with slit spectroscopy,
the observing targets are selected based on the photometric data
\citep[e.g.,][]{Raichoor2023}, and thus,
the selection is not purely based on line luminosity.
In our current modelling, the line contributions
from active galactic nuclei (AGNs) are not considered,
which hinders from accurately reproducing the galaxy colour distribution.
However, the line luminosity is the primary factor in the selection of ELGs,
and we adopt the line luminosity selection to construct mock ELG samples.
Figure~\ref{fig:Mstar_vs_SFR} shows the distribution of stellar mass
and SFR for the ELG samples.
The current selection based on line luminosity leads to
the minimum SFR of $\simeq 10^{0.5} \, \Msun \, \mathrm{yr}^{-1}$
and the minimum stellar mass of $\simeq 10^9 \, \Msun$ for both ELG samples.

To highlight the difference from ELGs, we also construct
control samples for \HA and \OII ELGs.
First, we sort the subhalos in the parent catalogue by their mass
at the redshift $z = 1.5$, and then select the most massive subhalos
until the number of selected subhalos reaches that of ELGs.
The resultant mass limit of the control samples for \HA and \OII ELGs is
$2.09 \times 10^{12} \, h^{-1} \, \Msun$ and
$1.51 \times 10^{12} \, h^{-1} \, \Msun$, respectively.
These samples exhibit the similar galaxy bias to the corresponding ELG samples
but are expected to indicate the different nature
in velocity and statistics in redshift space.
In addition, the theoretical model of clustering statistics
for subhalos is well established, and thus,
the control samples serve as the baseline for evaluating
the performance of the theoretical model when applied to ELG clustering.

\section{Theoretical model for anisotropic correlation functions}
\label{sec:theory}
Here, we describe the theoretical model of the anisotropic correlation functions.
In spectroscopic surveys, the line-of-sight (LoS) positions of galaxies are estimated
from the redshift.
The redshift estimation is distorted by the peculiar velocity of the galaxy,
which is referred to as redshift space distortion (RSD).
This effect induces the anisotropy in clustering statistics
and needs to be incorporated in the theoretical modelling of clustering statistics
for unbiased estimation of cosmological parameters.

\begin{figure*}
  \includegraphics[width=\columnwidth]{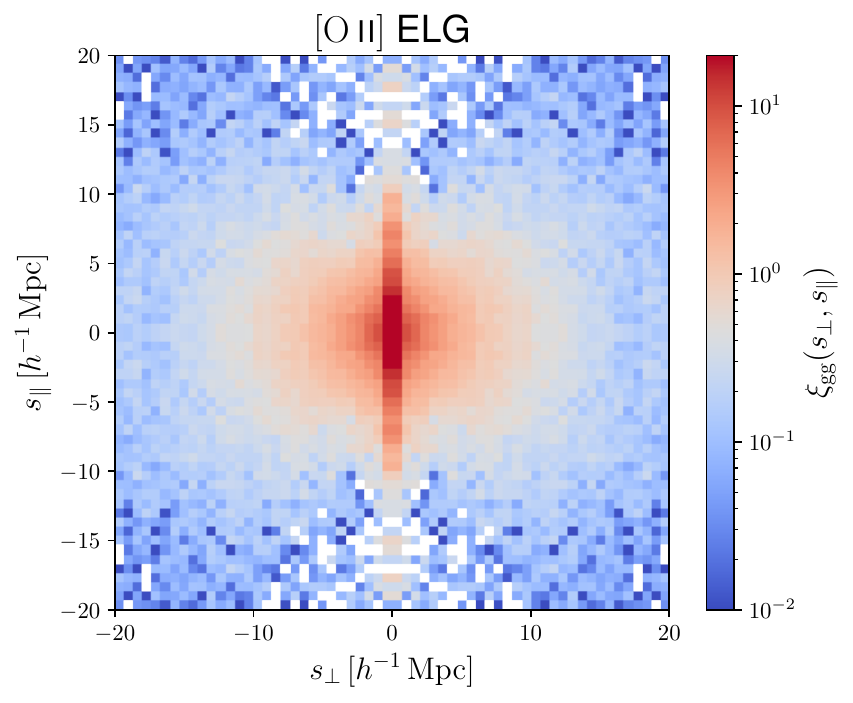}
  \includegraphics[width=\columnwidth]{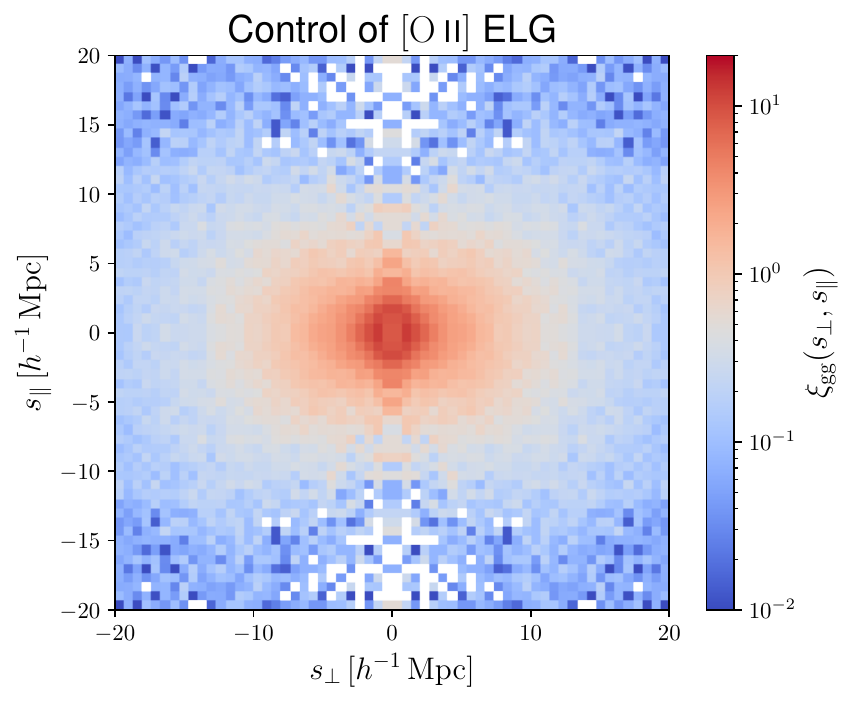}
  \caption{The 2D anisotropic correlation functions of \OII ELGs and
  the corresponding control sample.
  The white pixels correspond to the no pair in the measurement.}
  \label{fig:xi2D}
\end{figure*}

First, we begin with galaxy power spectra in redshift space.
We employ the extended Taruya--Nishimichi--Saito (eTNS) model
\citet{Taruya2010,Taruya2013}.
The galaxy power spectrum in this model is given as
\begin{multline}
  \label{eq:galaxy_power}
  P_\mathrm{gg} (k, \mu) = D_\mathrm{FoG}(k\mu \sigma_\mathrm{v})
  \left[ P_{\delta_\mathrm{g} \delta_\mathrm{g}} (k)
  + 2 f \mu^2 P_{\delta_\mathrm{g} \theta_\mathrm{g}} (k)
  \right. \\
  \left. + f^2 \mu^4 P_{\theta_\mathrm{g} \theta_\mathrm{g}} (k) +
  b_1^3 A (k, \mu, \beta) + b_1^4 B (k, \mu, \beta) \right] ,
\end{multline}
where $D_\mathrm{FoG}$ is the Finger-of-God (FoG) damping function,
$\delta_\mathrm{g}$ is the density contrast of galaxies,
$\theta_\mathrm{g}$ is the velocity divergence field of galaxies,
$P_{\delta_\mathrm{g} \delta_\mathrm{g}} (k)$,
$P_{\delta_\mathrm{g} \theta_\mathrm{g}} (k)$,
and $P_{\theta_\mathrm{g} \theta_\mathrm{g}} (k)$
are the density-density auto-, density-velocity cross-,
and velocity-velocity auto-power spectra in real space, respectively,
and $b_1$ is the linear bias parameter.
Here, $\mu$ is the cosine of the angle between the wave vector $\bm{k}$ and the LoS direction,
and the velocity divergence of galaxies $\theta_\mathrm{g}$
is expressed as $\theta_\mathrm{g} = \nabla \cdot \bm{v}_\mathrm{g} / (aH)$,
where $\bm{v}_\mathrm{g}$ is the peculiar velocity of galaxies, $a$ is the scale factor,
and $H$ is the Hubble parameter.
The second and third terms in Eq.~\eqref{eq:galaxy_power} correspond to the Kaiser effect \citep{Kaiser1987},
which enhances the clustering signal at the large scale due to coherent infall
towards the potential well.
The linear growth rate $f = d \log D_+ / d \log a$,
where $D_+$ is the linear growth factor,
regulates the enhancement and is sensitive to the underlying gravity theory.
In contrast to the Kaiser effect,
the FoG effect refers to smearing at small scales due to random motions of galaxies \citep{Jackson1972},
and the damping feature is well explained by the phenomenological model to suppress
the amplitude of the power spectrum.
We adopt the Lorentzian form of FoG damping function $D_\mathrm{FoG} (x) = (1+x^2/2)^{-2}$.
The last terms, $A(k, \mu, \beta)$ and $B(k, \mu, \beta)$ with $\beta = f/b_1$,
are non-linear coupling correction terms presented in \citet{Taruya2010} and the details will be described later.
All the expressions of power spectra and correction terms depend on the redshift $z$,
but we omit its dependence hereafter for notational simplicity.

For modelling of real-space galaxy power spectra,
$P_{\delta_\mathrm{g} \delta_\mathrm{g}} (k)$,
$P_{\delta_\mathrm{g} \theta_\mathrm{g}} (k)$,
and $P_{\theta_\mathrm{g} \theta_\mathrm{g}} (k)$,
we adopt the model presented in \citet{McDonald2009}
based on galaxy bias expansion.
Since the model requires the non-linear power spectra of matter density and velocity,
the non-linear matter power spectrum is computed with \texttt{HMCODE-2020} \citep{Mead2021},
and for velocity spectrum, the fitting formula by \citet{Bel2019} is employed.
The details of the model are presented in Appendix~\ref{sec:galaxy_bias}.
One of the key assumptions in this model is no velocity bias for galaxies,
i.e., $\theta_\mathrm{g} = \theta$, where $\theta$ is
the velocity divergence field of matter.
This relation may not hold for ELGs
and will be addressed through estimation of the linear growth rate
because the break of this relation appears as the bias in the inferred linear growth rate.
Note that these power spectra can be directly measured from simulations,
and thus, it is possible to use the measured power spectra
in the theoretical template of the galaxy power spectrum.
However, the measured power spectra are noisy due to the small box size of simulations, and thus, we adopt the modelled power spectra to obtain smooth templates.
In addition, similarly to the real cosmological analysis, we adopt the predictive model for the power spectra to infer cosmological parameters, which is the main purpose of this work.

Finally, let us consider the non-linear coupling terms $A(k, \mu, \beta)$ and $B(k, \mu, \beta)$
based on perturbation theory \citep{Taruya2010}.
The terms are computed at the 1-loop order with regularized perturbation theory \citep{Taruya2012,Taruya2013}.
In the presence of the linear bias, the coupling terms scale as
\begin{align}
  A(k, \mu, f) &\to b_1^3 A(k, \mu, \beta),\\
  B(k, \mu, f) &\to b_1^4 B(k, \mu, \beta).
\end{align}

In total, there are four bias parameters ($b_1, b_2, b_{s2}, b_\mathrm{3nl}$) in galaxy power spectrum.
However, these parameters are tightly degenerate with cosmological parameters,
especially for the limited survey volume.
In order to obtain converged results, we adopt the local Lagrangian relation for $b_{s2}$ and $b_{3\mathrm{nl}}$ \citep{Baldauf2012,Chan2012,Sheth2013,Saito2014,Mirbabayi2015}:
\begin{align}
  \label{eq:bs2}
  b_{s2} &= -\frac{4}{7} (b_1 - 1), \\
  \label{eq:b3nl}
  b_{3\mathrm{nl}} &= \frac{32}{315} (b_1 - 1) .
\end{align}
For the second order bias $b_2$, we adopt the fitting formula presented in \citet{Lazeyras2016}:
\begin{equation}
  \label{eq:b2}
  b_2 = 0.412 - 2.143 b_1 + 0.929 b_1^2 + 0.008 b_1^3 .
\end{equation}
The relation is calibrated with halos but holds well for galaxies selected with
star formation rate in hydrodynamical simulations \citep{Barreira2021}.
In summary, there is only one free bias parameter $b_1$
and other bias parameters are derived from Eqs.~\eqref{eq:bs2}, \eqref{eq:b3nl}, and \eqref{eq:b2}.

The multipole expansion of the galaxy power spectrum is given as
\begin{equation}
  \label{eq:multipole_expansion}
  P_\ell (k) = \frac{2 \ell+1}{2} \int_{-1}^{+1} \dd{\mu}
  P_\mathrm{gg}(k, \mu) L_\ell (\mu) ,
\end{equation}
where $L_\ell (\mu)$ is the $\ell$-th order Legendre polynomial.
The dependence on the directional cosine $\mu$ in the eTNS 2-loop model except the FoG damping function
is the polynomial at even degrees up to the octic.
The integration can be done analytically, and the conversion formulae are presented in Appendix~\ref{sec:conversion}.
Then, the multipoles of the anisotropic correlation functions are obtained through Hankel transformation:
\begin{equation}
  \xi_\ell (s) = i^\ell \int \frac{k^2 \dd{k}}{2\pi^2} P_\ell (k) j_\ell (k s) ,
\end{equation}
where $j_\ell (x)$ is the $\ell$-th order spherical Bessel function.
We utilise \texttt{FFTLog} \citep{Hamilton2000} algorithm to compute Hankel transformation efficiently.
Due to the symmetry in directions perpendicular to the LoS direction, the moments at odd orders vanish.
The moments at higher orders are expected to be noisy, and thus, we focus only on
the monopole ($\ell = 0$) and quadrupole ($\ell = 2$) moments for the subsequent analysis.

There is one caveat about Alcock--Paczynski (AP) effect \citep{Alcock1979}.
The largest scale in the correlation function is limited by the box size of simulations,
i.e., $205 \, h^{-1} \, \mathrm{Mpc}$,
and thus, the BAO scale cannot accurately be probed in the analysis.
Hence, we do not include the AP effect in the theoretical modelling of anisotropic correlation functions.

\section{Cosmology challenge with the anisotropic correlation functions}
\label{sec:cosmo_challenge}

\subsection{Mock measurements of the correlation functions}
Here, we present the scheme to measure the anisotropic correlation functions from the ELG catalogues.
In this Section, we focus on the \OII ELG sample
and the corresponding mass-limited control sample.
The results for \HA ELG sample and the control sample
are similar to \OII ELG sample, and the selected key results are presented
in Appendix~\ref{sec:ha_results}.
Since the random pair count is analytically computed under the periodic boundary condition,
we employ the natural estimator of the correlation function \citep{Peebles1974}:
\begin{equation}
  \hat{\xi}_\mathrm{gg} (s_\perp, s_\parallel) = \frac{DD}{RR} - 1 ,
\end{equation}
where $s_\perp$ ($s_\parallel$) is the separation distance of the pair
perpendicular (parallel) to the LoS direction in the redshift space,
$DD$ and $RR$ is the pair counts of ELGs and randoms normalised by the total pair counts,
respectively.
The perpendicular and parallel separation is binned linearly
in the range of $[0, 20] \, \hMpc$ with 20 bins.
The pair counts are computed with neighbour search algorithm using R-tree.
In the measurements from simulations with the periodic boundary condition,
the random pair counts are given as
\begin{equation}
  RR = \frac{\Delta V}{V_\mathrm{sim}} ,
\end{equation}
where $V_\mathrm{sim} = (205 \, \hMpc)^3$ is the simulation box volume,
and $\Delta V$ is the volume occupied in the bin range.

In Figure~\ref{fig:xi2D}, the measured 2D anisotropic correlation functions of \OII ELGs
and corresponding mass-limited control samples are shown.
On large scales, the structure of correlation functions is quite similar between ELG and control samples.
However, on small scales, the FoG effect appears differently
due to the different dynamics within halos. We will address this feature in more details later.

Next, the estimator of the multipole moment of anisotropic correlation function \citep{Valageas2012}
is given as
\begin{equation}
  \hat{\xi}_\ell (s) = \frac{2 \ell + 1}{RR} \sum_{dd} L_\ell (\mu_{dd})  - \delta_{\ell 0},
\end{equation}
where $s$ is the 3D separation distance in the redshift space,
the summation is performed for all pairs of ELGs,
$\mu_{dd}$ is the directional cosine with respect to the LoS direction,
and $\delta_{ij}$ is the Kronecker delta.
The separation is binned linearly in the range of $[5, 50] \, \hMpc$ with 30 bins.
For reference, the isotropic correlation function $\xi_\mathrm{gg} (r)$
is also computed in the same manner,
where the separation $r$ is measured in real space, i.e., without RSD.

The covariance matrix of the multipole moments of the anisotropic correlation functions
is estimated with bootstraping method \citep[e.g.,][]{Norberg2009,Pu2025}.
The bootstrap samples are constructed by resampling ELGs with duplications until the sample
contains the same number of ELGs as the parent catalogue.
Then, we repeat this procedure and $N_\mathrm{BS} = 1120$ bootstrap samples.
The covariance matrix is estimated as
\begin{align}
  \mathrm{Cov}(\ell_a, s_i ; \ell_b, s_j) =& \frac{1}{N_\mathrm{BS} - 1} \sum_{r = 1}^{N_\mathrm{BS}}
  \left( \xi^{(r)}_{\ell_a} (s_i) - \bar{\xi}_{\ell_a} (s_i) \right) \nonumber \\
  & \times 
  \left( \xi^{(r)}_{\ell_b} (s_j) - \bar{\xi}_{\ell_b} (s_j) \right),
  \label{eq:cov} \\
  \bar{\xi}_{\ell_a} (s_i) =& \frac{1}{N_\mathrm{BS}} \sum_{r = 1}^{N_\mathrm{BS}} \xi^{(r)}_{\ell_a} (s_i) .
\end{align}
The label $a$ and $b$ denotes the multipole order: monopole ($\ell = 0$) and quadrupole ($\ell = 2$)
and the index $i$ and $j$ represent the bin of the separation distance.
The covariance of the isotropic correlation functions is estimated in the same manner.

\begin{figure}
  \includegraphics[width=\columnwidth]{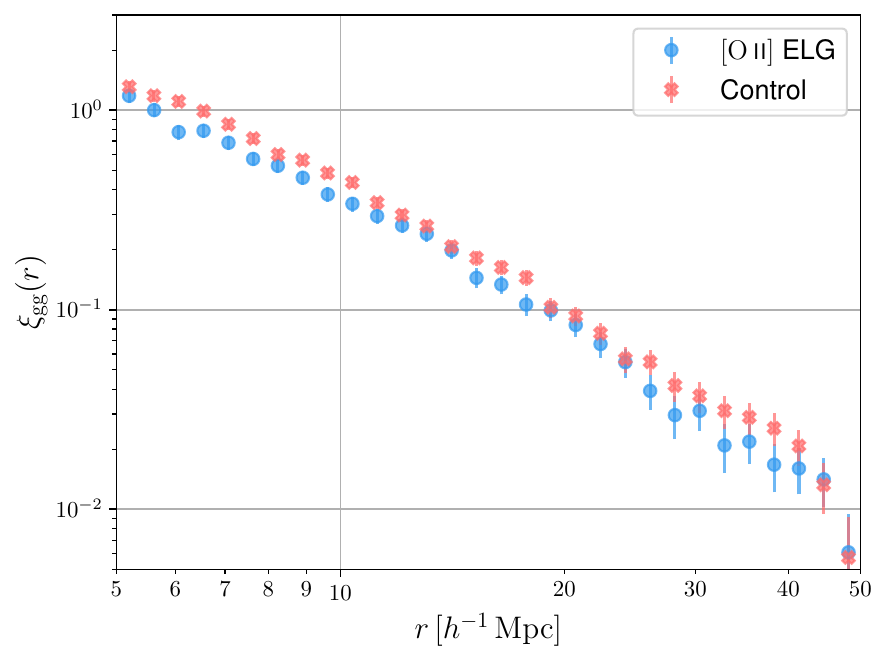}
  \caption{The isotropic correlation functions of \OII ELGs and
  the corresponding control sample.}
  \label{fig:xi_iso}
\end{figure}

\begin{figure}
  \includegraphics[width=\columnwidth]{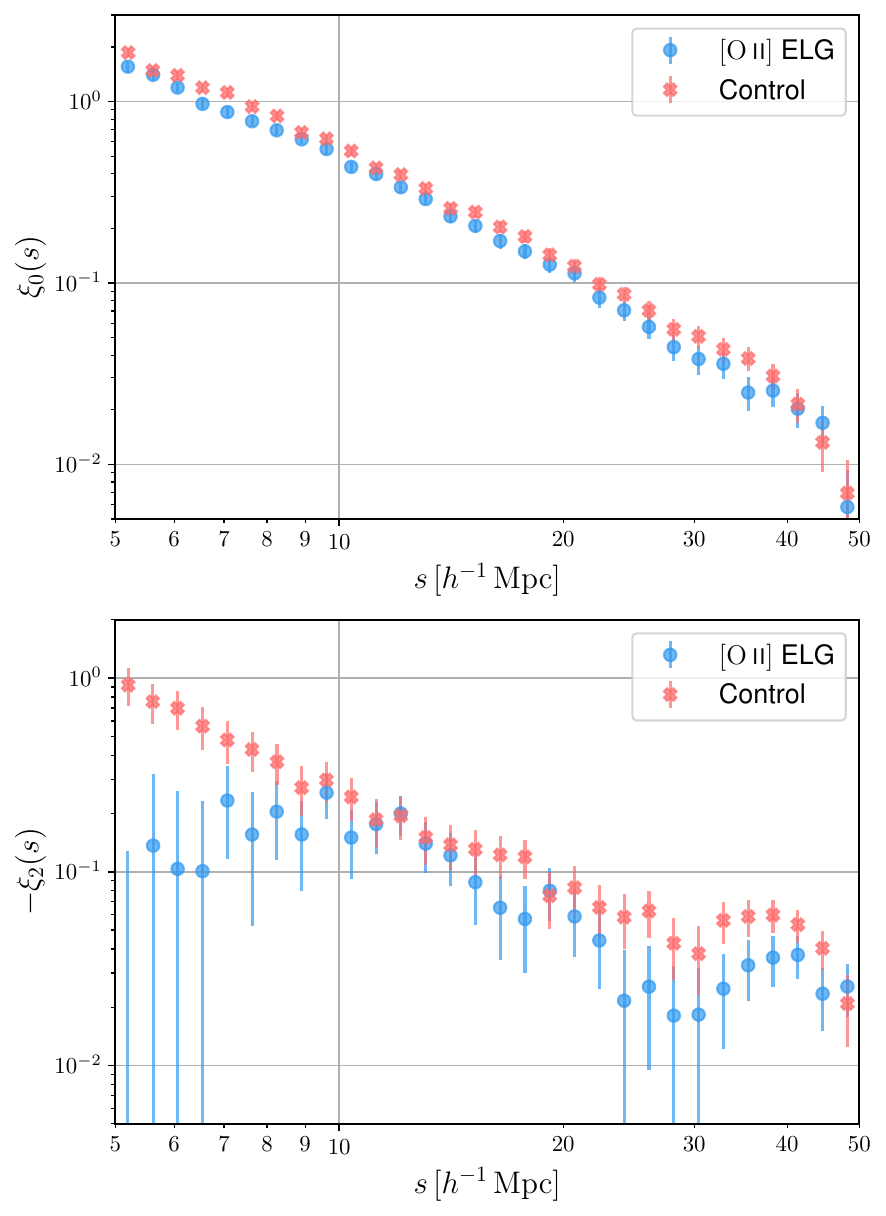}
  \caption{The monopole and quadrupole moments of correlation functions of
  \OII ELGs and the corresponding control sample.}
  \label{fig:xi_multi}
\end{figure}

Figures~\ref{fig:xi_iso} and \ref{fig:xi_multi} present
the isotropic correlation functions and monopole and quadrupole moments of anisotropic correlation functions
for \OII ELGs and control samples, respectively.
The amplitude of isotropic correlation functions of ELGs and control samples is similar
since the control sample is constructed so that the number density is the same as the ELG samples,
and thus, the sample exhibits a similar galaxy bias.
In contrast, the quadrupole moment shows a significant difference.
The small-scale amplitude of the ELG sample is suppressed, and
the overall amplitude is smaller than that of the control sample.
This feature suggests that the velocity of ELGs, i.e., dynamics within halos,
differs from that of the control sample.

\subsection{Cosmological parameter inference}

\begin{figure*}
  \includegraphics[width=\columnwidth]{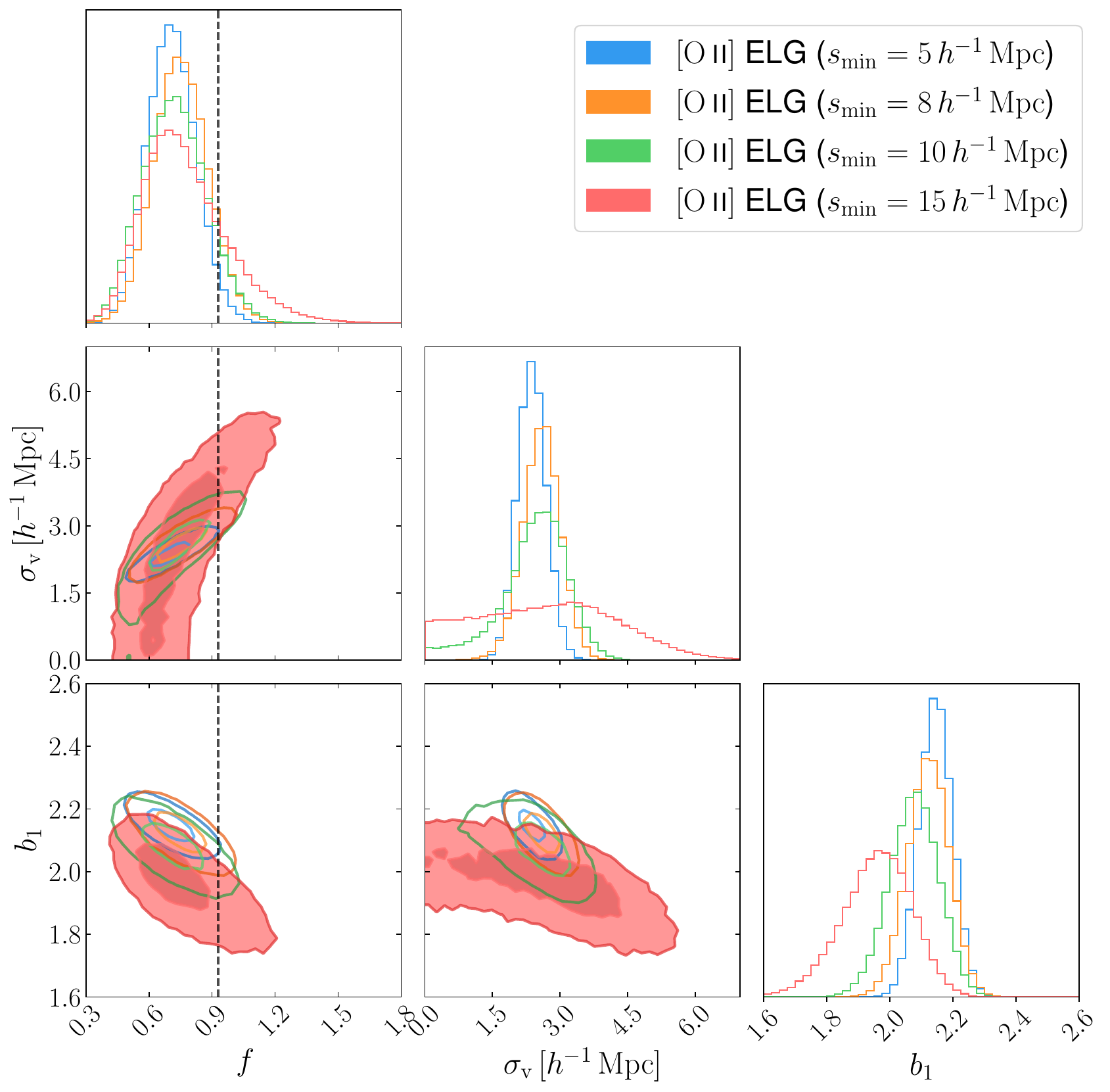}
  \includegraphics[width=\columnwidth]{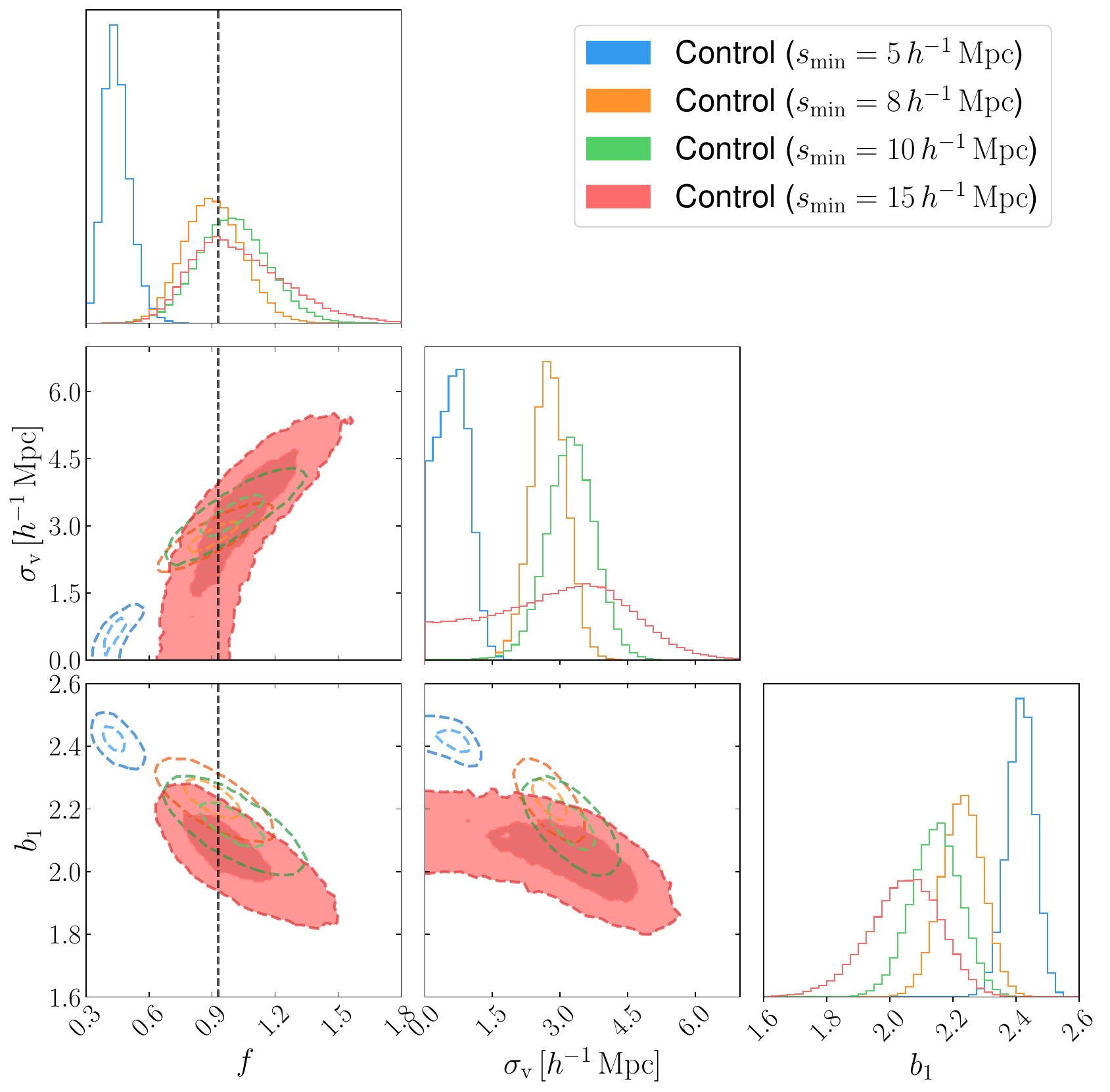}
  \caption{The parameter constraints with monopole and quadrupole moments of
  the anisotropic correlation function for \OII ELGs (left panels) and the control sample (right panels).
  The contours correspond to 1-$\sigma$ and 2-$\sigma$ levels.
  The minimum separation $s_\mathrm{min}$ is varied from $5 \, \hMpc$ to $15 \, \hMpc$.}
  \label{fig:triangle_OII}
\end{figure*}

The primary objective of this work is to quantify the possible bias in cosmological parameters,
especially the linear growth rate $f$,
inferred from the anisotropic correlation functions of ELGs.
As seen in the previous section,
the dynamics of ELGs within halos can lower the amplitude of the quadrupole moment,
which can be translated to an underestimation of the linear growth rate.
Here, we describe the parameter inference
from the mock measurements of anisotropic correlation functions.

In this analysis, the parameter space consists of
the linear growth rate $f$, the linear bias parameter $b_1$,
and the velocity dispersion $\sigma_\mathrm{v}$ in the FoG damping function.
The other cosmological parameters are fixed to the fiducial values (see Section~\ref{sec:simulations}).
We assume that the likelihood takes the multivariate Gaussian form:
\begin{align}
  \log L (\bm{\theta}) =& -\frac{1}{2} \sum_{a, b, i, j}
  \left( \hat{\xi}_{\ell_a} (s_i) - \xi_{\ell_a} (s_i; \bm{\theta}) \right)
  \mathrm{Cov}^{-1}(\ell_a, s_i ; \ell_b, s_j)
  \nonumber \\
  &\times \left( \hat{\xi}_{\ell_b} (s_j) - \xi_{\ell_b} (s_j; \bm{\theta}) \right) + \mathrm{const}.,
\end{align}
where $\bm{\theta} = (f, b_1, \sigma_\mathrm{v})$ is the parameter vector and
$\xi_\ell (s; \bm{\theta})$ is the prediction based on the theoretical model presented
in Section~\ref{sec:theory}.
The bin of the separation is the same as in the measurements but
we introduce the minimum separation $s_\mathrm{min}$,
i.e., excluding the data points at $s < s_\mathrm{min}$,
to investigate the impact of small-scale information,
which may suffer from less accuracy in the theoretical modelling.
The inverse of the covariance matrix estimated by inverting the covariance matrix in Eq.~\eqref{eq:cov}
is not unbiased but can be corrected with \citet{Hartlap2007} factor:
\begin{equation}
  \mathrm{Cov}^{-1} \to \frac{N_\mathrm{BS} - N_\mathrm{data} - 2}{N_\mathrm{BS} - 1} \mathrm{Cov}^{-1} ,
\end{equation}
where $N_\mathrm{data}$ is the number of data points used in the analysis.
Finally, the posterior distribution of parameters is explored
with the Markov Chain Monte-Carlo (MCMC) method with \texttt{nautilus} sampler \citep{Lange2023}.
Flat priors are assumed for all three parameters.
The live points are set to 10,000 and the sampling process runs
until the effective sample size reaches 1,000.

In Figure~\ref{fig:triangle_OII},
the posterior distribution of parameters by varying the minimum separation $s_\mathrm{min}$
from $5 \, \hMpc$ to $15 \, \hMpc$ are shown
for \OII ELGs and control samples, respectively.
As naively expected, increasing $s_\mathrm{min}$ leads to broader constraints on parameters.
Let us focus on the linear growth rate $f$, which is the only cosmological parameter in the analysis.
For the control sample, the true value of $f$ is recovered within the 1-$\sigma$ level
except for the case with $s_\mathrm{min} = 5 \, \hMpc$,
where the theoretical model is no longer accurate enough due to the strong non-linearity.
On the other hand, for \OII ELGs,
the inferred value of $f$ is biased low with all choices of $s_\mathrm{min}$,
consistent with the early study based on dark-matter-only simulations \citep{Okumura2011}.
This result indicates that the dynamics of ELGs within halos, which are not incorporated
in the theoretical modelling, propagate as systematic mismodeling.
The underestimation is qualitatively consistent with the dynamical nature of ELGs,
i.e., ELGs are likely to undergo infall motion towards the halo centre,
and thus, the velocity is lower than that of dark matter within halos.
In other words, the assumption of no velocity bias ($\theta_\mathrm{g} = \theta$)
does not hold for ELGs.
The parameter bias of the linear growth rate
at small $s_\mathrm{min}$ should be interpreted with care.
In this simplified analysis, most cosmological parameters are fixed and survey level
systematics are not included.
Therefore, the shift should be regarded as an upper bound on the
potential bias in a more realistic inference.
The parameter challenge analysis indicates
that the model misses part of the ELG velocity statistics,
motivating either larger scale cuts or model extensions that include
ELG velocity bias and infall components.

\begin{figure}
	\includegraphics[width=\columnwidth]{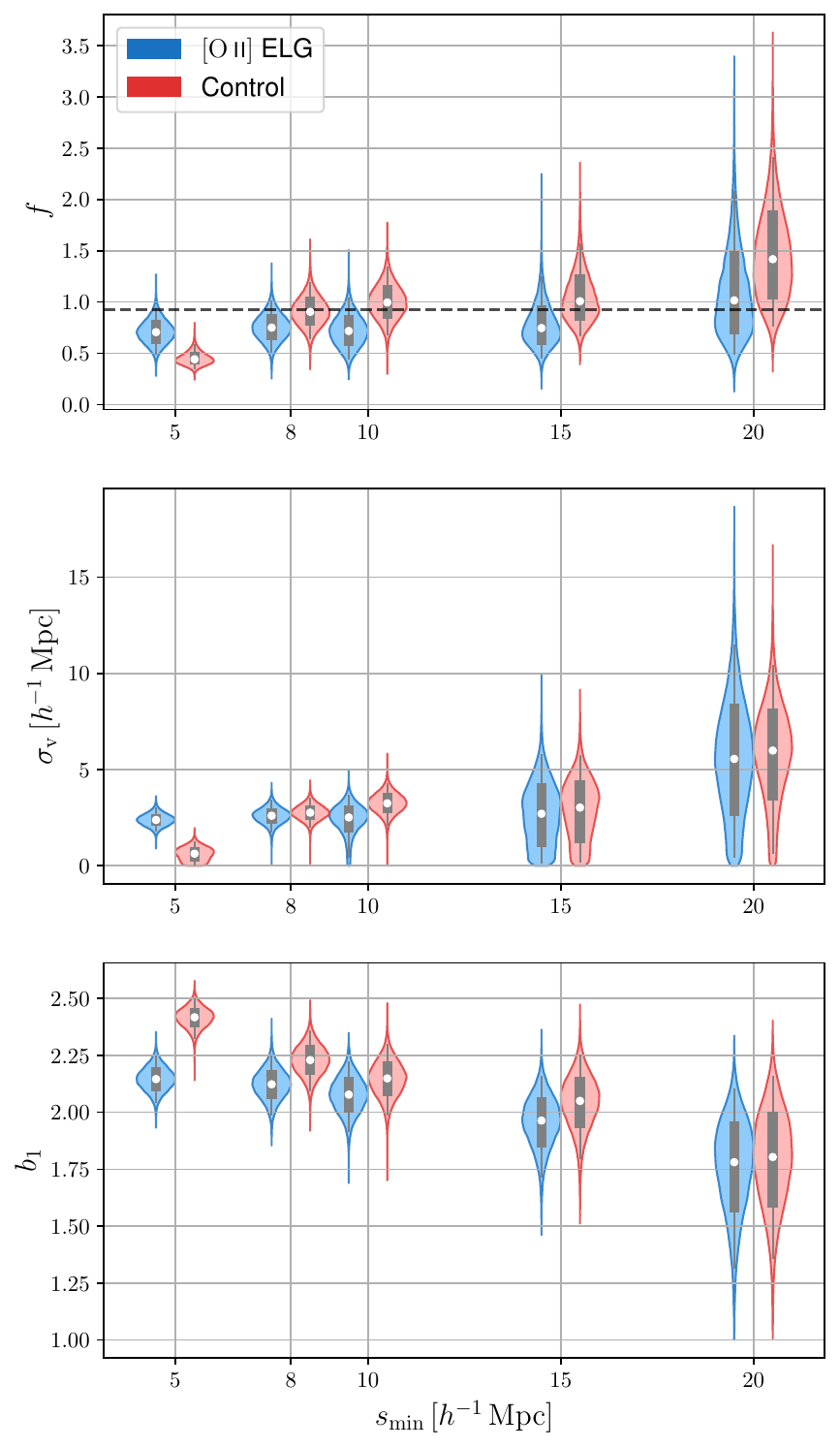}
  \caption{The marginalised posterior distributions with monopole and quadrupole moments of
  the anisotropic correlation function for \OII ELGs and the control sample.
  The minimum separation $s_\mathrm{min}$ is varied from $5 \, \hMpc$ to $20 \, \hMpc$.
  The thick (thin) grey bars correspond to $[15.87, 84.13]$ ($[2.28, 97.72]$) percentiles,
  and the white circles indicate the median values.
  The black dashed line corresponds to the true value of the linear growth rate $f$.
  The data points are slightly shifted horizontally for better visibility.}
  \label{fig:constraints_min_OII_multi}
\end{figure}

\begin{figure*}
  \includegraphics[width=0.9\textwidth]{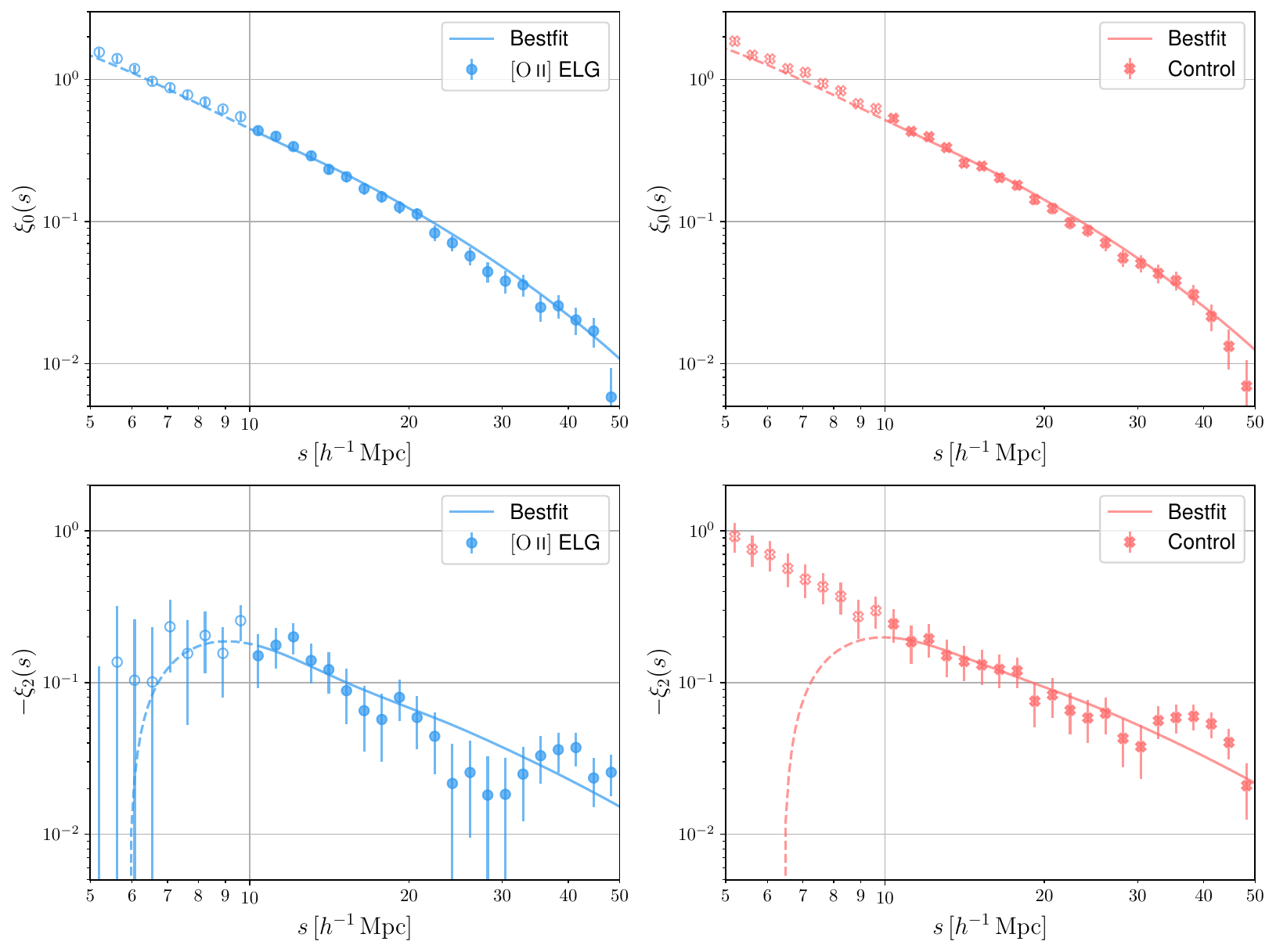}
  \caption{The best-fitting monopole and quadrupole moments of anisotropic correlation functions
  with $s_\mathrm{min} = 10 \, \hMpc$ for \OII ELGs and the control sample.
  The dashed lines and unfilled symbols represent the predictions and the data points
  not used in the MCMC analysis, respectively.}
  \label{fig:xi_multi_best}
\end{figure*}

The marginalised posterior distributions of parameters are shown
in Figure~\ref{fig:constraints_min_OII_multi}.
For \OII ELGs, the linear growth rate $f$ is finally recovered within a $1\text{-}\sigma$ level
with $s_\mathrm{min} = 20 \, \hMpc$.
The marginalized posterior distributions of the linear growth rate with $s_\mathrm{min} = 20 \, \hMpc$
for the control sample indicates the bias over the $1\text{-}\sigma$ level.
However, when the analysis is limited to large scales,
the degeneracy between $f$ and $b_1$ becomes severe,
and thus, the bias apparently appears due to the projection effect.
Figure~\ref{fig:xi_multi_best} presents an example of
the best-fitting multipole moments of anisotropic correlation functions
with $s_\mathrm{min} = 10 \, \hMpc$ for \OII ELGs and control samples.
The overall shapes of the best-fitting model are quite similar
between \OII ELGs and the control sample.
The only difference is the amplitude of the quadrupole moment and
that leads to a smaller inferred value of $f$ for \OII ELGs.

\begin{figure}
	\includegraphics[width=\columnwidth]{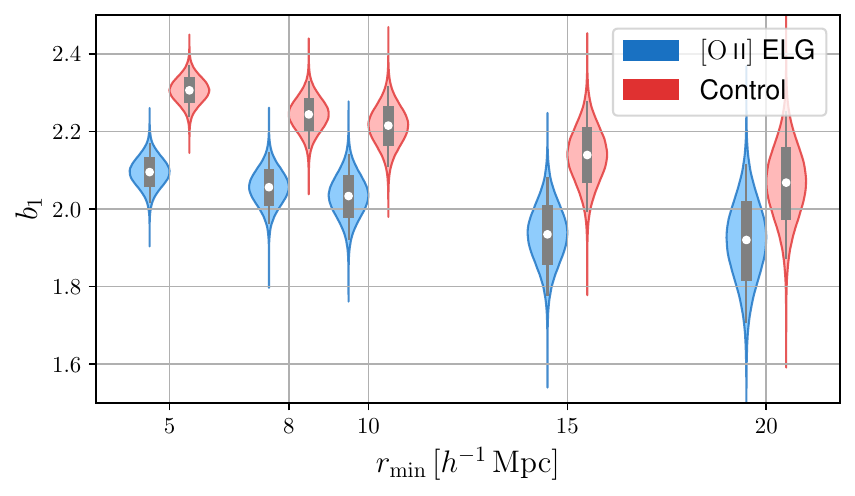}
  \caption{The marginalised posterior distributions with the isotropic correlation function for \OII ELGs and the control sample.
  The minimum separation $r_\mathrm{min}$ is varied from $5 \, \hMpc$ to $20 \, \hMpc$.
  The thick (thin) grey bars correspond to $[15.87, 84.13]$ ($[2.28, 97.72]$) percentiles,
  and the white circles indicate the median values.
  The data points are slightly shifted horizontally for better visibility.}
  \label{fig:constraints_min_OII_iso}
\end{figure}

\begin{figure*}
  \includegraphics[width=0.9\textwidth]{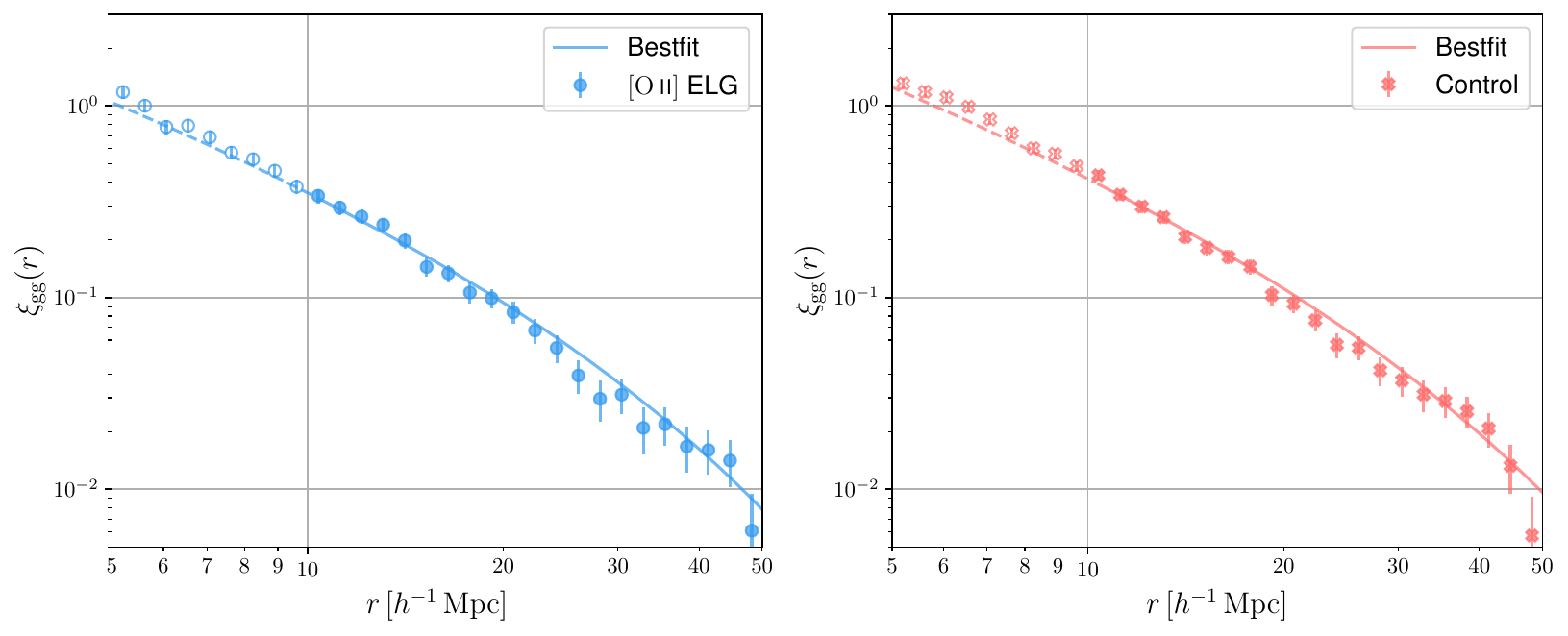}
  \caption{The best-fitting isotropic correlation functions
  with $r_\mathrm{min} = 10 \, \hMpc$ for \OII ELGs and the control sample.
  The dashed lines and unfilled symbols represent the predictions and the data points
  not used in the MCMC analysis, respectively.}
  \label{fig:xi_iso_best}
\end{figure*}

Similarly to the anisotropic correlation functions, we also perform the parameter inference
with the isotropic correlation functions.
In this case, the only free parameter is the linear bias $b_1$.
Hence, we do not run the MCMC sampling but directly compute the posterior distribution
by evaluating the likelihood on a grid of $b_1$.
The results of the marginalised posterior distribution
and the best-fitting isotropic correlation functions
are shown in Figures~\ref{fig:constraints_min_OII_iso} and \ref{fig:xi_iso_best}, respectively.
Both \OII ELGs and the control sample yield similar constraints on $b_1$
except for the case with $r_\mathrm{min} = 20 \, \hMpc$,
where the inferred value of $b_1$ is significantly lower compared with the isotropic case
due to the degeneracy with $f$ in the anisotropic analysis.

\section{ELG-halo connection}
\label{sec:connection}
Next, we address the connection between ELGs and hosting halos
to understand the origin of the systematic bias in the cosmological parameters
inferred from the anisotropic correlation functions of ELGs.

The most widely used scheme to connect galaxies and halos is the HOD model
and the HOD of simulated ELGs from IllustrisTNG has already been investigated
in \citetalias{Osato2023}.
Here, we summarise the main findings in that work.
The HOD of ELGs significantly differs from that of LRGs,
which are well described by the parametrised model proposed in \citet{Zheng2005}.
The ELG HOD exhibits the following features:
\begin{itemize}
  \item The central HOD has a peak at the halo mass of $M_\mathrm{vir} \sim 10^{12} \, h^{-1} \, \Msun$,
  which corresponds to the infalling halos along filamentary structures.
  That feature can be captured by the HOD model tailored to ELGs \citep{Geach2012}.
  \item The large fraction of the halos around this peak mass scale are disk-dominated galaxies.
  \item The slope of satellite HOD is shallower than unity,
  which indicates quenching of star formation in massive halos.
\end{itemize}

Here, we extend the analysis to study the connection between ELGs and hosting halos,
in particular for satellite ELGs.

\subsection{Phase space distribution of satellites in hosting halos}
\begin{figure}
	\includegraphics[width=\columnwidth]{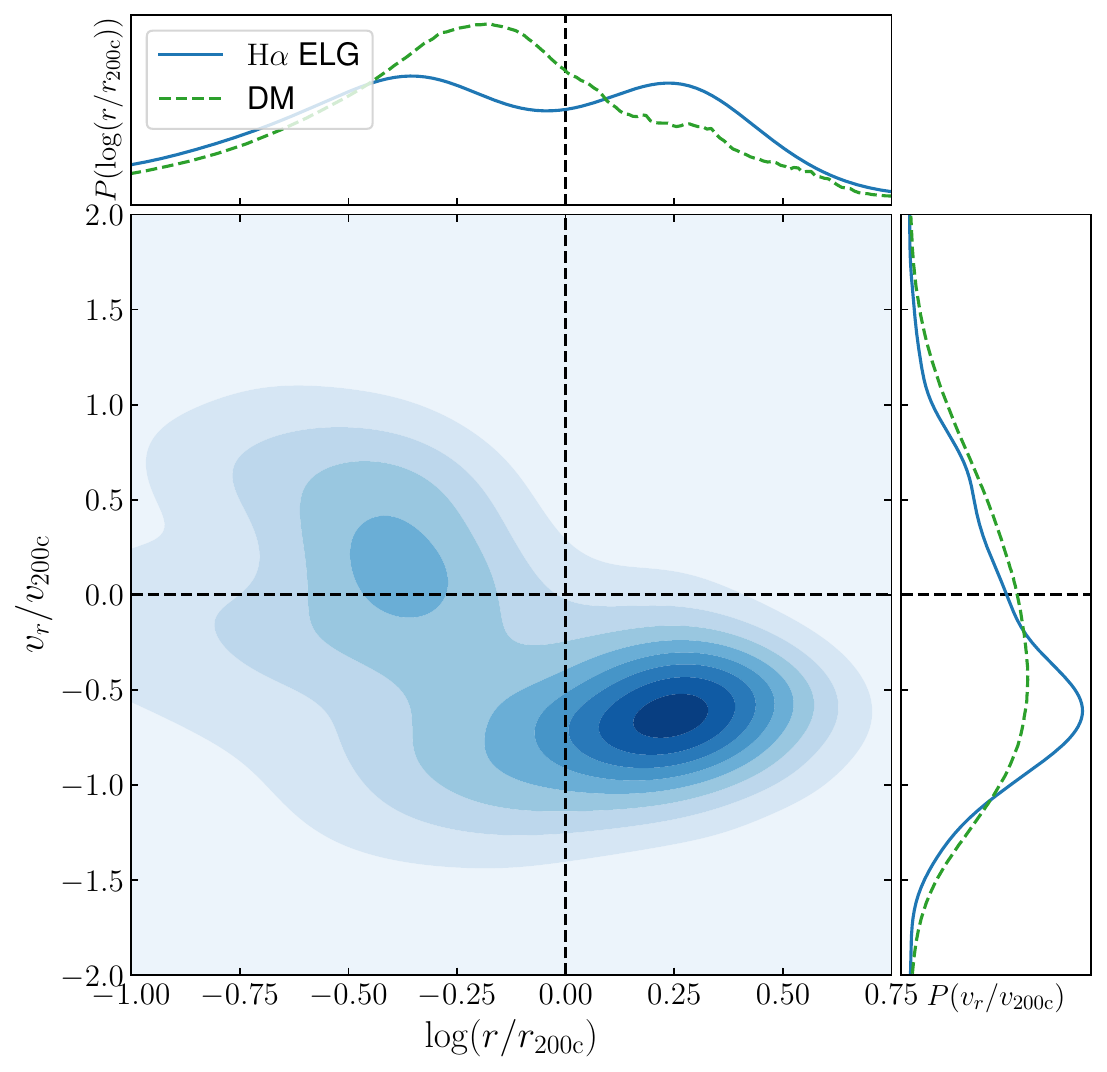}
  \includegraphics[width=\columnwidth]{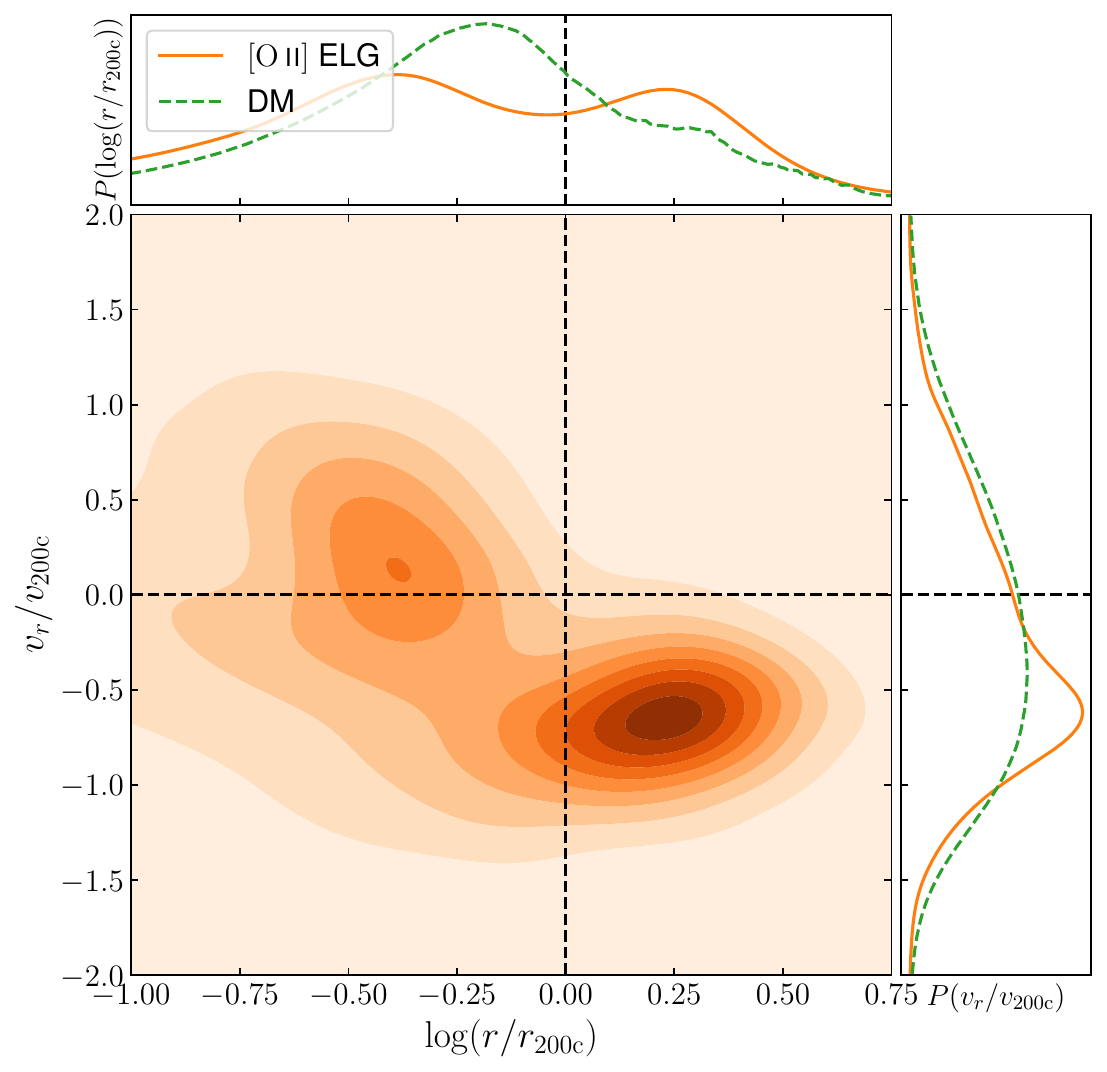}
  \caption{The phase space distribution of satellite \HA and \OII ELGs in hosting halos.
  The green dashed lines represent the distribution of dark matter particles
  in the same halos.
  The distribution is estimated with
  the Gaussian kernel density estimation method.}
  \label{fig:phase}
\end{figure}

To study the dynamics of satellite ELGs within halos,
we measure the phase space distribution, i.e.,
the joint distribution of the radial position $r$ and velocity $v_r$.
Figure~\ref{fig:phase} presents the phase space distribution of satellite \HA and \OII ELGs.
Here, the halo mass definition is the enclosed mass within the radius
where the mean density is $200$ times the critical density $\rho_\mathrm{cr}$,
which is denoted as $M_\mathrm{200c}$.
The radial position and velocity are normalised by the halo radius $r_\mathrm{200c}$
and the circular velocity at the halo radius $v_\mathrm{200c} = \sqrt{G M_\mathrm{200c} / r_\mathrm{200c}}$, respectively,
where $G$ is the gravitational constant.
For the radial velocity, the sign is defined so that
the positive (negative) value represents the outward (inward) motion
relative to the halo centre.
Both \HA and \OII ELGs exhibit two distinct components in the phase space:
the virialised component ($\log(r/r_\mathrm{200c}) \simeq -0.4$ and $v_r/v_\mathrm{200c} \simeq 0$),
and the infalling one ($\log(r/r_\mathrm{200c}) \simeq 0.2$ and $v_r/v_\mathrm{200c} \simeq -0.5$).
If galaxies are completely virialised within halos,
the 1D distribution of radial velocity is expected to be symmetric with a zero mean.
To illustrate this, the distribution of dark matter particles
in the same halos is overplotted as green dashed lines in Figure~\ref{fig:phase}.
There is a weak peak as the infalling component, but the dominant peak is
the virialised component for dark matter.
This feature indicates that satellite ELGs are preferentially
located in the outskirts of halos rather than the virialised one
and is consistent with the semi-analytic simulations \citep{Orsi2018}
and SHAM \citep{Ortega-Martinez2025}.
It is widely assumed that the radial satellite distribution follows
the Navarro--Frenk--White density profile \citep{Navarro1996,Navarro1997}
but the deviation requires modifications to the approach to modelling clustering statistics based on HOD
\citep[see, e.g.,][]{Rocher2023}.

The lifetime of line emission is $\sim 10 \, \mathrm{Myr}$ \citep{Byler2018}
and is much shorter than the dynamical time scale of halos $t_\mathrm{dyn}$,
which is given as
\begin{equation}
  t_\mathrm{dyn} = \frac{r_{200\mathrm{c}}}{v_{200\mathrm{c}}} =
  \sqrt{\frac{r_{200\mathrm{c}}^3}{G M_{200\mathrm{c}}}}
  = \left( \frac{4 \pi}{3} G (200 \rho_\mathrm{cr}) \right)^{-\frac{1}{2}}
  = 616 \, \mathrm{Myr} .
\end{equation}
Therefore, ELGs are likely to be recently accreted onto halos
and quenched by the time they reach the pericentre.
The observations \citep{Blanton2007,Wetzel2012} also show that
the quenched fraction of galaxies depends on the radius from the group or cluster centres.

\subsection{Conformity}
The galactic conformity refers to the correlation of properties of galaxies \citep{Weinmann2006}.
Here, we address the conformity of the star formation activity in the ELG population.
Table~\ref{tab:conformity_halos} presents the fraction of halos hosting ELGs\footnote{The `halos' here correspond to the groups based on friends-of-friends algorithm
in the simulations and contain `subhalos', which are also referred to as `galaxies'.}
and the mean halo mass with respect to the level of conformity.
The halos hosting central ELGs and at least one satellite ELG are referred to as `conformal halos'.
The ratio is roughly consistent with the results based on semi-analytic or
SHAM methods \citep[see Figure~8 of][]{Ortega-Martinez2025},
though a different selection is adopted in that work.
Next, Table~\ref{tab:conformity_ELGs} presents
the fraction of ELGs in each category of conformity.
Recently, \citet{Favole2026} studied the conformity of ELG
in the DESI Y1 data through clustering analysis.
Though the sample lies at the redshift range of $0.8 < z < 1.1$,
which is different from the ELG sample in this work,
the conformity of ELGs is detected: $7.02 \%$ of ELGs are satellite ELGs with conformity,
which is roughly consistent with the results in this work:
$9.3 \%$ for \HA ELGs and $12.1 \%$ for \OII ELGs.
The large conformity fraction can be attributed to the large satellite fraction,
which strongly depends on the galaxy-halo connection model
\citep{Yu2024,Yuan2025,Prada2025}.
In terms of the mean halo mass, the halos hosting only central ELG, only satellite ELGs,
and both central and satellite ELGs (conformal halos)
show the increasing trend in this order.
The halos hosting only central ELGs are the dominant population among ELGs,
and are likely to be independent halos.
On the other hand, the conformal halos, here both central and satellite ELGs are found,
are relatively massive halos.
Halos hosting ELGs are infalling towards these massive halos, and the central galaxy
undergoes star formation because the gas is still supplied to the galaxy in the dense core.
Subhalos around this dense environment can also host ELGs and can be identified as satellite ELGs
in the virialised component in the phase space distribution.

\begin{table}
  \centering
  \caption{
    The fraction of halos hosting ELGs and the mean halo mass with respect to the level of conformity.
    The halo mass definition is $M_\mathrm{200c}$.
  }
  \label{tab:conformity_halos}
  \begin{tabular}{ccc}
    \hline
    & Fraction (Number) & Mean halo mass $(h^{-1} \, \Msun)$ \\
    \hline \hline
    \multicolumn{3}{c}{\HA ELG} \\
    \hline
    Central only & 76.2\% (4,673) & $2.35 \times 10^{12}$ \\
    Satellite only & 15.9\% (975) & $5.76 \times 10^{12}$ \\
    Conformal & 7.9\% (485) & $1.24 \times 10^{13}$ \\
    \hline
    \multicolumn{3}{c}{\OII ELG} \\
    \hline
    Central only & 74.0\% (6,985) & $1.75 \times 10^{12}$ \\
    Satellite only & 16.1\% (1,517) & $5.03 \times 10^{12}$ \\
    Conformal & 9.9\% (933) & $9.60 \times 10^{12}$ \\
    \hline
  \end{tabular}
\end{table}

\begin{table}
  \centering
  \caption{
    The fraction of ELGs with respect to the level of conformity.
  }
  \label{tab:conformity_ELGs}
  \begin{tabular}{cc}
    \hline
    & Fraction (Number) \\
    \hline \hline
    \multicolumn{2}{c}{\HA ELG} \\
    \hline
    Central & 75.1\% (5,158) \\
    Satellite & 24.9\% (1,706) \\
    Satellite with conformity & 9.3\% (639) \\
    \hline
    \multicolumn{2}{c}{\OII ELG} \\
    \hline
    Central & 72.0\% (7,918) \\
    Satellite & 28.0\% (3,083) \\
    Satellite with conformity & 12.1\% (1,332) \\
    \hline
  \end{tabular}
\end{table}

\section{Conclusions}
\label{sec:conclusions}
ELGs are the main target population of galaxies in the upcoming and ongoing Stage-IV
spectroscopic surveys such as DESI, PFS, \textit{Euclid}, and Roman Space Telescope.
Previous works utilising semi-analytic models and hydrodynamical simulations
have revealed that ELGs are likely to be located in the filaments
or undergo infall towards massive halos.
The dynamical nature specific to ELGs can have a significant impact on
the clustering statistics in redshift space.
The theoretical modelling of the clustering statistics has been proven to be
accurate enough for LRGs.
However, it is not clear whether the same modelling can be applied to ELGs
without introducing any systematic bias in the inferred cosmological parameters.

In this work, we have performed the cosmology challenge
with the anisotropic correlation functions of ELGs
using the realistic ELG catalogues \citepalias{Osato2023} constructed from
the IllustrisTNG hydrodynamical simulations.
In order to highlight the impact due to the dynamics of ELGs,
we also construct the mass-limited control samples
which have the same number density and a similar large-scale bias.
Then, we have measured the anisotropic correlation functions of ELGs
and performed the MCMC parameter inference
with the theoretical model based on eTNS model at the 2-loop order.
We have found that the inferred value of the linear growth rate $f$
from monopole and quadrupole moments of anisotropic correlation functions of \OII ELGs
is biased low compared with the true value,
though unbiased estimation is possible with the control sample.
That underestimation can be attributed to the infalling motion of satellite ELGs
towards the halo centre, which lowers the amplitude of the quadrupole moment.
That is also an indication that the no velocity bias is not valid for ELGs.

In \citetalias{Osato2023}, it has been shown that
the HOD of ELGs significantly differs from that of LRGs
and the central ELGs are more likely to be disk-dominated isolated galaxies.
To extend the analysis into the ELG-halo connection for satellite ELGs,
we have investigated the phase space distribution of satellite ELGs within halos.
A large fraction of satellite ELGs are found to be
infalling towards the halo centre rather than virialised.
This feature indicates that ELGs are likely to be recently accreted onto halos
and quenched by the time they reach the pericentre.
This feature justifies the hypothesis that ELGs have the specific dynamics within halos,
which leads to the systematic bias in the inferred linear growth rate $f$.
Next, we addressed the galactic conformity of star formation activity in ELG population.
The conformity analysis has also supported this picture:
the halos hosting both central and satellite ELGs are more massive
than those hosting only central or satellite ELGs.
75\% of halos hosting ELGs contain only central ELGs,
15\% contain only satellite ELGs, and 10\% contain both central and satellite ELGs.
Most of halos hosting ELGs are central only ELG,
and this is consistent with the fact that most of ELGs reside
in the filamentary structures.
The mean halo mass is the largest for the conformal halos,
which is consistent with a scenario in which satellite ELGs are recently accreted
and not yet fully quenched, while some central ELGs may still sustain star formation
through continued gas supply.
In terms of the number of ELGs, the satellite ELGs with conformity account
for about 10\% of the total ELG population, which is consistent
with the recent observational results based on DESI Y1 data \citep{Favole2026}.

In summary, we have demonstrated that the dynamics of ELGs within halos
can lead to systematic bias in the cosmological parameters,
in particular, the linear growth rate, inferred
from the anisotropic correlation functions of ELGs at small scales.
The size of this bias depends on the analysis setup
and should not be read as a direct survey level prediction,
but it suggests that the simplified eTNS model
without an explicit ELG velocity bias degree of freedom
is incomplete for small scales. Therefore, our cautionary conclusion
should be interpreted as specific to the inference framework adopted in this work,
rather than as a generic statement about all ELG clustering models.
More flexible galaxy--halo connection frameworks,
such as decorated HODs \citep{Hearin2016} or
extended SHAM models including assembly and velocity bias
\citep{Contreras2021,Ortega-Martinez2025,Yu2024},
may reduce or remove the inferred bias if they provide sufficient freedom
to capture the non-virialized infall component seen in the simulations.
The accurate theoretical modelling taking into account the specific dynamics of ELGs
is essential to fully exploit the cosmological information in the upcoming surveys.
The analysis presented in this work provides useful insights into the small-scale clustering ELG.
On the other hand, the large-scale clustering can also be probed
by benefiting from the wide survey area in Stage-IV surveys.
The large-scale clustering, especially BAO and AP effect,
is not addressed in this work due to the limited box size
of hydrodynamical simulations.
Recently, the large-volume hydrodynamical simulations have become available
\citep[e.g.,][]{HernandezAguayo2023,Schaye2023}.
The impact at the large scales due to the specific dynamics of ELGs
will be addressed in future work utilising those simulations.

\section*{Acknowledgements}
K.~O. thanks Shun Saito for fruitiful discussions
and Volker Springel for hosting the author at Max Planck Institute for Astrophysics,
where part of this work was done.
This work was supported in part by JSPS KAKENHI Grant Number 
JP24H00215, JP25H00662, JP25H01513, JP25K17380, JP22K21349, JP19KK0076,
and JGC-Saneyoshi Scholarship Foundation (K.~O.).
T.~O. acknowledges support from the Taiwan National Science and Technology Council under Grants 
Nos. NSTC 112-2112-M-001-034-, NSTC 113-2112-M-001-011- and NSTC 114-2112-M-001-004-,
and the Academia Sinica Investigator Project Grant No. AS-IV-114-M03 for the period of 2025--2029.
The numerical calculations were carried out on Yukawa-21 and Heian at YITP in Kyoto University.

\section*{Data Availability}
The catalogues of simulated ELGs are available
upon reasonable request to the author.



\bibliographystyle{mnras}
\bibliography{main} 

@ARTICLE{Contreras2021,
       author = {{Contreras}, S. and {Angulo}, R.~E. and {Zennaro}, M.},
        title = "{A flexible subhalo abundance matching model for galaxy clustering in redshift space}",
      journal = {\mnras},
         year = 2021,
        month = nov,
       volume = {508},
       number = {1},
        pages = {175-189},
          doi = {10.1093/mnras/stab2560},
archivePrefix = {arXiv},
       eprint = {2012.06596},
 primaryClass = {astro-ph.CO},
       adsurl = {https://ui.adsabs.harvard.edu/abs/2021MNRAS.508..175C}
}

@ARTICLE{Hearin2016,
       author = {{Hearin}, Andrew P. and {Zentner}, Andrew R. and {van den Bosch}, Frank C. and {Campbell}, Duncan and {Tollerud}, Erik},
        title = "{Introducing decorated HODs: modelling assembly bias in the galaxy-halo connection}",
      journal = {\mnras},
         year = 2016,
        month = aug,
       volume = {460},
       number = {3},
        pages = {2552-2570},
          doi = {10.1093/mnras/stw840},
archivePrefix = {arXiv},
       eprint = {1512.03050},
 primaryClass = {astro-ph.CO},
       adsurl = {https://ui.adsabs.harvard.edu/abs/2016MNRAS.460.2552H}
}

@ARTICLE{Prada2025,
       author = {{Prada}, F. and {Ereza}, J. and {Smith}, A. and {Lasker}, J. and {Vaisakh}, R. and {Kehoe}, R. and {Dong-P{\'a}ez}, C.~A. and {Siudek}, M. and {Wang}, M.~S. and {Alam}, S. and {Beutler}, F. and {Bianchi}, D. and {Cole}, S. and {Dey}, B. and {Kirkby}, D. and {Norberg}, P. and {Aguilar}, J. and {Ahlen}, S. and {Brooks}, D. and {Claybaugh}, T. and {Dawson}, K. and {de la Macorra}, A. and {Fanning}, K. and {Forero-Romero}, J.~E. and {Gontcho A Gontcho}, S. and {Hahn}, C. and {Honscheid}, K. and {Ishak}, M. and {Kisner}, T. and {Landriau}, M. and {Manera}, M. and {Meisner}, A. and {Miquel}, R. and {Moustakas}, J. and {Mueller}, E. and {Nie}, J. and {Percival}, W.~J. and {Poppett}, C. and {Rezaie}, M. and {Rossi}, G. and {Sanchez}, E. and {Schubnell}, M. and {Tarl{\'e}}, G. and {Vargas-Maga{\~n}a}, M. and {Weaver}, B.~A. and {Yuan}, S. and {Zhou}, Z.},
        title = "{The DESI One-Percent Survey: Modelling the clustering and halo occupation of all four DESI tracers with UCHUU}",
      journal = {\aap},
         year = 2025,
        month = jun,
       volume = {698},
          eid = {A170},
        pages = {A170},
          doi = {10.1051/0004-6361/202451022},
archivePrefix = {arXiv},
       eprint = {2306.06315},
 primaryClass = {astro-ph.CO},
       adsurl = {https://ui.adsabs.harvard.edu/abs/2025A&A...698A.170P}
}

@ARTICLE{Yu2024,
       author = {{Yu}, Jiaxi and {Zhao}, Cheng and {Gonzalez-Perez}, Violeta and {Chuang}, Chia-Hsun and {Brodzeller}, Allyson and {de Mattia}, Arnaud and {Kneib}, Jean-Paul and {Krolewski}, Alex and {Rocher}, Antoine and {Ross}, Ashley and {Wang}, Yunchong and {Yuan}, Sihan and {Zhang}, Hanyu and {Zhou}, Rongpu and {Aguilar}, Jessica Nicole and {Ahlen}, Steven and {Brooks}, David and {Dawson}, Kyle and {de la Macorra}, Alex and {Doel}, Peter and {Fanning}, Kevin and {Font-Ribera}, Andreu and {Forero-Romero}, Jaime and {Gontcho}, Satya Gontcho A. and {Honscheid}, Klaus and {Kehoe}, Robert and {Kisner}, Theodore and {Kremin}, Anthony and {Landriau}, Martin and {Manera}, Marc and {Martini}, Paul and {Meisner}, Aaron and {Miquel}, Ramon and {Moustakas}, John and {Nie}, Jundan and {Percival}, Will and {Poppett}, Claire and {Raichoor}, Anand and {Rossi}, Graziano and {Seo}, Hee-Jong and {Tarl{\'e}}, Gregory and {Zhou}, Zhimin and {Zou}, Hu},
        title = "{The DESI One-Percent Survey: exploring a generalized SHAM for multiple tracers with the UNIT simulation}",
      journal = {\mnras},
         year = 2024,
        month = jan,
       volume = {527},
       number = {3},
        pages = {6950-6969},
          doi = {10.1093/mnras/stad3559},
archivePrefix = {arXiv},
       eprint = {2306.06313},
 primaryClass = {astro-ph.CO},
       adsurl = {https://ui.adsabs.harvard.edu/abs/2024MNRAS.527.6950Y}
}

@ARTICLE{Favole2026,
       author = {{Favole}, Ginevra and {Kitaura}, Francisco-Shu and {Hadzhiyska}, Boryana and {Eisenstein}, Daniel and {Garrison}, Lehman H. and {Bose}, Sownak},
        title = "{ELG{\texttimes}LRG Distribution through Dark Matter Halo Dynamics}",
      journal = {\apj},
         year = 2026,
        month = may,
       volume = {1002},
       number = {1},
          eid = {90},
        pages = {90},
          doi = {10.3847/1538-4357/ae592a},
archivePrefix = {arXiv},
       eprint = {2512.04362},
 primaryClass = {astro-ph.GA},
       adsurl = {https://ui.adsabs.harvard.edu/abs/2026ApJ..1002...90F}
}

@ARTICLE{Raichoor2023,
       author = {{Raichoor}, A. and {Moustakas}, J. and {Newman}, Jeffrey A. and {Karim}, T. and {Ahlen}, S. and {Alam}, Shadab and {Bailey}, S. and {Brooks}, D. and {Dawson}, K. and {de la Macorra}, A. and {de Mattia}, A. and {Dey}, A. and {Dey}, Biprateep and {Dhungana}, G. and {Eftekharzadeh}, S. and {Eisenstein}, D.~J. and {Fanning}, K. and {Font-Ribera}, A. and {Garc{\'\i}a-Bellido}, J. and {Gazta{\~n}aga}, E. and {A Gontcho}, S. Gontcho and {Guy}, J. and {Honscheid}, K. and {Ishak}, M. and {Kehoe}, R. and {Kisner}, T. and {Kremin}, Anthony and {Lan}, Ting-Wen and {Landriau}, M. and {Le Guillou}, L. and {Levi}, Michael E. and {Magneville}, C. and {Manera}, M. and {Martini}, P. and {Meisner}, Aaron M. and {Myers}, Adam D. and {Nie}, Jundan and {Palanque-Delabrouille}, N. and {Percival}, W.~J. and {Poppett}, C. and {Prada}, F. and {Ross}, A.~J. and {Ruhlmann-Kleider}, V. and {Sabiu}, C.~G. and {Schlafly}, E.~F. and {Schlegel}, D. and {Tarl{\'e}}, Gregory and {Weaver}, B.~A. and {Y{\`e}che}, Christophe and {Zhou}, Rongpu and {Zhou}, Zhimin and {Zou}, H.},
        title = "{Target Selection and Validation of DESI Emission Line Galaxies}",
      journal = {\aj},
         year = 2023,
        month = mar,
       volume = {165},
       number = {3},
          eid = {126},
        pages = {126},
          doi = {10.3847/1538-3881/acb213},
archivePrefix = {arXiv},
       eprint = {2208.08513},
 primaryClass = {astro-ph.CO},
       adsurl = {https://ui.adsabs.harvard.edu/abs/2023AJ....165..126R}
}

@ARTICLE{Zehavi2004,
       author = {{Zehavi}, Idit and {Weinberg}, David H. and {Zheng}, Zheng and {Berlind}, Andreas A. and {Frieman}, Joshua A. and {Scoccimarro}, Rom{\'a}n and {Sheth}, Ravi K. and {Blanton}, Michael R. and {Tegmark}, Max and {Mo}, Houjun J. and {Bahcall}, Neta A. and {Brinkmann}, Jon and {Burles}, Scott and {Csabai}, Istv{\'a}n and {Fukugita}, Masataka and {Gunn}, James E. and {Lamb}, Don Q. and {Loveday}, Jon and {Lupton}, Robert H. and {Meiksin}, Avery and {Munn}, Jeffrey A. and {Nichol}, Robert C. and {Schlegel}, David and {Schneider}, Donald P. and {SubbaRao}, Mark and {Szalay}, Alexander S. and {Uomoto}, Alan and {York}, Donald G. and {SDSS Collaboration}},
        title = "{On Departures from a Power Law in the Galaxy Correlation Function}",
      journal = {\apj},
         year = 2004,
        month = jun,
       volume = {608},
       number = {1},
        pages = {16-24},
          doi = {10.1086/386535},
archivePrefix = {arXiv},
       eprint = {astro-ph/0301280},
 primaryClass = {astro-ph},
       adsurl = {https://ui.adsabs.harvard.edu/abs/2004ApJ...608...16Z}
}

@article{Okumura2011,
	adsurl = {https://ui.adsabs.harvard.edu/abs/2011ApJ...726....5O},
	archiveprefix = {arXiv},
	author = {{Okumura}, Teppei and {Jing}, Y.~P.},
	doi = {10.1088/0004-637X/726/1/5},
	eid = {5},
	eprint = {1004.3548},
	journal = {\apj},
	month = jan,
	number = {1},
	pages = {5},
	primaryclass = {astro-ph.CO},
	title = {{Systematic Effects on Determination of the Growth Factor from Redshift-space Distortions}},
	volume = {726},
	year = 2011}

@article{Ishikawa2025,
	adsurl = {https://ui.adsabs.harvard.edu/abs/2025MNRAS.539.2323I},
	archiveprefix = {arXiv},
	author = {{Ishikawa}, Shogo and {Okumura}, Teppei and {Hayashi}, Masao and {Takeuchi}, Tsutomu T.},
	doi = {10.1093/mnras/staf577},
	eprint = {2412.19898},
	journal = {\mnras},
	month = may,
	number = {3},
	pages = {2323-2346},
	primaryclass = {astro-ph.GA},
	title = {{A new constraint on galaxy─halo connections of [O II] emitters via HOD modelling with angular clustering and luminosity functions from the Subaru HSC survey}},
	volume = {539},
	year = 2025}

@ARTICLE{Navarro1997,
       author = {{Navarro}, Julio F. and {Frenk}, Carlos S. and {White}, Simon D.~M.},
        title = "{A Universal Density Profile from Hierarchical Clustering}",
      journal = {\apj},
         year = 1997,
        month = dec,
       volume = {490},
       number = {2},
        pages = {493-508},
          doi = {10.1086/304888},
archivePrefix = {arXiv},
       eprint = {astro-ph/9611107},
 primaryClass = {astro-ph},
       adsurl = {https://ui.adsabs.harvard.edu/abs/1997ApJ...490..493N}
}

@ARTICLE{Navarro1996,
       author = {{Navarro}, Julio F. and {Frenk}, Carlos S. and {White}, Simon D.~M.},
        title = "{The Structure of Cold Dark Matter Halos}",
      journal = {\apj},
         year = 1996,
        month = may,
       volume = {462},
        pages = {563},
          doi = {10.1086/177173},
archivePrefix = {arXiv},
       eprint = {astro-ph/9508025},
 primaryClass = {astro-ph},
       adsurl = {https://ui.adsabs.harvard.edu/abs/1996ApJ...462..563N}
}

@ARTICLE{Wetzel2012,
       author = {{Wetzel}, Andrew R. and {Tinker}, Jeremy L. and {Conroy}, Charlie},
        title = "{Galaxy evolution in groups and clusters: star formation rates, red sequence fractions and the persistent bimodality}",
      journal = {\mnras},
         year = 2012,
        month = jul,
       volume = {424},
       number = {1},
        pages = {232-243},
          doi = {10.1111/j.1365-2966.2012.21188.x},
archivePrefix = {arXiv},
       eprint = {1107.5311},
 primaryClass = {astro-ph.CO},
       adsurl = {https://ui.adsabs.harvard.edu/abs/2012MNRAS.424..232W}
}

@ARTICLE{Blanton2007,
       author = {{Blanton}, Michael R. and {Berlind}, Andreas A.},
        title = "{What Aspects of Galaxy Environment Matter?}",
      journal = {\apj},
         year = 2007,
        month = aug,
       volume = {664},
       number = {2},
        pages = {791-803},
          doi = {10.1086/512478},
archivePrefix = {arXiv},
       eprint = {astro-ph/0608353},
 primaryClass = {astro-ph},
       adsurl = {https://ui.adsabs.harvard.edu/abs/2007ApJ...664..791B}
}

@ARTICLE{Norberg2009,
       author = {{Norberg}, P. and {Baugh}, C.~M. and {Gazta{\~n}aga}, E. and {Croton}, D.~J.},
        title = "{Statistical analysis of galaxy surveys - I. Robust error estimation for two-point clustering statistics}",
      journal = {\mnras},
         year = 2009,
        month = jun,
       volume = {396},
       number = {1},
        pages = {19-38},
          doi = {10.1111/j.1365-2966.2009.14389.x},
archivePrefix = {arXiv},
       eprint = {0810.1885},
 primaryClass = {astro-ph},
       adsurl = {https://ui.adsabs.harvard.edu/abs/2009MNRAS.396...19N}
}

@ARTICLE{Pu2025,
       author = {{Pu}, Sy-Yun and {Okumura}, Teppei and {Chen}, Chian-Chou and {Nishimichi}, Takahiro and {Akitsu}, Kazuyuki},
        title = "{Constraining cosmology with N-body simulations for future spectroscopic galaxy surveys at 2{\ensuremath{\leq}}z{\ensuremath{\leq}}3}",
      journal = {\prd},
         year = 2025,
        month = aug,
       volume = {112},
       number = {4},
          eid = {L041302},
        pages = {L041302},
          doi = {10.1103/n9cp-vkz7},
archivePrefix = {arXiv},
       eprint = {2410.02524},
 primaryClass = {astro-ph.CO},
       adsurl = {https://ui.adsabs.harvard.edu/abs/2025PhRvD.112d1302P}
}

@ARTICLE{Schaye2023,
       author = {{Schaye}, Joop and {Kugel}, Roi and {Schaller}, Matthieu and {Helly}, John C. and {Braspenning}, Joey and {Elbers}, Willem and {McCarthy}, Ian G. and {van Daalen}, Marcel P. and {Vandenbroucke}, Bert and {Frenk}, Carlos S. and {Kwan}, Juliana and {Salcido}, Jaime and {Bah{\'e}}, Yannick M. and {Borrow}, Josh and {Chaikin}, Evgenii and {Hahn}, Oliver and {Hu{\v{s}}ko}, Filip and {Jenkins}, Adrian and {Lacey}, Cedric G. and {Nobels}, Folkert S.~J.},
        title = "{The FLAMINGO project: cosmological hydrodynamical simulations for large-scale structure and galaxy cluster surveys}",
      journal = {\mnras},
         year = 2023,
        month = dec,
       volume = {526},
       number = {4},
        pages = {4978-5020},
          doi = {10.1093/mnras/stad2419},
archivePrefix = {arXiv},
       eprint = {2306.04024},
 primaryClass = {astro-ph.CO},
       adsurl = {https://ui.adsabs.harvard.edu/abs/2023MNRAS.526.4978S}
}

@ARTICLE{HernandezAguayo2023,
       author = {{Hern{\'a}ndez-Aguayo}, C{\'e}sar and {Springel}, Volker and {Pakmor}, R{\"u}diger and {Barrera}, Monica and {Ferlito}, Fulvio and {White}, Simon D.~M. and {Hernquist}, Lars and {Hadzhiyska}, Boryana and {Delgado}, Ana Maria and {Kannan}, Rahul and {Bose}, Sownak and {Frenk}, Carlos},
        title = "{The MillenniumTNG Project: high-precision predictions for matter clustering and halo statistics}",
      journal = {\mnras},
         year = 2023,
        month = sep,
       volume = {524},
       number = {2},
        pages = {2556-2578},
          doi = {10.1093/mnras/stad1657},
archivePrefix = {arXiv},
       eprint = {2210.10059},
 primaryClass = {astro-ph.CO},
       adsurl = {https://ui.adsabs.harvard.edu/abs/2023MNRAS.524.2556H}
}

@ARTICLE{Weinmann2006,
       author = {{Weinmann}, Simone M. and {van den Bosch}, Frank C. and {Yang}, Xiaohu and {Mo}, H.~J. and {Croton}, Darren J. and {Moore}, Ben},
        title = "{Properties of galaxy groups in the Sloan Digital Sky Survey - II. Active galactic nucleus feedback and star formation truncation}",
      journal = {\mnras},
         year = 2006,
        month = nov,
       volume = {372},
       number = {3},
        pages = {1161-1174},
          doi = {10.1111/j.1365-2966.2006.10932.x},
archivePrefix = {arXiv},
       eprint = {astro-ph/0606458},
 primaryClass = {astro-ph},
       adsurl = {https://ui.adsabs.harvard.edu/abs/2006MNRAS.372.1161W}
}

@ARTICLE{Lange2023,
       author = {{Lange}, Johannes U.},
        title = "{NAUTILUS: boosting Bayesian importance nested sampling with deep learning}",
      journal = {\mnras},
         year = 2023,
        month = oct,
       volume = {525},
       number = {2},
        pages = {3181-3194},
          doi = {10.1093/mnras/stad2441},
archivePrefix = {arXiv},
       eprint = {2306.16923},
 primaryClass = {astro-ph.IM},
       adsurl = {https://ui.adsabs.harvard.edu/abs/2023MNRAS.525.3181L}
}

@ARTICLE{Hartlap2007,
       author = {{Hartlap}, J. and {Simon}, P. and {Schneider}, P.},
        title = "{Why your model parameter confidences might be too optimistic. Unbiased estimation of the inverse covariance matrix}",
      journal = {\aap},
         year = 2007,
        month = mar,
       volume = {464},
       number = {1},
        pages = {399-404},
          doi = {10.1051/0004-6361:20066170},
archivePrefix = {arXiv},
       eprint = {astro-ph/0608064},
 primaryClass = {astro-ph},
       adsurl = {https://ui.adsabs.harvard.edu/abs/2007A&A...464..399H}
}

@ARTICLE{Alcock1979,
       author = {{Alcock}, C. and {Paczynski}, B.},
        title = "{An evolution free test for non-zero cosmological constant}",
      journal = {\nat},
         year = 1979,
        month = oct,
       volume = {281},
        pages = {358},
          doi = {10.1038/281358a0},
       adsurl = {https://ui.adsabs.harvard.edu/abs/1979Natur.281..358A}
}

@ARTICLE{Kaiser1987,
       author = {{Kaiser}, Nick},
        title = "{Clustering in real space and in redshift space}",
      journal = {\mnras},
         year = 1987,
        month = jul,
       volume = {227},
        pages = {1-21},
          doi = {10.1093/mnras/227.1.1},
       adsurl = {https://ui.adsabs.harvard.edu/abs/1987MNRAS.227....1K}
}

@ARTICLE{Taruya2012,
       author = {{Taruya}, Atsushi and {Bernardeau}, Francis and {Nishimichi}, Takahiro and {Codis}, Sandrine},
        title = "{Direct and fast calculation of regularized cosmological power spectrum at two-loop order}",
      journal = {\prd},
         year = 2012,
        month = nov,
       volume = {86},
       number = {10},
          eid = {103528},
        pages = {103528},
          doi = {10.1103/PhysRevD.86.103528},
archivePrefix = {arXiv},
       eprint = {1208.1191},
 primaryClass = {astro-ph.CO},
       adsurl = {https://ui.adsabs.harvard.edu/abs/2012PhRvD..86j3528T}
}

@ARTICLE{Baldauf2012,
       author = {{Baldauf}, Tobias and {Seljak}, Uro{\v{s}} and {Desjacques}, Vincent and {McDonald}, Patrick},
        title = "{Evidence for quadratic tidal tensor bias from the halo bispectrum}",
      journal = {\prd},
         year = 2012,
        month = oct,
       volume = {86},
       number = {8},
          eid = {083540},
        pages = {083540},
          doi = {10.1103/PhysRevD.86.083540},
archivePrefix = {arXiv},
       eprint = {1201.4827},
 primaryClass = {astro-ph.CO},
       adsurl = {https://ui.adsabs.harvard.edu/abs/2012PhRvD..86h3540B}
}

@ARTICLE{GilMarin2015,
       author = {{Gil-Mar{\'\i}n}, H{\'e}ctor and {Nore{\~n}a}, Jorge and {Verde}, Licia and {Percival}, Will J. and {Wagner}, Christian and {Manera}, Marc and {Schneider}, Donald P.},
        title = "{The power spectrum and bispectrum of SDSS DR11 BOSS galaxies - I. Bias and gravity}",
      journal = {\mnras},
         year = 2015,
        month = jul,
       volume = {451},
       number = {1},
        pages = {539-580},
          doi = {10.1093/mnras/stv961},
archivePrefix = {arXiv},
       eprint = {1407.5668},
 primaryClass = {astro-ph.CO},
       adsurl = {https://ui.adsabs.harvard.edu/abs/2015MNRAS.451..539G}
}

@ARTICLE{Beutler2014,
       author = {{Beutler}, Florian and {Saito}, Shun and {Seo}, Hee-Jong and {Brinkmann}, Jon and {Dawson}, Kyle S. and {Eisenstein}, Daniel J. and {Font-Ribera}, Andreu and {Ho}, Shirley and {McBride}, Cameron K. and {Montesano}, Francesco and {Percival}, Will J. and {Ross}, Ashley J. and {Ross}, Nicholas P. and {Samushia}, Lado and {Schlegel}, David J. and {S{\'a}nchez}, Ariel G. and {Tinker}, Jeremy L. and {Weaver}, Benjamin A.},
        title = "{The clustering of galaxies in the SDSS-III Baryon Oscillation Spectroscopic Survey: testing gravity with redshift space distortions using the power spectrum multipoles}",
      journal = {\mnras},
         year = 2014,
        month = sep,
       volume = {443},
       number = {2},
        pages = {1065-1089},
          doi = {10.1093/mnras/stu1051},
archivePrefix = {arXiv},
       eprint = {1312.4611},
 primaryClass = {astro-ph.CO},
       adsurl = {https://ui.adsabs.harvard.edu/abs/2014MNRAS.443.1065B}
}

@ARTICLE{Chan2012,
       author = {{Chan}, Kwan Chuen and {Scoccimarro}, Rom{\'a}n and {Sheth}, Ravi K.},
        title = "{Gravity and large-scale nonlocal bias}",
      journal = {\prd},
         year = 2012,
        month = apr,
       volume = {85},
       number = {8},
          eid = {083509},
        pages = {083509},
          doi = {10.1103/PhysRevD.85.083509},
archivePrefix = {arXiv},
       eprint = {1201.3614},
 primaryClass = {astro-ph.CO},
       adsurl = {https://ui.adsabs.harvard.edu/abs/2012PhRvD..85h3509C}
}

@ARTICLE{Sheth2013,
       author = {{Sheth}, Ravi K. and {Chan}, Kwan Chuen and {Scoccimarro}, Rom{\'a}n},
        title = "{Nonlocal Lagrangian bias}",
      journal = {\prd},
         year = 2013,
        month = apr,
       volume = {87},
       number = {8},
          eid = {083002},
        pages = {083002},
          doi = {10.1103/PhysRevD.87.083002},
archivePrefix = {arXiv},
       eprint = {1207.7117},
 primaryClass = {astro-ph.CO},
       adsurl = {https://ui.adsabs.harvard.edu/abs/2013PhRvD..87h3002S}
}

@ARTICLE{Saito2014,
       author = {{Saito}, Shun and {Baldauf}, Tobias and {Vlah}, Zvonimir and {Seljak}, Uro{\v{s}} and {Okumura}, Teppei and {McDonald}, Patrick},
        title = "{Understanding higher-order nonlocal halo bias at large scales by combining the power spectrum with the bispectrum}",
      journal = {\prd},
         year = 2014,
        month = dec,
       volume = {90},
       number = {12},
          eid = {123522},
        pages = {123522},
          doi = {10.1103/PhysRevD.90.123522},
archivePrefix = {arXiv},
       eprint = {1405.1447},
 primaryClass = {astro-ph.CO},
       adsurl = {https://ui.adsabs.harvard.edu/abs/2014PhRvD..90l3522S}
}

@ARTICLE{Mirbabayi2015,
       author = {{Mirbabayi}, Mehrdad and {Schmidt}, Fabian and {Zaldarriaga}, Matias},
        title = "{Biased tracers and time evolution}",
      journal = {\jcap},
         year = 2015,
        month = jul,
       volume = {2015},
       number = {7},
        pages = {030-030},
          doi = {10.1088/1475-7516/2015/07/030},
archivePrefix = {arXiv},
       eprint = {1412.5169},
 primaryClass = {astro-ph.CO},
       adsurl = {https://ui.adsabs.harvard.edu/abs/2015JCAP...07..030M}
}

@ARTICLE{Barreira2021,
       author = {{Barreira}, Alexandre and {Lazeyras}, Titouan and {Schmidt}, Fabian},
        title = "{Galaxy bias from forward models: linear and second-order bias of IllustrisTNG galaxies}",
      journal = {\jcap},
         year = 2021,
        month = aug,
       volume = {2021},
       number = {8},
          eid = {029},
        pages = {029},
          doi = {10.1088/1475-7516/2021/08/029},
archivePrefix = {arXiv},
       eprint = {2105.02876},
 primaryClass = {astro-ph.CO},
       adsurl = {https://ui.adsabs.harvard.edu/abs/2021JCAP...08..029B}
}

@ARTICLE{Jackson1972,
       author = {{Jackson}, J.~C.},
        title = "{A critique of Rees's theory of primordial gravitational radiation}",
      journal = {\mnras},
         year = 1972,
        month = jan,
       volume = {156},
        pages = {1P},
          doi = {10.1093/mnras/156.1.1P},
archivePrefix = {arXiv},
       eprint = {0810.3908},
 primaryClass = {astro-ph},
       adsurl = {https://ui.adsabs.harvard.edu/abs/1972MNRAS.156P...1J}
}

@ARTICLE{Taruya2013,
       author = {{Taruya}, Atsushi and {Nishimichi}, Takahiro and {Bernardeau}, Francis},
        title = "{Precision modeling of redshift-space distortions from a multipoint propagator expansion}",
      journal = {\prd},
         year = 2013,
        month = apr,
       volume = {87},
       number = {8},
          eid = {083509},
        pages = {083509},
          doi = {10.1103/PhysRevD.87.083509},
archivePrefix = {arXiv},
       eprint = {1301.3624},
 primaryClass = {astro-ph.CO},
       adsurl = {https://ui.adsabs.harvard.edu/abs/2013PhRvD..87h3509T}
}

@ARTICLE{Springel2001,
       author = {{Springel}, Volker and {White}, Simon D.~M. and {Tormen}, Giuseppe and {Kauffmann}, Guinevere},
        title = "{Populating a cluster of galaxies - I. Results at z=0}",
      journal = {\mnras},
         year = 2001,
        month = dec,
       volume = {328},
       number = {3},
        pages = {726-750},
          doi = {10.1046/j.1365-8711.2001.04912.x},
archivePrefix = {arXiv},
       eprint = {astro-ph/0012055},
 primaryClass = {astro-ph},
       adsurl = {https://ui.adsabs.harvard.edu/abs/2001MNRAS.328..726S}
}

@ARTICLE{Weinberger2017,
       author = {{Weinberger}, Rainer and {Springel}, Volker and {Hernquist}, Lars and {Pillepich}, Annalisa and {Marinacci}, Federico and {Pakmor}, R{\"u}diger and {Nelson}, Dylan and {Genel}, Shy and {Vogelsberger}, Mark and {Naiman}, Jill and {Torrey}, Paul},
        title = "{Simulating galaxy formation with black hole driven thermal and kinetic feedback}",
      journal = {\mnras},
         year = 2017,
        month = mar,
       volume = {465},
       number = {3},
        pages = {3291-3308},
          doi = {10.1093/mnras/stw2944},
archivePrefix = {arXiv},
       eprint = {1607.03486},
 primaryClass = {astro-ph.GA},
       adsurl = {https://ui.adsabs.harvard.edu/abs/2017MNRAS.465.3291W}
}

@ARTICLE{Pillepich2018a,
       author = {{Pillepich}, Annalisa and {Springel}, Volker and {Nelson}, Dylan and {Genel}, Shy and {Naiman}, Jill and {Pakmor}, R{\"u}diger and {Hernquist}, Lars and {Torrey}, Paul and {Vogelsberger}, Mark and {Weinberger}, Rainer and {Marinacci}, Federico},
        title = "{Simulating galaxy formation with the IllustrisTNG model}",
      journal = {\mnras},
         year = 2018,
        month = jan,
       volume = {473},
       number = {3},
        pages = {4077-4106},
          doi = {10.1093/mnras/stx2656},
archivePrefix = {arXiv},
       eprint = {1703.02970},
 primaryClass = {astro-ph.GA},
       adsurl = {https://ui.adsabs.harvard.edu/abs/2018MNRAS.473.4077P}
}

@ARTICLE{Rapoport2025,
       author = {{Rapoport}, Ivan and {Desjacques}, Vincent and {Parimbelli}, Gabriele and {Behar}, Ehud and {Crocce}, Martin},
        title = "{Spatially Resolved Modeling of Galactic H{\ensuremath{\alpha}} Emission for Galaxy Clustering}",
      journal = {\apj},
         year = 2025,
        month = jul,
       volume = {988},
       number = {1},
          eid = {44},
        pages = {44},
          doi = {10.3847/1538-4357/adde4c},
archivePrefix = {arXiv},
       eprint = {2502.08778},
 primaryClass = {astro-ph.GA},
       adsurl = {https://ui.adsabs.harvard.edu/abs/2025ApJ...988...44R}
}

@ARTICLE{Nelson2019,
       author = {{Nelson}, Dylan and {Springel}, Volker and {Pillepich}, Annalisa and {Rodriguez-Gomez}, Vicente and {Torrey}, Paul and {Genel}, Shy and {Vogelsberger}, Mark and {Pakmor}, Ruediger and {Marinacci}, Federico and {Weinberger}, Rainer and {Kelley}, Luke and {Lovell}, Mark and {Diemer}, Benedikt and {Hernquist}, Lars},
        title = "{The IllustrisTNG simulations: public data release}",
      journal = {Computational Astrophysics and Cosmology},
         year = 2019,
        month = may,
       volume = {6},
       number = {1},
          eid = {2},
        pages = {2},
          doi = {10.1186/s40668-019-0028-x},
archivePrefix = {arXiv},
       eprint = {1812.05609},
 primaryClass = {astro-ph.GA},
       adsurl = {https://ui.adsabs.harvard.edu/abs/2019ComAC...6....2N}
}

@ARTICLE{Ferland2017,
       author = {{Ferland}, G.~J. and {Chatzikos}, M. and {Guzm{\'a}n}, F. and {Lykins}, M.~L. and {van Hoof}, P.~A.~M. and {Williams}, R.~J.~R. and {Abel}, N.~P. and {Badnell}, N.~R. and {Keenan}, F.~P. and {Porter}, R.~L. and {Stancil}, P.~C.},
        title = "{The 2017 Release Cloudy}",
      journal = {\rmxaa},
         year = 2017,
        month = oct,
       volume = {53},
        pages = {385-438},
          doi = {10.48550/arXiv.1705.10877},
archivePrefix = {arXiv},
       eprint = {1705.10877},
 primaryClass = {astro-ph.GA},
       adsurl = {https://ui.adsabs.harvard.edu/abs/2017RMxAA..53..385F}
}

@ARTICLE{Salim2020,
       author = {{Salim}, Samir and {Narayanan}, Desika},
        title = "{The Dust Attenuation Law in Galaxies}",
      journal = {\araa},
         year = 2020,
        month = aug,
       volume = {58},
        pages = {529-575},
          doi = {10.1146/annurev-astro-032620-021933},
archivePrefix = {arXiv},
       eprint = {2001.03181},
 primaryClass = {astro-ph.GA},
       adsurl = {https://ui.adsabs.harvard.edu/abs/2020ARA&A..58..529S}
}

@ARTICLE{Charlot2000,
       author = {{Charlot}, St{\'e}phane and {Fall}, S. Michael},
        title = "{A Simple Model for the Absorption of Starlight by Dust in Galaxies}",
      journal = {\apj},
         year = 2000,
        month = aug,
       volume = {539},
       number = {2},
        pages = {718-731},
          doi = {10.1086/309250},
archivePrefix = {arXiv},
       eprint = {astro-ph/0003128},
 primaryClass = {astro-ph},
       adsurl = {https://ui.adsabs.harvard.edu/abs/2000ApJ...539..718C}
}

@ARTICLE{Gutkin2016,
       author = {{Gutkin}, Julia and {Charlot}, St{\'e}phane and {Bruzual}, Gustavo},
        title = "{Modelling the nebular emission from primeval to present-day star-forming galaxies}",
      journal = {\mnras},
         year = 2016,
        month = oct,
       volume = {462},
       number = {2},
        pages = {1757-1774},
          doi = {10.1093/mnras/stw1716},
archivePrefix = {arXiv},
       eprint = {1607.06086},
 primaryClass = {astro-ph.GA},
       adsurl = {https://ui.adsabs.harvard.edu/abs/2016MNRAS.462.1757G}
}

@ARTICLE{Osato2023,
       author = {{Osato}, Ken and {Okumura}, Teppei},
        title = "{Clustering of emission line galaxies with IllustrisTNG - I. Fundamental properties and halo occupation distribution}",
      journal = {\mnras},
         year = 2023,
        month = feb,
       volume = {519},
       number = {2},
        pages = {1771-1791},
          doi = {10.1093/mnras/stac3582},
archivePrefix = {arXiv},
       eprint = {2206.08678},
 primaryClass = {astro-ph.GA},
       adsurl = {https://ui.adsabs.harvard.edu/abs/2023MNRAS.519.1771O}
}

@ARTICLE{Favole2020,
       author = {{Favole}, G. and {Gonzalez-Perez}, V. and {Stoppacher}, D. and {Orsi}, {\'A}. and {Comparat}, J. and {Cora}, S.~A. and {Vega-Mart{\'\i}nez}, C.~A. and {Stevens}, A.~R.~H. and {Maraston}, C. and {Croton}, D. and {Knebe}, A. and {Benson}, A.~J. and {Montero-Dorta}, A.~D. and {Padilla}, N. and {Prada}, F. and {Thomas}, D.},
        title = "{[O II] emitters in MultiDark-Galaxies and DEEP2}",
      journal = {\mnras},
         year = 2020,
        month = oct,
       volume = {497},
       number = {4},
        pages = {5432-5453},
          doi = {10.1093/mnras/staa2292},
archivePrefix = {arXiv},
       eprint = {1908.05626},
 primaryClass = {astro-ph.GA},
       adsurl = {https://ui.adsabs.harvard.edu/abs/2020MNRAS.497.5432F}
}

@ARTICLE{Favole2022,
       author = {{Favole}, Ginevra and {Montero-Dorta}, Antonio D. and {Artale}, M. Celeste and {Contreras}, Sergio and {Zehavi}, Idit and {Xu}, Xiaoju},
        title = "{Subhalo abundance matching through the lens of a hydrodynamical simulation}",
      journal = {\mnras},
         year = 2022,
        month = jan,
       volume = {509},
       number = {2},
        pages = {1614-1625},
          doi = {10.1093/mnras/stab3006},
archivePrefix = {arXiv},
       eprint = {2101.10733},
 primaryClass = {astro-ph.GA},
       adsurl = {https://ui.adsabs.harvard.edu/abs/2022MNRAS.509.1614F}
}

@ARTICLE{Hadzhiyska2021,
       author = {{Hadzhiyska}, Boryana and {Tacchella}, Sandro and {Bose}, Sownak and {Eisenstein}, Daniel J.},
        title = "{The galaxy-halo connection of emission-line galaxies in IllustrisTNG}",
      journal = {\mnras},
         year = 2021,
        month = apr,
       volume = {502},
       number = {3},
        pages = {3599-3617},
          doi = {10.1093/mnras/stab243},
archivePrefix = {arXiv},
       eprint = {2011.05331},
 primaryClass = {astro-ph.GA},
       adsurl = {https://ui.adsabs.harvard.edu/abs/2021MNRAS.502.3599H}
}

@ARTICLE{Hadzhiyska2023b,
       author = {{Hadzhiyska}, Boryana and {Eisenstein}, Daniel and {Hernquist}, Lars and {Pakmor}, R{\"u}diger and {Bose}, Sownak and {Delgado}, Ana Maria and {Contreras}, Sergio and {Kannan}, Rahul and {White}, Simon D.~M. and {Springel}, Volker and {Frenk}, Carlos and {Hern{\'a}ndez-Aguayo}, C{\'e}sar and {Barrera}, Fulvio Ferlito and {Monica}},
        title = "{The MillenniumTNG Project: an improved two-halo model for the galaxy-halo connection of red and blue galaxies}",
      journal = {\mnras},
         year = 2023,
        month = sep,
       volume = {524},
       number = {2},
        pages = {2507-2523},
          doi = {10.1093/mnras/stad731},
archivePrefix = {arXiv},
       eprint = {2210.10072},
 primaryClass = {astro-ph.CO},
       adsurl = {https://ui.adsabs.harvard.edu/abs/2023MNRAS.524.2507H}
}

@ARTICLE{Hadzhiyska2023a,
       author = {{Hadzhiyska}, Boryana and {Hernquist}, Lars and {Eisenstein}, Daniel and {Delgado}, Ana Maria and {Bose}, Sownak and {Kannan}, Rahul and {Pakmor}, R{\"u}diger and {Springel}, Volker and {Contreras}, Sergio and {Barrera}, Monica and {Ferlito}, Fulvio and {Hern{\'a}ndez-Aguayo}, C{\'e}sar and {White}, Simon D.~M. and {Frenk}, Carlos},
        title = "{The MillenniumTNG Project: refining the one-halo model of red and blue galaxies at different redshifts}",
      journal = {\mnras},
         year = 2023,
        month = sep,
       volume = {524},
       number = {2},
        pages = {2524-2538},
          doi = {10.1093/mnras/stad279},
archivePrefix = {arXiv},
       eprint = {2210.10068},
 primaryClass = {astro-ph.CO},
       adsurl = {https://ui.adsabs.harvard.edu/abs/2023MNRAS.524.2524H}
}

@ARTICLE{Gao2022,
       author = {{Gao}, Hongyu and {Jing}, Y.~P. and {Zheng}, Yun and {Xu}, Kun},
        title = "{Constructing the Emission-line Galaxy-Host Halo Connection through Auto and Cross Correlations}",
      journal = {\apj},
         year = 2022,
        month = mar,
       volume = {928},
       number = {1},
          eid = {10},
        pages = {10},
          doi = {10.3847/1538-4357/ac501b},
archivePrefix = {arXiv},
       eprint = {2111.11657},
 primaryClass = {astro-ph.GA},
       adsurl = {https://ui.adsabs.harvard.edu/abs/2022ApJ...928...10G}
}

@ARTICLE{Yuan2025,
       author = {{Yuan}, Sihan and {Wechsler}, Risa H. and {Wang}, Yunchong and {de los Reyes}, Mithi A.~C. and {Myles}, Justin and {Rocher}, Antoine and {Hadzhiyska}, Boryana and {Aguilar}, Jessica Nicole and {Ahlen}, Steven and {Brooks}, David and {Claybaugh}, Todd and {Cole}, Shaun and {de la Macorra}, Axel and {Forero-Romero}, Jaime E. and {Gontcho}, Satya Gontcho A. and {Guy}, Julien and {Honscheid}, Klaus and {Kisner}, Theodore and {Levi}, Michael and {Manera}, Marc and {Meisner}, Aaron and {Miquel}, Ramon and {Moustakas}, John and {Nie}, Jundan and {Palanque-Delabrouille}, Nathalie and {Poppett}, Claire and {Rezaie}, Mehdi and {Ross}, Ashley J. and {Rossi}, Graziano and {Sanchez}, Eusebio and {Schubnell}, Michael and {Seo}, Hee-Jong and {Tarl{\'e}}, Gregory and {Weaver}, Benjamin Alan and {Zhou}, Zhimin},
        title = "{Unraveling emission line galaxy conformity at z \raisebox{-0.5ex}\textasciitilde 1 with DESI early data}",
      journal = {\mnras},
         year = 2025,
        month = apr,
       volume = {538},
       number = {2},
        pages = {1216-1240},
          doi = {10.1093/mnras/staf368},
archivePrefix = {arXiv},
       eprint = {2310.09329},
 primaryClass = {astro-ph.CO},
       adsurl = {https://ui.adsabs.harvard.edu/abs/2025MNRAS.538.1216Y}
}

@ARTICLE{Yuan2022,
       author = {{Yuan}, Sihan and {Hadzhiyska}, Boryana and {Bose}, Sownak and {Eisenstein}, Daniel J.},
        title = "{Illustrating galaxy-halo connection in the DESI era with ILLUSTRISTNG}",
      journal = {\mnras},
         year = 2022,
        month = jun,
       volume = {512},
       number = {4},
        pages = {5793-5811},
          doi = {10.1093/mnras/stac830},
archivePrefix = {arXiv},
       eprint = {2202.12911},
 primaryClass = {astro-ph.CO},
       adsurl = {https://ui.adsabs.harvard.edu/abs/2022MNRAS.512.5793Y}
}

@ARTICLE{Avila2020,
       author = {{Avila}, S. and {Gonzalez-Perez}, V. and {Mohammad}, F.~G. and {de Mattia}, A. and {Zhao}, C. and {Raichoor}, A. and {Tamone}, A. and {Alam}, S. and {Bautista}, J. and {Bianchi}, D. and {Burtin}, E. and {Chapman}, M.~J. and {Chuang}, C. -H. and {Comparat}, J. and {Dawson}, K. and {Divers}, T. and {du Mas des Bourboux}, H. and {Gil-Marin}, H. and {Mueller}, E.~M. and {Habib}, S. and {Heitmann}, K. and {Ruhlmann-Kleider}, V. and {Padilla}, N. and {Percival}, W.~J. and {Ross}, A.~J. and {Seo}, H.~J. and {Schneider}, D.~P. and {Zhao}, G.},
        title = "{The Completed SDSS-IV extended Baryon Oscillation Spectroscopic Survey: exploring the halo occupation distribution model for emission line galaxies}",
      journal = {\mnras},
         year = 2020,
        month = dec,
       volume = {499},
       number = {4},
        pages = {5486-5507},
          doi = {10.1093/mnras/staa2951},
archivePrefix = {arXiv},
       eprint = {2007.09012},
 primaryClass = {astro-ph.CO},
       adsurl = {https://ui.adsabs.harvard.edu/abs/2020MNRAS.499.5486A}
}

@ARTICLE{Ortega-Martinez2025,
       author = {{Ortega-Martinez}, Sara and {Contreras}, Sergio and {Angulo}, Raul E. and {Chaves-Montero}, Jon{\'a}s},
        title = "{Investigating the galaxy{\textendash}halo connection of DESI emission-line galaxies with SHAMe-SF}",
      journal = {\aap},
         year = 2025,
        month = may,
       volume = {697},
          eid = {A226},
        pages = {A226},
          doi = {10.1051/0004-6361/202453086},
archivePrefix = {arXiv},
       eprint = {2411.11830},
 primaryClass = {astro-ph.CO},
       adsurl = {https://ui.adsabs.harvard.edu/abs/2025A&A...697A.226O}
}

@ARTICLE{Okumura2021,
       author = {{Okumura}, Teppei and {Hayashi}, Masao and {Chiu}, I. -Non and {Lin}, Yen-Ting and {Osato}, Ken and {Hsieh}, Bau-Ching and {Lin}, Sheng-Chieh},
        title = "{Angular clustering and host halo properties of [O II] emitters at z > 1 in the Subaru HSC survey}",
      journal = {\pasj},
         year = 2021,
        month = aug,
       volume = {73},
       number = {4},
        pages = {1186-1207},
          doi = {10.1093/pasj/psab068},
archivePrefix = {arXiv},
       eprint = {2012.12224},
 primaryClass = {astro-ph.GA},
       adsurl = {https://ui.adsabs.harvard.edu/abs/2021PASJ...73.1186O}
}

@ARTICLE{Hayashi2018,
       author = {{Hayashi}, Masao and {Tanaka}, Masayuki and {Shimakawa}, Rhythm and {Furusawa}, Hisanori and {Momose}, Rieko and {Koyama}, Yusei and {Silverman}, John D. and {Kodama}, Tadayuki and {Komiyama}, Yutaka and {Leauthaud}, Alexie and {Lin}, Yen-Ting and {Miyazaki}, Satoshi and {Nagao}, Tohru and {Nishizawa}, Atsushi J. and {Ouchi}, Masami and {Shibuya}, Takatoshi and {Tadaki}, Ken-ichi and {Yabe}, Kiyoto},
        title = "{A 16 deg$^{2}$ survey of emission-line galaxies at z < 1.5 in HSC-SSP Public Data Release 1}",
      journal = {\pasj},
         year = 2018,
        month = jan,
       volume = {70},
          eid = {S17},
        pages = {S17},
          doi = {10.1093/pasj/psx088},
archivePrefix = {arXiv},
       eprint = {1704.05978},
 primaryClass = {astro-ph.GA},
       adsurl = {https://ui.adsabs.harvard.edu/abs/2018PASJ...70S..17H}
}

@ARTICLE{Hayashi2020,
       author = {{Hayashi}, Masao and {Shimakawa}, Rhythm and {Tanaka}, Masayuki and {Onodera}, Masato and {Koyama}, Yusei and {Inoue}, Akio K. and {Komiyama}, Yutaka and {Lee}, Chien-Hsiu and {Lin}, Yen-Ting and {Yabe}, Kiyoto},
        title = "{A 16 deg$^{2}$ survey of emission-line galaxies at z < 1.6 from HSC-SSP PDR2 and CHORUS}",
      journal = {\pasj},
         year = 2020,
        month = oct,
       volume = {72},
       number = {5},
          eid = {86},
        pages = {86},
          doi = {10.1093/pasj/psaa076},
archivePrefix = {arXiv},
       eprint = {2007.07413},
 primaryClass = {astro-ph.GA},
       adsurl = {https://ui.adsabs.harvard.edu/abs/2020PASJ...72...86H}
}

@ARTICLE{DESI2024VII,
       author = {{DESI Collaboration}},
        title = "{DESI 2024 VII: cosmological constraints from the full-shape modeling of clustering measurements}",
      journal = {\jcap},
         year = 2025,
        month = jul,
       volume = {2025},
       number = {7},
          eid = {028},
        pages = {028},
          doi = {10.1088/1475-7516/2025/07/028},
archivePrefix = {arXiv},
       eprint = {2411.12022},
 primaryClass = {astro-ph.CO},
       adsurl = {https://ui.adsabs.harvard.edu/abs/2025JCAP...07..028A}
}

@ARTICLE{DESI2024VI,
       author = {{DESI Collaboration}},
        title = "{DESI 2024 VI: cosmological constraints from the measurements of baryon acoustic oscillations}",
      journal = {\jcap},
         year = 2025,
        month = feb,
       volume = {2025},
       number = {2},
          eid = {021},
        pages = {021},
          doi = {10.1088/1475-7516/2025/02/021},
archivePrefix = {arXiv},
       eprint = {2404.03002},
 primaryClass = {astro-ph.CO},
       adsurl = {https://ui.adsabs.harvard.edu/abs/2025JCAP...02..021A}
}

@ARTICLE{DESI2024V,
       author = {{DESI Collaboration}},
        title = "{DESI 2024 V: Full-Shape galaxy clustering from galaxies and quasars}",
      journal = {\jcap},
         year = 2025,
        month = sep,
       volume = {2025},
       number = {9},
          eid = {008},
        pages = {008},
          doi = {10.1088/1475-7516/2025/09/008},
archivePrefix = {arXiv},
       eprint = {2411.12021},
 primaryClass = {astro-ph.CO},
       adsurl = {https://ui.adsabs.harvard.edu/abs/2025JCAP...09..008A}
}

@ARTICLE{DESI2024III,
       author = {{DESI Collaboration}},
        title = "{DESI 2024 III: baryon acoustic oscillations from galaxies and quasars}",
      journal = {\jcap},
         year = 2025,
        month = apr,
       volume = {2025},
       number = {4},
          eid = {012},
        pages = {012},
          doi = {10.1088/1475-7516/2025/04/012},
archivePrefix = {arXiv},
       eprint = {2404.03000},
 primaryClass = {astro-ph.CO},
       adsurl = {https://ui.adsabs.harvard.edu/abs/2025JCAP...04..012A}
}

@ARTICLE{Tamone2020,
       author = {{Tamone}, Am{\'e}lie and {Raichoor}, Anand and {Zhao}, Cheng and {de Mattia}, Arnaud and {Gorgoni}, Claudio and {Burtin}, Etienne and {Ruhlmann-Kleider}, Vanina and {Ross}, Ashley J. and {Alam}, Shadab and {Percival}, Will J. and {Avila}, Santiago and {Chapman}, Michael J. and {Chuang}, Chia-Hsun and {Comparat}, Johan and {Dawson}, Kyle S. and {de la Torre}, Sylvain and {du Mas des Bourboux}, H{\'e}lion and {Escoffier}, Stephanie and {Gonzalez-Perez}, Violeta and {Hou}, Jiamin and {Kneib}, Jean-Paul and {Mohammad}, Faizan G. and {Mueller}, Eva-Maria and {Paviot}, Romain and {Rossi}, Graziano and {Schneider}, Donald P. and {Wang}, Yuting and {Zhao}, Gong-Bo},
        title = "{The completed SDSS-IV extended baryon oscillation spectroscopic survey: growth rate of structure measurement from anisotropic clustering analysis in configuration space between redshift 0.6 and 1.1 for the emission-line galaxy sample}",
      journal = {\mnras},
         year = 2020,
        month = dec,
       volume = {499},
       number = {4},
        pages = {5527-5546},
          doi = {10.1093/mnras/staa3050},
archivePrefix = {arXiv},
       eprint = {2007.09009},
 primaryClass = {astro-ph.CO},
       adsurl = {https://ui.adsabs.harvard.edu/abs/2020MNRAS.499.5527T}
}

@ARTICLE{deMattia2021,
       author = {{de Mattia}, Arnaud and {Ruhlmann-Kleider}, Vanina and {Raichoor}, Anand and {Ross}, Ashley J. and {Tamone}, Am{\'e}lie and {Zhao}, Cheng and {Alam}, Shadab and {Avila}, Santiago and {Burtin}, Etienne and {Bautista}, Julian and {Beutler}, Florian and {Brinkmann}, Jonathan and {Brownstein}, Joel R. and {Chapman}, Michael J. and {Chuang}, Chia-Hsun and {Comparat}, Johan and {du Mas des Bourboux}, H{\'e}lion and {Dawson}, Kyle S. and {de la Macorra}, Axel and {Gil-Mar{\'\i}n}, H{\'e}ctor and {Gonzalez-Perez}, Violeta and {Gorgoni}, Claudio and {Hou}, Jiamin and {Kong}, Hui and {Lin}, Sicheng and {Nadathur}, Seshadri and {Newman}, Jeffrey A. and {Mueller}, Eva-Maria and {Percival}, Will J. and {Rezaie}, Mehdi and {Rossi}, Graziano and {Schneider}, Donald P. and {Tiwari}, Prabhakar and {Vivek}, M. and {Wang}, Yuting and {Zhao}, Gong-Bo},
        title = "{The completed SDSS-IV extended Baryon Oscillation Spectroscopic Survey: measurement of the BAO and growth rate of structure of the emission line galaxy sample from the anisotropic power spectrum between redshift 0.6 and 1.1}",
      journal = {\mnras},
         year = 2021,
        month = mar,
       volume = {501},
       number = {4},
        pages = {5616-5645},
          doi = {10.1093/mnras/staa3891},
archivePrefix = {arXiv},
       eprint = {2007.09008},
 primaryClass = {astro-ph.CO},
       adsurl = {https://ui.adsabs.harvard.edu/abs/2021MNRAS.501.5616D}
}

@ARTICLE{Raichoor2021,
       author = {{Raichoor}, Anand and {de Mattia}, Arnaud and {Ross}, Ashley J. and {Zhao}, Cheng and {Alam}, Shadab and {Avila}, Santiago and {Bautista}, Julian and {Brinkmann}, Jonathan and {Brownstein}, Joel R. and {Burtin}, Etienne and {Chapman}, Michael J. and {Chuang}, Chia-Hsun and {Comparat}, Johan and {Dawson}, Kyle S. and {Dey}, Arjun and {du Mas des Bourboux}, H{\'e}lion and {Elvin-Poole}, Jack and {Gonzalez-Perez}, Violeta and {Gorgoni}, Claudio and {Kneib}, Jean-Paul and {Kong}, Hui and {Lang}, Dustin and {Moustakas}, John and {Myers}, Adam D. and {M{\"u}ller}, Eva-Maria and {Nadathur}, Seshadri and {Newman}, Jeffrey A. and {Percival}, Will J. and {Rezaie}, Mehdi and {Rossi}, Graziano and {Ruhlmann-Kleider}, Vanina and {Schlegel}, David J. and {Schneider}, Donald P. and {Seo}, Hee-Jong and {Tamone}, Am{\'e}lie and {Tinker}, Jeremy L. and {Tojeiro}, Rita and {Vivek}, M. and {Y{\`e}che}, Christophe and {Zhao}, Gong-Bo},
        title = "{The completed SDSS-IV extended Baryon Oscillation Spectroscopic Survey: large-scale structure catalogues and measurement of the isotropic BAO between redshift 0.6 and 1.1 for the Emission Line Galaxy Sample}",
      journal = {\mnras},
         year = 2021,
        month = jan,
       volume = {500},
       number = {3},
        pages = {3254-3274},
          doi = {10.1093/mnras/staa3336},
archivePrefix = {arXiv},
       eprint = {2007.09007},
 primaryClass = {astro-ph.CO},
       adsurl = {https://ui.adsabs.harvard.edu/abs/2021MNRAS.500.3254R}
}

@ARTICLE{Okumura2016,
       author = {{Okumura}, Teppei and {Hikage}, Chiaki and {Totani}, Tomonori and {Tonegawa}, Motonari and {Okada}, Hiroyuki and {Glazebrook}, Karl and {Blake}, Chris and {Ferreira}, Pedro G. and {More}, Surhud and {Taruya}, Atsushi and {Tsujikawa}, Shinji and {Akiyama}, Masayuki and {Dalton}, Gavin and {Goto}, Tomotsugu and {Ishikawa}, Takashi and {Iwamuro}, Fumihide and {Matsubara}, Takahiko and {Nishimichi}, Takahiro and {Ohta}, Kouji and {Shimizu}, Ikkoh and {Takahashi}, Ryuichi and {Takato}, Naruhisa and {Tamura}, Naoyuki and {Yabe}, Kiyoto and {Yoshida}, Naoki},
        title = "{The Subaru FMOS galaxy redshift survey (FastSound). IV. New constraint on gravity theory from redshift space distortions at z {\ensuremath{\sim}} 1.4}",
      journal = {\pasj},
         year = 2016,
        month = jun,
       volume = {68},
       number = {3},
          eid = {38},
        pages = {38},
          doi = {10.1093/pasj/psw029},
archivePrefix = {arXiv},
       eprint = {1511.08083},
 primaryClass = {astro-ph.CO},
       adsurl = {https://ui.adsabs.harvard.edu/abs/2016PASJ...68...38O}
}

@ARTICLE{Okada2016,
       author = {{Okada}, Hiroyuki and {Totani}, Tomonori and {Tonegawa}, Motonari and {Akiyama}, Masayuki and {Dalton}, Gavin and {Glazebrook}, Karl and {Iwamuro}, Fumihide and {Ohta}, Kouji and {Takato}, Naruhisa and {Tamura}, Naoyuki and {Yabe}, Kiyoto and {Bunker}, Andrew J. and {Goto}, Tomotsugu and {Hikage}, Chiaki and {Ishikawa}, Takashi and {Okumura}, Teppei and {Shimizu}, Ikkoh},
        title = "{The Subaru FMOS galaxy redshift survey (FastSound). II. The emission line catalog and properties of emission line galaxies}",
      journal = {\pasj},
         year = 2016,
        month = jun,
       volume = {68},
       number = {3},
          eid = {47},
        pages = {47},
          doi = {10.1093/pasj/psw043},
archivePrefix = {arXiv},
       eprint = {1504.05592},
 primaryClass = {astro-ph.GA},
       adsurl = {https://ui.adsabs.harvard.edu/abs/2016PASJ...68...47O}
}

@ARTICLE{Tonegawa2015,
       author = {{Tonegawa}, Motonari and {Totani}, Tomonori and {Okada}, Hiroyuki and {Akiyama}, Masayuki and {Dalton}, Gavin and {Glazebrook}, Karl and {Iwamuro}, Fumihide and {Maihara}, Toshinori and {Ohta}, Kouji and {Shimizu}, Ikkoh and {Takato}, Naruhisa and {Tamura}, Naoyuki and {Yabe}, Kiyoto and {Bunker}, Andrew J. and {Coupon}, Jean and {Ferreira}, Pedro G. and {Frenk}, Carlos S. and {Goto}, Tomotsugu and {Hikage}, Chiaki and {Ishikawa}, Takashi and {Matsubara}, Takahiko and {More}, Surhud and {Okumura}, Teppei and {Percival}, Will J. and {Spitler}, Lee R. and {Szapudi}, Istvan},
        title = "{The Subaru FMOS galaxy redshift survey (FastSound). I. Overview of the survey targeting H{\ensuremath{\alpha}} emitters at z {\ensuremath{\sim}} 1.4}",
      journal = {\pasj},
         year = 2015,
        month = oct,
       volume = {67},
       number = {5},
          eid = {81},
        pages = {81},
          doi = {10.1093/pasj/psv044},
archivePrefix = {arXiv},
       eprint = {1502.07900},
 primaryClass = {astro-ph.CO},
       adsurl = {https://ui.adsabs.harvard.edu/abs/2015PASJ...67...81T}
}

@ARTICLE{Geach2008,
       author = {{Geach}, J.~E. and {Smail}, Ian and {Best}, P.~N. and {Kurk}, J. and {Casali}, M. and {Ivison}, R.~J. and {Coppin}, K.},
        title = "{HiZELS: a high-redshift survey of H{\ensuremath{\alpha}} emitters - I. The cosmic star formation rate and clustering at z = 2.23}",
      journal = {\mnras},
         year = 2008,
        month = aug,
       volume = {388},
       number = {4},
        pages = {1473-1486},
          doi = {10.1111/j.1365-2966.2008.13481.x},
archivePrefix = {arXiv},
       eprint = {0805.2861},
 primaryClass = {astro-ph},
       adsurl = {https://ui.adsabs.harvard.edu/abs/2008MNRAS.388.1473G}
}

@ARTICLE{Sobral2012,
       author = {{Sobral}, David and {Best}, Philip N. and {Matsuda}, Yuichi and {Smail}, Ian and {Geach}, James E. and {Cirasuolo}, Michele},
        title = "{Star formation at z=1.47 from HiZELS: an H>{\ensuremath{\alpha}}+[O II] double-blind study}",
      journal = {\mnras},
         year = 2012,
        month = mar,
       volume = {420},
       number = {3},
        pages = {1926-1945},
          doi = {10.1111/j.1365-2966.2011.19977.x},
archivePrefix = {arXiv},
       eprint = {1109.1830},
 primaryClass = {astro-ph.CO},
       adsurl = {https://ui.adsabs.harvard.edu/abs/2012MNRAS.420.1926S}
}

@ARTICLE{Rocher2023,
       author = {{Rocher}, Antoine and {Ruhlmann-Kleider}, Vanina and {Burtin}, Etienne and {Yuan}, Sihan and {de Mattia}, Arnaud and {Ross}, Ashley J. and {Aguilar}, Jessica and {Ahlen}, Steven and {Alam}, Shadab and {Bianchi}, Davide and {Brooks}, David and {Cole}, Shaun and {Dawson}, Kyle and {de la Macorra}, Axel and {Doel}, Peter and {Eisenstein}, Daniel J. and {Fanning}, Kevin and {Forero-Romero}, Jaime E. and {Garrison}, Lehman H. and {Gontcho A Gontcho}, Satya and {Gonzalez-Perez}, Violeta and {Guy}, Julien and {Hadzhiyska}, Boryana and {Hahn}, ChangHoon and {Honscheid}, Klaus and {Kisner}, Theodore and {Landriau}, Martin and {Lasker}, James and {E. Levi}, Michael and {Manera}, Marc and {Meisner}, Aaron and {Miquel}, Ramon and {Moustakas}, John and {Mueller}, Eva-Maria and {Newman}, Jeffrey A. and {Nie}, Jundan and {Percival}, Will J. and {Poppett}, Claire and {Qin}, Fei and {Rossi}, Graziano and {Samushia}, Lado and {Sanchez}, Eusebio and {Schlegel}, David and {Schubnell}, Michael and {Seo}, Hee-Jong and {Tarl{\'e}}, Gregory and {Vargas-Maga{\~n}a}, Mariana and {Weaver}, Benjamin A. and {Yu}, Jiaxi and {Zhang}, Hanyu and {Zheng}, Zheng and {Zhou}, Zhimin and {Zou}, Hu},
        title = "{The DESI One-Percent survey: exploring the Halo Occupation Distribution of Emission Line Galaxies with ABACUSSUMMIT simulations}",
      journal = {\jcap},
         year = 2023,
        month = oct,
       volume = {2023},
       number = {10},
          eid = {016},
        pages = {016},
          doi = {10.1088/1475-7516/2023/10/016},
archivePrefix = {arXiv},
       eprint = {2306.06319},
 primaryClass = {astro-ph.CO},
       adsurl = {https://ui.adsabs.harvard.edu/abs/2023JCAP...10..016R}
}

@ARTICLE{Ortega-Martinez2024,
       author = {{Ortega-Martinez}, S. and {Contreras}, S. and {Angulo}, R.},
        title = "{SHAMe-SF: Predicting the clustering of star-forming galaxies with an enhanced abundance matching model}",
      journal = {\aap},
         year = 2024,
        month = sep,
       volume = {689},
          eid = {A66},
        pages = {A66},
          doi = {10.1051/0004-6361/202449597},
archivePrefix = {arXiv},
       eprint = {2401.17374},
 primaryClass = {astro-ph.CO},
       adsurl = {https://ui.adsabs.harvard.edu/abs/2024A&A...689A..66O}
}

@ARTICLE{Spergel2015,
       author = {{Spergel}, D. and {Gehrels}, N. and {Baltay}, C. and {Bennett}, D. and {Breckinridge}, J. and {Donahue}, M. and {Dressler}, A. and {Gaudi}, B.~S. and {Greene}, T. and {Guyon}, O. and {Hirata}, C. and {Kalirai}, J. and {Kasdin}, N.~J. and {Macintosh}, B. and {Moos}, W. and {Perlmutter}, S. and {Postman}, M. and {Rauscher}, B. and {Rhodes}, J. and {Wang}, Y. and {Weinberg}, D. and {Benford}, D. and {Hudson}, M. and {Jeong}, W. -S. and {Mellier}, Y. and {Traub}, W. and {Yamada}, T. and {Capak}, P. and {Colbert}, J. and {Masters}, D. and {Penny}, M. and {Savransky}, D. and {Stern}, D. and {Zimmerman}, N. and {Barry}, R. and {Bartusek}, L. and {Carpenter}, K. and {Cheng}, E. and {Content}, D. and {Dekens}, F. and {Demers}, R. and {Grady}, K. and {Jackson}, C. and {Kuan}, G. and {Kruk}, J. and {Melton}, M. and {Nemati}, B. and {Parvin}, B. and {Poberezhskiy}, I. and {Peddie}, C. and {Ruffa}, J. and {Wallace}, J.~K. and {Whipple}, A. and {Wollack}, E. and {Zhao}, F.},
        title = "{Wide-Field InfrarRed Survey Telescope-Astrophysics Focused Telescope Assets WFIRST-AFTA 2015 Report}",
      journal = {arXiv e-prints},
         year = 2015,
        month = mar,
          eid = {arXiv:1503.03757},
        pages = {arXiv:1503.03757},
          doi = {10.48550/arXiv.1503.03757},
archivePrefix = {arXiv},
       eprint = {1503.03757},
 primaryClass = {astro-ph.IM},
       adsurl = {https://ui.adsabs.harvard.edu/abs/2015arXiv150303757S}
}

@ARTICLE{Akeson2019,
       author = {{Akeson}, Rachel and {Armus}, Lee and {Bachelet}, Etienne and {Bailey}, Vanessa and {Bartusek}, Lisa and {Bellini}, Andrea and {Benford}, Dominic and {Bennett}, David and {Bhattacharya}, Aparna and {Bohlin}, Ralph and {Boyer}, Martha and {Bozza}, Valerio and {Bryden}, Geoffrey and {Calchi Novati}, Sebastiano and {Carpenter}, Kenneth and {Casertano}, Stefano and {Choi}, Ami and {Content}, David and {Dayal}, Pratika and {Dressler}, Alan and {Dor{\'e}}, Olivier and {Fall}, S. Michael and {Fan}, Xiaohui and {Fang}, Xiao and {Filippenko}, Alexei and {Finkelstein}, Steven and {Foley}, Ryan and {Furlanetto}, Steven and {Kalirai}, Jason and {Gaudi}, B. Scott and {Gilbert}, Karoline and {Girard}, Julien and {Grady}, Kevin and {Greene}, Jenny and {Guhathakurta}, Puragra and {Heinrich}, Chen and {Hemmati}, Shoubaneh and {Hendel}, David and {Henderson}, Calen and {Henning}, Thomas and {Hirata}, Christopher and {Ho}, Shirley and {Huff}, Eric and {Hutter}, Anne and {Jansen}, Rolf and {Jha}, Saurabh and {Johnson}, Samson and {Jones}, David and {Kasdin}, Jeremy and {Kelly}, Patrick and {Kirshner}, Robert and {Koekemoer}, Anton and {Kruk}, Jeffrey and {Lewis}, Nikole and {Macintosh}, Bruce and {Madau}, Piero and {Malhotra}, Sangeeta and {Mandel}, Kaisey and {Massara}, Elena and {Masters}, Daniel and {McEnery}, Julie and {McQuinn}, Kristen and {Melchior}, Peter and {Melton}, Mark and {Mennesson}, Bertrand and {Peeples}, Molly and {Penny}, Matthew and {Perlmutter}, Saul and {Pisani}, Alice and {Plazas}, Andr{\'e}s and {Poleski}, Radek and {Postman}, Marc and {Ranc}, Cl{\'e}ment and {Rauscher}, Bernard and {Rest}, Armin and {Roberge}, Aki and {Robertson}, Brant and {Rodney}, Steven and {Rhoads}, James and {Rhodes}, Jason and {Ryan}, Jr., Russell and {Sahu}, Kailash and {Sand}, David and {Scolnic}, Dan and {Seth}, Anil and {Shvartzvald}, Yossi and {Siellez}, Karelle and {Smith}, Arfon and {Spergel}, David and {Stassun}, Keivan and {Street}, Rachel and {Strolger}, Louis-Gregory and {Szalay}, Alexander and {Trauger}, John and {Troxel}, M.~A. and {Turnbull}, Margaret and {van der Marel}, Roeland and {von der Linden}, Anja and {Wang}, Yun and {Weinberg}, David and {Williams}, Benjamin and {Windhorst}, Rogier and {Wollack}, Edward and {Wu}, Hao-Yi and {Yee}, Jennifer and {Zimmerman}, Neil},
        title = "{The Wide Field Infrared Survey Telescope: 100 Hubbles for the 2020s}",
      journal = {arXiv e-prints},
         year = 2019,
        month = feb,
          eid = {arXiv:1902.05569},
        pages = {arXiv:1902.05569},
          doi = {10.48550/arXiv.1902.05569},
archivePrefix = {arXiv},
       eprint = {1902.05569},
 primaryClass = {astro-ph.IM},
       adsurl = {https://ui.adsabs.harvard.edu/abs/2019arXiv190205569A}
}

@ARTICLE{Takada2014,
       author = {{Takada}, Masahiro and {Ellis}, Richard S. and {Chiba}, Masashi and {Greene}, Jenny E. and {Aihara}, Hiroaki and {Arimoto}, Nobuo and {Bundy}, Kevin and {Cohen}, Judith and {Dor{\'e}}, Olivier and {Graves}, Genevieve and {Gunn}, James E. and {Heckman}, Timothy and {Hirata}, Christopher M. and {Ho}, Paul and {Kneib}, Jean-Paul and {Le F{\`e}vre}, Olivier and {Lin}, Lihwai and {More}, Surhud and {Murayama}, Hitoshi and {Nagao}, Tohru and {Ouchi}, Masami and {Seiffert}, Michael and {Silverman}, John D. and {Sodr{\'e}}, Laerte and {Spergel}, David N. and {Strauss}, Michael A. and {Sugai}, Hajime and {Suto}, Yasushi and {Takami}, Hideki and {Wyse}, Rosemary},
        title = "{Extragalactic science, cosmology, and Galactic archaeology with the Subaru Prime Focus Spectrograph}",
      journal = {\pasj},
         year = 2014,
        month = feb,
       volume = {66},
       number = {1},
          eid = {R1},
        pages = {R1},
          doi = {10.1093/pasj/pst019},
archivePrefix = {arXiv},
       eprint = {1206.0737},
 primaryClass = {astro-ph.CO},
       adsurl = {https://ui.adsabs.harvard.edu/abs/2014PASJ...66R...1T}
}

@ARTICLE{EuclidI,
       author = {{Euclid Collaboration}},
        title = "{Euclid: I. Overview of the Euclid mission}",
      journal = {\aap},
         year = 2025,
        month = may,
       volume = {697},
          eid = {A1},
        pages = {A1},
          doi = {10.1051/0004-6361/202450810},
archivePrefix = {arXiv},
       eprint = {2405.13491},
 primaryClass = {astro-ph.CO},
       adsurl = {https://ui.adsabs.harvard.edu/abs/2025A&A...697A...1E}
}

@ARTICLE{DESI2022,
       author = {{DESI Collaboration}},
        title = "{Overview of the Instrumentation for the Dark Energy Spectroscopic Instrument}",
      journal = {\aj},
         year = 2022,
        month = nov,
       volume = {164},
       number = {5},
          eid = {207},
        pages = {207},
          doi = {10.3847/1538-3881/ac882b},
archivePrefix = {arXiv},
       eprint = {2205.10939},
 primaryClass = {astro-ph.IM},
       adsurl = {https://ui.adsabs.harvard.edu/abs/2022AJ....164..207D}
}

@ARTICLE{Padmanabhan2007,
       author = {{Padmanabhan}, Nikhil and {Schlegel}, David J. and {Seljak}, Uro{\v{s}} and {Makarov}, Alexey and {Bahcall}, Neta A. and {Blanton}, Michael R. and {Brinkmann}, Jonathan and {Eisenstein}, Daniel J. and {Finkbeiner}, Douglas P. and {Gunn}, James E. and {Hogg}, David W. and {Ivezi{\'c}}, {\v{Z}}eljko and {Knapp}, Gillian R. and {Loveday}, Jon and {Lupton}, Robert H. and {Nichol}, Robert C. and {Schneider}, Donald P. and {Strauss}, Michael A. and {Tegmark}, Max and {York}, Donald G.},
        title = "{The clustering of luminous red galaxies in the Sloan Digital Sky Survey imaging data}",
      journal = {\mnras},
         year = 2007,
        month = jul,
       volume = {378},
       number = {3},
        pages = {852-872},
          doi = {10.1111/j.1365-2966.2007.11593.x},
archivePrefix = {arXiv},
       eprint = {astro-ph/0605302},
 primaryClass = {astro-ph},
       adsurl = {https://ui.adsabs.harvard.edu/abs/2007MNRAS.378..852P}
}

@ARTICLE{Dawson2013,
       author = {{Dawson}, Kyle S. and {Schlegel}, David J. and {Ahn}, Christopher P. and {Anderson}, Scott F. and {Aubourg}, {\'E}ric and {Bailey}, Stephen and {Barkhouser}, Robert H. and {Bautista}, Julian E. and {Beifiori}, Alessandra and {Berlind}, Andreas A. and {Bhardwaj}, Vaishali and {Bizyaev}, Dmitry and {Blake}, Cullen H. and {Blanton}, Michael R. and {Blomqvist}, Michael and {Bolton}, Adam S. and {Borde}, Arnaud and {Bovy}, Jo and {Brandt}, W.~N. and {Brewington}, Howard and {Brinkmann}, Jon and {Brown}, Peter J. and {Brownstein}, Joel R. and {Bundy}, Kevin and {Busca}, N.~G. and {Carithers}, William and {Carnero}, Aurelio R. and {Carr}, Michael A. and {Chen}, Yanmei and {Comparat}, Johan and {Connolly}, Natalia and {Cope}, Frances and {Croft}, Rupert A.~C. and {Cuesta}, Antonio J. and {da Costa}, Luiz N. and {Davenport}, James R.~A. and {Delubac}, Timoth{\'e}e and {de Putter}, Roland and {Dhital}, Saurav and {Ealet}, Anne and {Ebelke}, Garrett L. and {Eisenstein}, Daniel J. and {Escoffier}, S. and {Fan}, Xiaohui and {Filiz Ak}, N. and {Finley}, Hayley and {Font-Ribera}, Andreu and {G{\'e}nova-Santos}, R. and {Gunn}, James E. and {Guo}, Hong and {Haggard}, Daryl and {Hall}, Patrick B. and {Hamilton}, Jean-Christophe and {Harris}, Ben and {Harris}, David W. and {Ho}, Shirley and {Hogg}, David W. and {Holder}, Diana and {Honscheid}, Klaus and {Huehnerhoff}, Joe and {Jordan}, Beatrice and {Jordan}, Wendell P. and {Kauffmann}, Guinevere and {Kazin}, Eyal A. and {Kirkby}, David and {Klaene}, Mark A. and {Kneib}, Jean-Paul and {Le Goff}, Jean-Marc and {Lee}, Khee-Gan and {Long}, Daniel C. and {Loomis}, Craig P. and {Lundgren}, Britt and {Lupton}, Robert H. and {Maia}, Marcio A.~G. and {Makler}, Martin and {Malanushenko}, Elena and {Malanushenko}, Viktor and {Mandelbaum}, Rachel and {Manera}, Marc and {Maraston}, Claudia and {Margala}, Daniel and {Masters}, Karen L. and {McBride}, Cameron K. and {McDonald}, Patrick and {McGreer}, Ian D. and {McMahon}, Richard G. and {Mena}, Olga and {Miralda-Escud{\'e}}, Jordi and {Montero-Dorta}, Antonio D. and {Montesano}, Francesco and {Muna}, Demitri and {Myers}, Adam D. and {Naugle}, Tracy and {Nichol}, Robert C. and {Noterdaeme}, Pasquier and {Nuza}, Sebasti{\'a}n E. and {Olmstead}, Matthew D. and {Oravetz}, Audrey and {Oravetz}, Daniel J. and {Owen}, Russell and {Padmanabhan}, Nikhil and {Palanque-Delabrouille}, Nathalie and {Pan}, Kaike and {Parejko}, John K. and {P{\^a}ris}, Isabelle and {Percival}, Will J. and {P{\'e}rez-Fournon}, Ismael and {P{\'e}rez-R{\`a}fols}, Ignasi and {Petitjean}, Patrick and {Pfaffenberger}, Robert and {Pforr}, Janine and {Pieri}, Matthew M. and {Prada}, Francisco and {Price-Whelan}, Adrian M. and {Raddick}, M. Jordan and {Rebolo}, Rafael and {Rich}, James and {Richards}, Gordon T. and {Rockosi}, Constance M. and {Roe}, Natalie A. and {Ross}, Ashley J. and {Ross}, Nicholas P. and {Rossi}, Graziano and {Rubi{\~n}o-Martin}, J.~A. and {Samushia}, Lado and {S{\'a}nchez}, Ariel G. and {Sayres}, Conor and {Schmidt}, Sarah J. and {Schneider}, Donald P. and {Sc{\'o}ccola}, C.~G. and {Seo}, Hee-Jong and {Shelden}, Alaina and {Sheldon}, Erin and {Shen}, Yue and {Shu}, Yiping and {Slosar}, An{\v{z}}e and {Smee}, Stephen A. and {Snedden}, Stephanie A. and {Stauffer}, Fritz and {Steele}, Oliver and {Strauss}, Michael A. and {Streblyanska}, Alina and {Suzuki}, Nao and {Swanson}, Molly E.~C. and {Tal}, Tomer and {Tanaka}, Masayuki and {Thomas}, Daniel and {Tinker}, Jeremy L. and {Tojeiro}, Rita and {Tremonti}, Christy A. and {Vargas Maga{\~n}a}, M. and {Verde}, Licia and {Viel}, Matteo and {Wake}, David A. and {Watson}, Mike and {Weaver}, Benjamin A. and {Weinberg}, David H. and {Weiner}, Benjamin J. and {West}, Andrew A. and {White}, Martin and {Wood-Vasey}, W.~M. and {Yeche}, Christophe and {Zehavi}, Idit and {Zhao}, Gong-Bo and {Zheng}, Zheng},
        title = "{The Baryon Oscillation Spectroscopic Survey of SDSS-III}",
      journal = {\aj},
         year = 2013,
        month = jan,
       volume = {145},
       number = {1},
          eid = {10},
        pages = {10},
          doi = {10.1088/0004-6256/145/1/10},
archivePrefix = {arXiv},
       eprint = {1208.0022},
 primaryClass = {astro-ph.CO},
       adsurl = {https://ui.adsabs.harvard.edu/abs/2013AJ....145...10D}
}

@ARTICLE{Anderson2012,
       author = {{Anderson}, Lauren and {Aubourg}, Eric and {Bailey}, Stephen and {Bizyaev}, Dmitry and {Blanton}, Michael and {Bolton}, Adam S. and {Brinkmann}, J. and {Brownstein}, Joel R. and {Burden}, Angela and {Cuesta}, Antonio J. and {da Costa}, Luiz A.~N. and {Dawson}, Kyle S. and {de Putter}, Roland and {Eisenstein}, Daniel J. and {Gunn}, James E. and {Guo}, Hong and {Hamilton}, Jean-Christophe and {Harding}, Paul and {Ho}, Shirley and {Honscheid}, Klaus and {Kazin}, Eyal and {Kirkby}, David and {Kneib}, Jean-Paul and {Labatie}, Antoine and {Loomis}, Craig and {Lupton}, Robert H. and {Malanushenko}, Elena and {Malanushenko}, Viktor and {Mandelbaum}, Rachel and {Manera}, Marc and {Maraston}, Claudia and {McBride}, Cameron K. and {Mehta}, Kushal T. and {Mena}, Olga and {Montesano}, Francesco and {Muna}, Demetri and {Nichol}, Robert C. and {Nuza}, Sebasti{\'a}n E. and {Olmstead}, Matthew D. and {Oravetz}, Daniel and {Padmanabhan}, Nikhil and {Palanque-Delabrouille}, Nathalie and {Pan}, Kaike and {Parejko}, John and {P{\^a}ris}, Isabelle and {Percival}, Will J. and {Petitjean}, Patrick and {Prada}, Francisco and {Reid}, Beth and {Roe}, Natalie A. and {Ross}, Ashley J. and {Ross}, Nicholas P. and {Samushia}, Lado and {S{\'a}nchez}, Ariel G. and {Schlegel}, David J. and {Schneider}, Donald P. and {Sc{\'o}ccola}, Claudia G. and {Seo}, Hee-Jong and {Sheldon}, Erin S. and {Simmons}, Audrey and {Skibba}, Ramin A. and {Strauss}, Michael A. and {Swanson}, Molly E.~C. and {Thomas}, Daniel and {Tinker}, Jeremy L. and {Tojeiro}, Rita and {Maga{\~n}a}, Mariana Vargas and {Verde}, Licia and {Wagner}, Christian and {Wake}, David A. and {Weaver}, Benjamin A. and {Weinberg}, David H. and {White}, Martin and {Xu}, Xiaoying and {Y{\`e}che}, Christophe and {Zehavi}, Idit and {Zhao}, Gong-Bo},
        title = "{The clustering of galaxies in the SDSS-III Baryon Oscillation Spectroscopic Survey: baryon acoustic oscillations in the Data Release 9 spectroscopic galaxy sample}",
      journal = {\mnras},
         year = 2012,
        month = dec,
       volume = {427},
       number = {4},
        pages = {3435-3467},
          doi = {10.1111/j.1365-2966.2012.22066.x},
archivePrefix = {arXiv},
       eprint = {1203.6594},
 primaryClass = {astro-ph.CO},
       adsurl = {https://ui.adsabs.harvard.edu/abs/2012MNRAS.427.3435A}
}

@ARTICLE{Zheng2009,
       author = {{Zheng}, Zheng and {Zehavi}, Idit and {Eisenstein}, Daniel J. and {Weinberg}, David H. and {Jing}, Y.~P.},
        title = "{Halo Occupation Distribution Modeling of Clustering of Luminous Red Galaxies}",
      journal = {\apj},
         year = 2009,
        month = dec,
       volume = {707},
       number = {1},
        pages = {554-572},
          doi = {10.1088/0004-637X/707/1/554},
archivePrefix = {arXiv},
       eprint = {0809.1868},
 primaryClass = {astro-ph},
       adsurl = {https://ui.adsabs.harvard.edu/abs/2009ApJ...707..554Z}
}



\appendix

\section{Galaxy power spectra at the leading order of bias expansion}
\label{sec:galaxy_bias}
Here, we describe the real-space galaxy spectra. Following \citet{McDonald2009},
we employ the galaxy power spectra with next-to-leading order bias terms:
\begin{align}
  P_{\delta_g \delta_g} (k) =& b_1^2 P_{\delta \delta} (k)
  + 2 b_1 b_2 P_{b2, \delta} (k) + 2 b_{s2} b_1 P_{bs2, \delta} (k)
  \nonumber \\
  & + 2 b_{3 \mathrm{nl}} b_1 \sigma_3^2 (k) P_\mathrm{L} (k)
  + b_2^2 P_{b22} (k)
  \nonumber \\
  & + 2 b_2 b_{s2} P_{b2s2} (k) + b_{s2}^2 P_{bs22} (k),
  \\
  P_{\delta_g \theta_g} (k) =& b_1 P_{\delta \theta} (k) + b_2 P_{b2, \theta} (k)
  +b_{s2} P_{bs2, \theta} (k)
  \nonumber \\
  & + b_{3 \mathrm{nl}} \sigma_3^2 (k) P_\mathrm{L} (k),
\end{align}
where $\delta$ is the density contrast of matter,
and $\theta$ is the velocity divergence field of matter,
$P_{\delta \delta} (k)$ and $P_{\delta \theta} (k)$
is the matter density-density auto-, and density-velocity cross-power spectra, respectively,
and $P_\mathrm{L} (k)$ is the linear matter power spectrum.
By assuming no velocity bias, the velocity-velocity auto-power spectrum is given as
\begin{equation}
  P_{\theta_g \theta_g} (k) = P_{\theta \theta} (k) .
\end{equation}
The non-linear matter power spectrum $P_{\delta \delta} (k)$ is
computed with \texttt{HMCODE-2020} \citep{Mead2021}.
For the density-velocity and velocity-velocity power spectra,
we adopt the fitting formulas in \citet{Bel2019}:
\begin{align}
  P_{\theta \theta} =& P_\mathrm{L} (k) e^{-k (a_1 + a_2 k + a_3 k^2)}, \nonumber \\
  P_{\delta \theta} =& \left[ P_{\delta \delta} (k) P_\mathrm{L} (k) \right]^{\frac{1}{2}}
  e^{-k/k_\delta -b k^6},
\end{align}
where the parameters are given as
\begin{align}
  a_1 =& -0.817 + 3.198 \sigma_8 (z) , \\
  a_2 =& 0.877 - 4.191 \sigma_8 (z) , \\
  a_3 =& -1.199 + 4.629 \sigma_8 (z) , \\
  k_\delta/(h\,\mathrm{Mpc}^{-1})  =& \left( -0.017 + 1.496 \sigma_8^2 (z) \right)^{-1}, \\
  b =& 0.091 + 0.702 \sigma_8^2 (z) ,
\end{align}
and $\sigma_8(z)$ is the amplitude of density fluctuation at the scale of $8 \, \hMpc$ at the redshift $z$.
The other bias terms are given as \citep{Beutler2014,GilMarin2015,Beutler2017}:
\begin{align}
  P_{b2,\delta} (k) =& \int \frac{\dd[3]{q}}{(2\pi)^3}
  P_\mathrm{L} (q) P_\mathrm{L} (|\bm{k} - \bm{q}|) F_2 (\bm{q}, \bm{k}-\bm{q}), \\
  P_{b2,\theta} (k) =& \int \frac{\dd[3]{q}}{(2\pi)^3}
  P_\mathrm{L} (q) P_\mathrm{L} (|\bm{k} - \bm{q}|) G_2 (\bm{q}, \bm{k}-\bm{q}), \\
  P_{bs2,\delta} (k) =& \int \frac{\dd[3]{q}}{(2\pi)^3}
  P_\mathrm{L} (q) P_\mathrm{L} (|\bm{k} - \bm{q}|) \nonumber \\
  & \times F_2 (\bm{q}, \bm{k}-\bm{q}) S_2 (\bm{q}, \bm{k}-\bm{q}), \\
  P_{bs2,\theta} (k) =& \int \frac{\dd[3]{q}}{(2\pi)^3}
  P_\mathrm{L} (q) P_\mathrm{L} (|\bm{k} - \bm{q}|) \nonumber \\
  & \times G_2 (\bm{q}, \bm{k}-\bm{q}) S_2 (\bm{q}, \bm{k}-\bm{q}), \\
  P_{b22} (k) =& \frac{1}{2} \int \frac{\dd[3]{q}}{(2\pi)^3}
  P_\mathrm{L} (q) \left[ P_\mathrm{L} (|\bm{k} - \bm{q}|) - P_\mathrm{L} (q) \right], \\
  P_{b2s2} (k) =& -\frac{1}{2} \int \frac{\dd[3]{q}}{(2\pi)^3} P_\mathrm{L} (q) \nonumber \\
  & \times \left[ \frac{2}{3} P_\mathrm{L} (q) - P_\mathrm{L} (|\bm{k} - \bm{q}|)
  S_2 (\bm{q}, \bm{k}-\bm{q})\right], \\
  P_{bs22} (k) =& -\frac{1}{2} \int \frac{\dd[3]{q}}{(2\pi)^3} P_\mathrm{L} (q) \nonumber \\
  & \times \left[ \frac{4}{9} P_\mathrm{L} (q) - P_\mathrm{L} (|\bm{k} - \bm{q}|)
  S_2 (\bm{q}, \bm{k}-\bm{q})^2 \right]. \\
\end{align}
$F_2 (\bm{q}_1, \bm{q}_2)$, $G_2 (\bm{q}_1, \bm{q}_2)$, and $S_2 (\bm{q}_1, \bm{q}_2)$
are symmetrised second-order perturbation theory kernels:
\begin{align}
  F_2 (\bm{q}_1, \bm{q}_2) &= \frac{5}{7} + \frac{\bm{q}_1 \cdot \bm{q}_2}{2 q_1 q_2}
  \left( \frac{q_1}{q_2} + \frac{q_2}{q_1} \right) +
  \frac{2}{7} \left( \frac{\bm{q}_1 \cdot \bm{q}_2}{q_1 q_2} \right)^2, \\
  G_2 (\bm{q}_1, \bm{q}_2) &= \frac{3}{7} + \frac{\bm{q}_1 \cdot \bm{q}_2}{2 q_1 q_2}
  \left( \frac{q_1}{q_2} + \frac{q_2}{q_1} \right) +
  \frac{4}{7} \left( \frac{\bm{q}_1 \cdot \bm{q}_2}{q_1 q_2} \right)^2, \\
  S_2 (\bm{q}_1, \bm{q}_2) &= \left( \frac{\bm{q}_1 \cdot \bm{q}_2}{q_1 q_2} \right)^2
  -\frac{1}{3}.
\end{align}
$\sigma_3^2 (k)$ is given as
\begin{equation}
  \sigma_3^2 (k) = \frac{105}{16} \int \frac{\dd[3]{q}}{(2\pi)^3}
  P_\mathrm{L} (q) \left[ D_2 (-\bm{q}, \bm{k}) S_2 (\bm{q}, \bm{k}-\bm{q})
  +\frac{8}{63} \right],
\end{equation}
where
\begin{equation}
  D_2 (\bm{q}_1 ,\bm{q}_2) = \frac{2}{7} \left[ S_2 (\bm{q}_1, \bm{q}_2) - \frac{2}{3} \right] .
\end{equation}

\section{Conversion functions for multipole expansion}
\label{sec:conversion}
The dependence on the directional cosine $\mu$ in the eTNS 2-loop model except FoG damping the function
is the polynomial at even degress up to the octic, i.e.,
$D_\mathrm{FoG}(k\mu \sigma_\mathrm{v}) \mu^{2n} \ (n = 0, 1, \ldots, 4)$.
Thus, the integration in Eq.~\eqref{eq:multipole_expansion} can be done analytically.
Here, we present the conversion formulae:
\begin{equation}
  p_\ell^\mathrm{L} (n, k) \equiv \frac{2 \ell + 1}{2} \int_{-1}^{+1} \mu^{2n}
  \left( 1+\frac{\alpha \mu^2}{2} \right)^{-2} L_\ell (\mu) \dd{\mu} ,
\end{equation}
where $\alpha = (k \sigma_\mathrm{v})^2$.
This function is analytically computed as
\begin{align}
  p_0^\mathrm{L} (n, k) =& \frac{1}{2}\left[ - \frac{n-1/2}{n+1/2}
  {}_2F_1 \left(1,n+1/2;n+3/2;-\alpha/2 \right) + \frac{2}{\alpha+2} \right] \\
  p_2^\mathrm{L} (n, k) =& \frac{5}{4} \left[ \frac{n-1/2}{n+1/2}
   {}_2F_1 \left(1,n+1/2;n+3/2;-\alpha/2 \right) \right. \nonumber \\
   & \left. -3 \frac{n+1/2}{n+3/2} {}_2F_1 \left(1,n+3/2;n+5/2;-\alpha/2 \right)
   + \frac{4}{\alpha+2} \right] , \\
  p_4^\mathrm{L} (n, k) =& \frac{9}{16} \left[ -3 \frac{n-1/2}{n+1/2}
   {}_2F_1 \left(1,n+1/2;n+3/2;-\alpha/2 \right) \right. \nonumber \\
   & +30 \frac{n+1/2}{n+3/2} {}_2F_1 \left(1,n+3/2;n+5/2;-\alpha/2 \right) \nonumber \\
   & \left. - 35 \frac{n+3/2}{n+5/2} {}_2F_1 \left(1,n+5/2;n+7/2;-\alpha/2 \right)
   + \frac{16}{\alpha+2} \right],
\end{align}
where ${}_2F_1 (a, b, c; z)$ is the hypergeometric function.

\section{Selected results for $\mathrm{H}\alpha$ ELGs}
\label{sec:ha_results}

\begin{figure}[b]
  \includegraphics[width=\columnwidth]{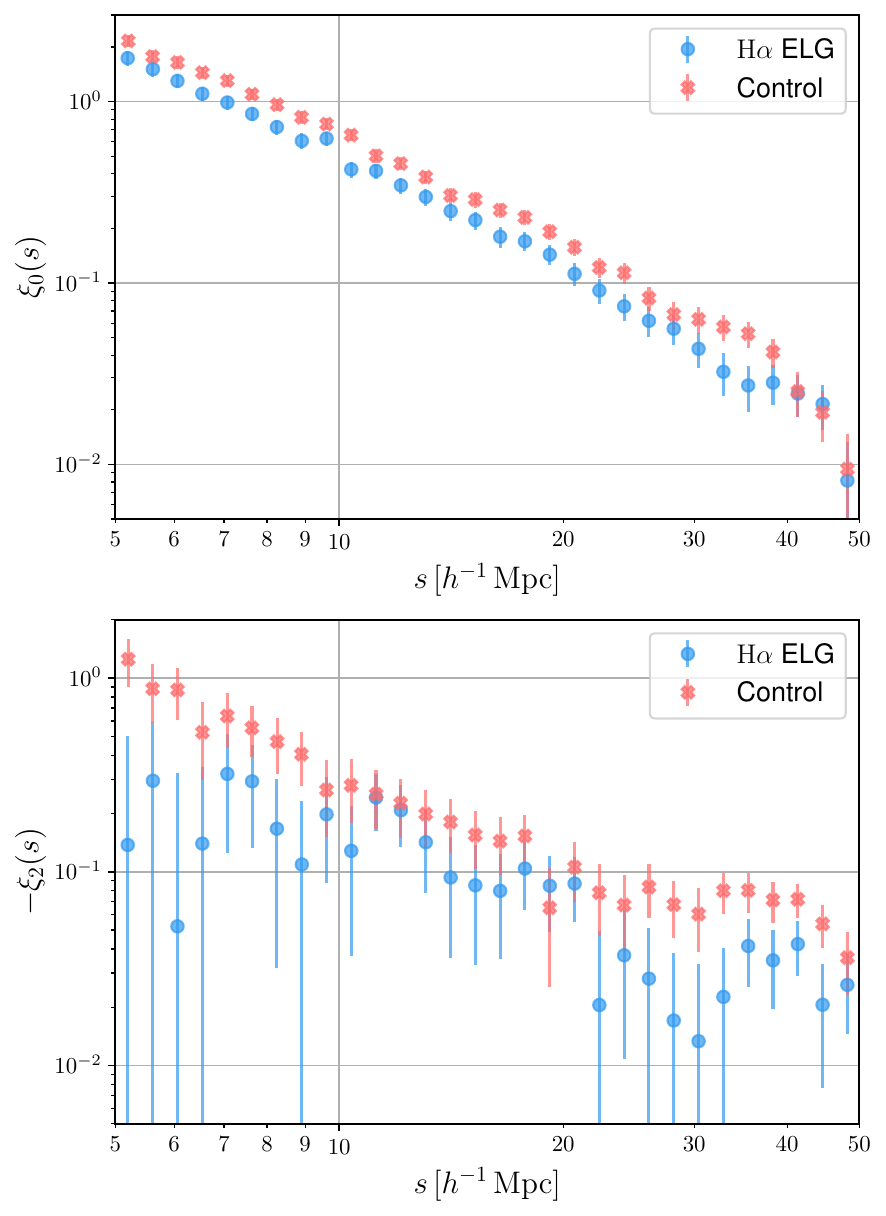}
  \caption{The multipole moments of correlation functions of
  \HA ELGs and the corresponding control sample.}
  \label{fig:xi_multi_HA}
\end{figure}
\begin{figure*}
	\includegraphics[width=\columnwidth]{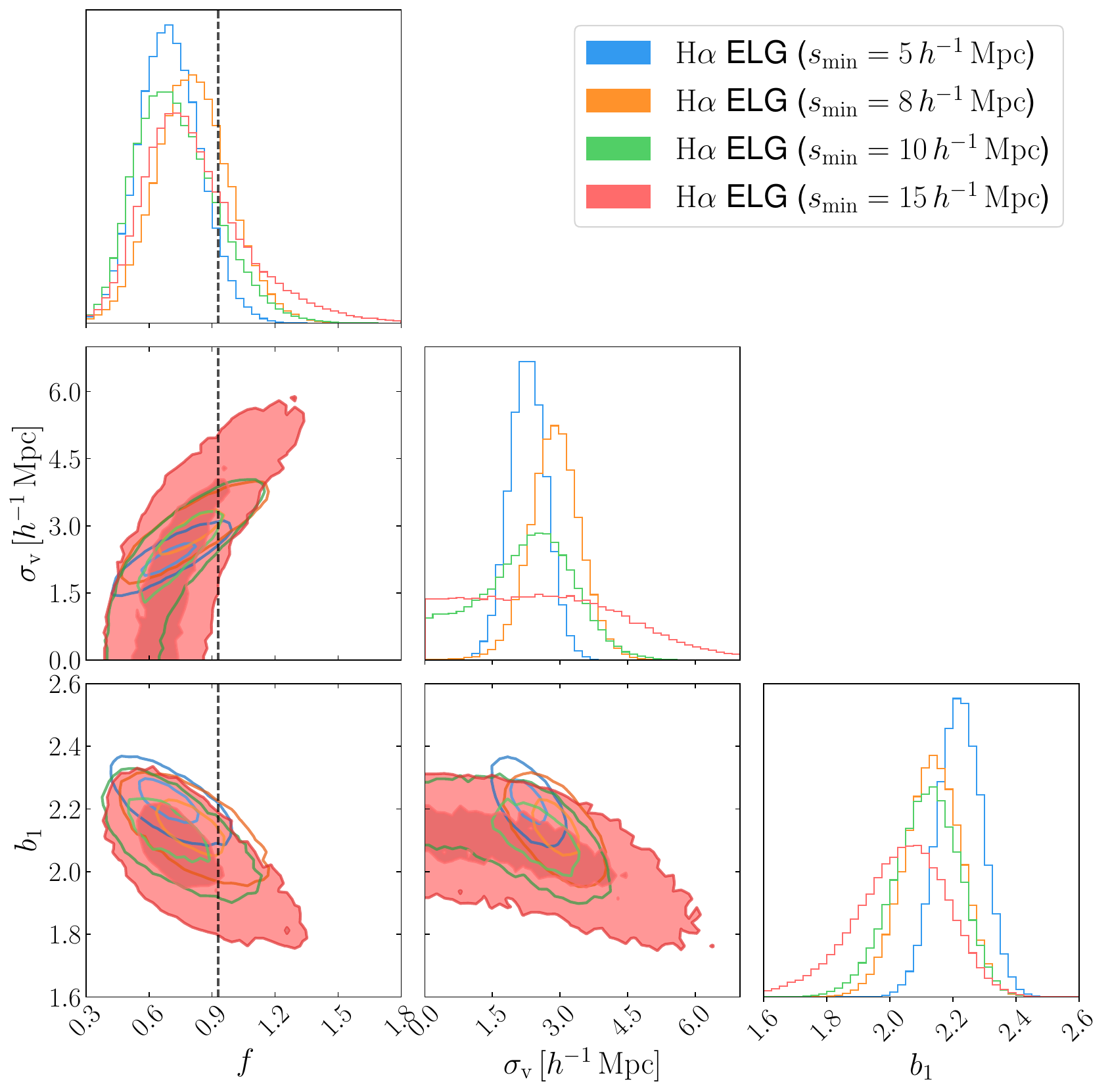}
  \includegraphics[width=\columnwidth]{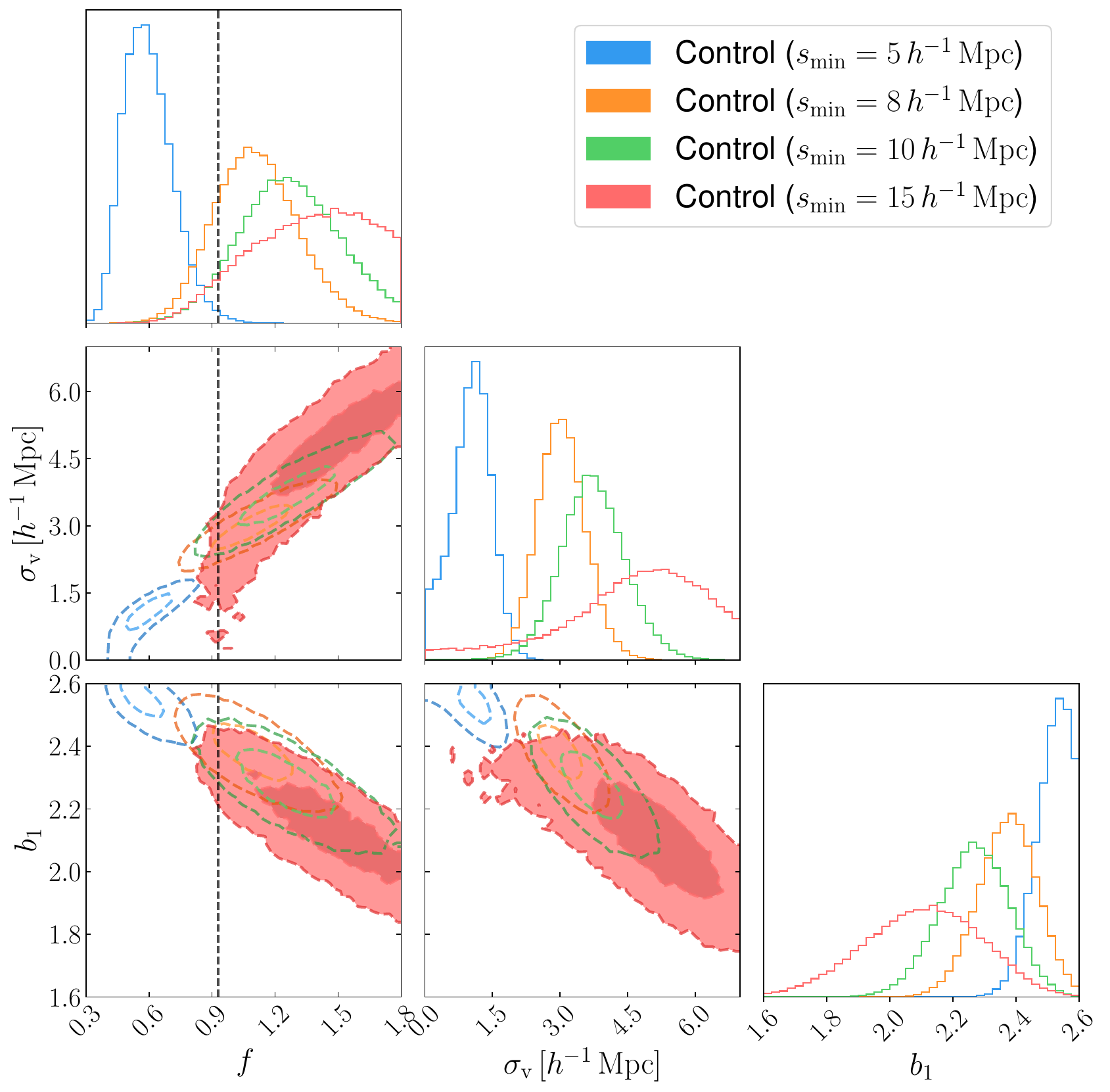}
  \caption{The parameter constraints with monopole and quadrupole moments of
  the anisotropic correlation function for \HA ELGs (left panels) and the control sample (right panels).
  The contours correspond to 1-$\sigma$ and 2-$\sigma$ levels.
  The minimum separation $s_\mathrm{min}$ is varied from $5 \, \hMpc$ to $15 \, \hMpc$.}
  \label{fig:triangle_HA_multi}
\end{figure*}

\begin{figure}
	\includegraphics[width=\columnwidth]{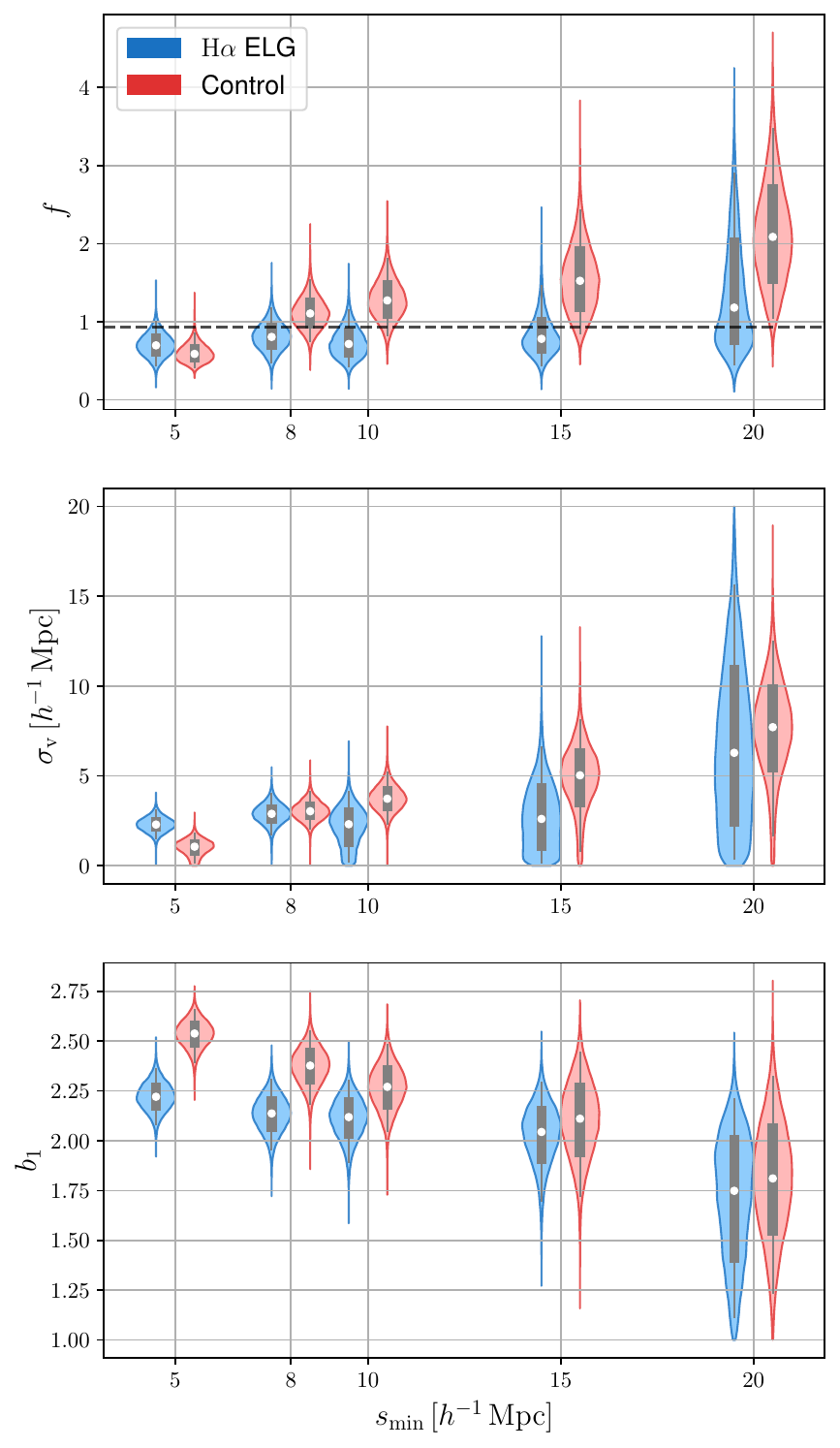}
  \caption{The marginalised posterior distributions with monopole and quadrupole moments of
  the anisotropic correlation function for \OII ELGs and the control sample.
  The minimum separation $s_\mathrm{min}$ is varied from $5 \, \hMpc$ to $20 \, \hMpc$.
  The thick (thin) grey bars correspond to $[15.87, 84.13]$ ($[2.28, 97.72]$) percentiles,
  and the white circles indicate the median values.
  The black dashed line corresponds to the true value of the linear growth rate $f$.
  The data points are slightly shifted horizontally for better visibility.}
  \label{fig:constraints_min_multi_HA}
\end{figure}

Here, we present the results of the anisotropic correlation functions
and the cosmological parameter inference
for \HA ELGs and the corresponding mass-limited control sample.
The measurements of the multipole moments of anisotropic correlation functions
are shown in Figure~\ref{fig:xi_multi_HA}.
The parameter inference results with varying the minimum separation $s_\mathrm{min}$
are presented in Figures~\ref{fig:triangle_HA_multi}.
The marginalised posterior distributions of parameters are shown
in Figure~\ref{fig:constraints_min_multi_HA}.
The qualitative features are similar to those of \OII ELGs
presented in Section~\ref{sec:cosmo_challenge}.
However, the \HA ELG sample contains more galaxies than \OII ELG sample,
and thus, the control sample consists of less massive subhalos.
For the control sample, the inferred linear growth rate is biased even for large $s_\mathrm{max}$
due to the lower accuracy of modelling for such less massive subhalos.
Nevertheless, comparing the inferred value between \HA ELGs and the control sample,
the values from \HA ELGs are always smaller than that of the control sample,
which indicate the velocity bias in \HA ELGs as well as \OII ELGs.

\bsp	
\label{lastpage}
\end{document}